\documentclass[traditabstract]{aa}  
\usepackage{graphicx}
\usepackage{siunitx}
\usepackage[]{natbib}
\usepackage{savesym}
\usepackage{amsmath}
\savesymbol{iint}
\usepackage{txfonts}
\restoresymbol{TXF}{iint}

\begin{document} 

\title{Inverse-Compton Emission from Clusters of Galaxies:\\ Predictions for ASTRO-H}
   
\author{Richard Bartels\inst{} \and 
          Fabio Zandanel\inst{} \and
          Shin'ichiro Ando\inst{} }
\institute{GRAPPA Institute, University of Amsterdam, 1098XH Amsterdam, The Netherlands\\
              \email{r.t.bartels@uva.nl}}

\date{Received 00 00, 0000; accepted 00 00, 0000}

\abstract{The intra-cluster medium of several galaxy clusters hosts large-scale regions of diffuse synchrotron radio emission, known as radio halos and relics, which demonstrate the presence of magnetic fields and relativistic electrons in clusters. These relativistic electrons should also emit X-rays through inverse-Compton scattering off of cosmic microwave background photons. The detection of such a non-thermal X-ray component, together with the radio measurement, would permit to clearly separate the magnetic field from the relativistic electron distribution as the inverse-Compton emission is independent from the magnetic field in the cluster. However, non-thermal X-rays have not been conclusively detected from any cluster of galaxies so far. In this paper, for the first time, we model the synchrotron and inverse-Compton emission of all clusters hosting radio halos and relics for which the spectral index can be determined. We provide constraints on the volume-average magnetic field by comparing with current X-ray measurements. We then estimate the maximum volume-average magnetic field that will allow the detection of inverse-Compton hard X-rays by the ASTRO-H satellite. We found that several clusters are good targets for ASTRO-H to detect their inverse-Compton emission, in particular that corresponding to radio relics, and propose a list of promising targets for which ASTRO-H can test $\ge1$~$\mu$G magnetic fields. We conclude that future hard X-ray observations by the already-operating NuSTAR and the soon-to-be-launched ASTRO-H definitely have the potential to shed light on the long-sought non-thermal hard-X-ray emission in clusters of galaxies.
}


\titlerunning{Inverse-Compton Emission from Clusters of Galaxies}
\authorrunning{R. Bartels et al.}

\maketitle

\section{Introduction}
The observation of diffuse synchrotron emission in clusters of galaxies proves the presence of magnetic fields and relativistic electrons in the intra-cluster medium (ICM). 
This diffuse radio emission is observationally classified into two phenomena: peripheral radio relics that show irregular morphology and appear to trace merger and 
accretion shocks, and radio halos centred on clusters showing a regular morphology resembling that of the thermal X-ray emission (e.g., \citealp{feretti2012}).

The electrons generating the observed radio emission can also produce X-rays by inverse-Compton (IC) scattering off of cosmic microwave background (CMB) photons. IC emission has been searched for extensively in the past, with a few claims of detection (see, e.g., \citealp{rephaeli2002,fusco-femiano2004,rephaeli2006,eckert2008,rephaeli2008}). However, more recent observations did not confirm most of the earlier claims \citep{molendi2008,ajello2009,ajello2010,wik2012,ota2012,ota2013,wik2014, gastaldello2014} and IC emission from clusters remains elusive.

The dominant emission in clusters of galaxies is the thermal bremsstrahlung from the ICM. Since this falls quickly above $\sim10\mathrm{\, keV}$, hard X-rays (HXRs) offer the best prospect for detecting IC emission from clusters. In the coming years, next generation X-ray satellites will increase the chances. Especially, ASTRO-H \citep{takahashi2012,takahashi2014,kitayama2014}, to be launched this year (2015), and the recently launched NuSTAR \citep{harrison2013} are excellent instruments to probe IC emission in HXRs.

In this work, we model, for the first time, the synchrotron and IC emission of {\it all galaxy clusters hosting radio halos and relics} for which the spectral index can be determined. The radio-emitting electrons in clusters can be of different origin (see, e.g., \citealp{brunetti2014} for a review). Our approach is phenomenological, we do not make any assumption on the injection and acceleration history of the relativistic electrons, and we only assume that the same electron distribution generating the observed synchrotron radio emission, IC scatter off of the CMB photons. We compare with current X-ray observations, where available, and provide detailed predictions for the ASTRO-H satellite. In particular, we estimate cap values for the volume-averaged magnetic field below which an IC signal would be detectable by ASTRO-H.

The detection of IC emission from clusters of galaxies is crucial to break the degeneracy in the determination of the electron distribution and magnetic field value in clusters. As the synchrotron radio emission depends on both, while the IC emission is independent of the cluster's magnetic field, the detection of the latter in HXR is of fundamental importance. This not only will shed new light on non-thermal emission in clusters, but also on the impact of non-thermal phenomena on the thermal content of galaxy clusters, which is fundamental in order to robustly use galaxy clusters for cosmological studies \citep{voit2005}.

This paper is organised as follows. We introduce our sample of radio halos and relics in Section~\ref{sec:sample} and describe how we treat the relativistic electrons in Section~\ref{sec:electrons}. In Section~\ref{sec:processes}, we briefly discuss how the non-thermal synchrotron and IC emission is modelled, while, in Section~\ref{sec:backs}, we discuss the considered background emissions. We describe the main characteristics of the ASTRO-H satellite that are of importance for this work in Section~\ref{sec:astroh}, and  explain the procedures adopted for determining the IC detectability by ASTRO-H in Section~\ref{sec:detectability}. Eventually, we present all results in Section~\ref{sec:results}. We present our conclusions in Section~\ref{sec:conclusions}. We work in cgs units and adopt a cosmological model with $\Omega_\mathrm{m} = 0.27$, $\Omega_\Lambda = 0.73$ and $H_0 = 70\mathrm{\,km\, s^{-1}\, Mpc^{-1}}$.

\section{Galaxy cluster sample}
\label{sec:sample}
We analyse the radio halos and relics from the \emph{September2011-Halo} and \emph{September2011-Relic} collections of \cite{feretti2012}. We limit our final sample to objects that have at least two radio measurements at different wavelengths in order to be able to determine the corresponding photon spectral index $\alpha$ and reduce the degeneracy in our modelling. There are a few sources for which the spectral index was estimated from neighbouring wavelengths being subject to a uncertainty and are therefore excluded from our sample (A545 halo, A115 and A548 relics). Additionally, we searched the literature for radio data published after \cite{feretti2012} which led to the inclusion of the Toothbrush (1RXS J0603.3+4214), A3376, A3411, A3667 south-eastern relic, El-Gordo (ACT-CL J0102-4915), MACSJ1149.5+2223, MACSJ1752.0+4440, PLCK G171.9-40.7 and ZwCl2341.1+0000 in our final sample. 

We exclude from the present analysis radio mini-halos hosted in cool-core clusters as the magnetic field estimates in the centre of these environments are high, up to $\sim 10$~$\mu$G \citep{clarke2004,ensslin2006}, implying very low relativistic electron densities for which it would be extremely difficult to aim for a IC detection.\footnote{See, however, \cite{eckert2008}, \cite{pereztorres2009} and \cite{murgia2010} for the case of the Ophiuchus galaxy cluster, hosting a radio mini-halo, for which the detection of non-thermal HXR emission has been claimed.} Table~\ref{tab:sample_halos} and \ref{tab:sample_relics} contain the information regarding our sample of radio halos and relics, respectively, including current X-ray upper limits where available. A note on the latter: most of current HXR upper limits come from observations pointed at the center of clusters; there are a just few works dedicated to radio relics. In most cases, these HXR upper limits contain contributions from both halo and relic regions. 

\begin{table*}[hbt!]
\begin{center}
\caption{\label{tab:sample_halos} Radio halo sample}
\resizebox{\textwidth}{!}{
\begin{tabular}{l  
	S[round-mode = places, round-precision=2] l S[round-mode=places, round-precision=2] l 	
	S[round-mode = places, round-precision=1, scientific-notation = false, round-integer-to-decimal] 
	S[table-format=1.2e1, scientific-notation = true]  
	S[round-mode = places, round-precision=1, scientific-notation = false, round-integer-to-decimal] 
	rrr}
\hline\hline
\phantom{\Big|}
Cluster  & $z$ & Size  & $F_\mathrm{X,~UL}$ & $\Delta E_\mathrm{X,~UL}$ & $\Gamma$ &  \multicolumn{1}{c}{APEC normalisation}& \multicolumn{1}{c}{$kT$}    & Ref.~1& Ref.~2 & Ref.~3\\
\hline\\[-0.5em]
    Bullet & 0.296 & 19    & 1.1   & 50-100 & 1.86  & 7.99E-03 & 11.69 & 1, 2  & 38    & 45 \\
    A0520 & 0.199 & 25    &       &       &       & 1.96E-03 & 4.41  & 3     &       & 45 \\
    A0521 & 0.2533 & 20    &       &       &       & 2.58E-03 & 4.9   & 4, 5, 6 &       & 45 \\
    A0665 & 0.1819 & 79    & 4.2   & 0.6-7 & 1.63  & 8.48E-03 & 6.81  & 7     & 39    & 45 \\
    A0697 & 0.282 & 20    &       &       &       & 1.37E-01 & 8.81  & 6, 8  &       & 45 \\
    A0754 & 0.0542 & 57    & 6.5   & 50-100 & 2     & 3.73E-02 & 9     & 9, 10 & 40    & 46 \\
    A1300 & 0.308 & 18    &       &       &       & 2.78E-03 & 6.27  & 11    &       & 45 \\
    Coma  & 0.0231 & 710   & 4.2   & 20-80 & 2     & 2.18E-01 & 8.38  & 12    & 41    & 46 \\
    A1758a & 0.279 & 28    &       &       &       & 2.53E-03 & 7.06  & 5, 11 &       & 45 \\
    A1914 & 0.1712 & 30    & 1.09  & 50-100 & 2     & 2.16E-02 & 10.53 & 9, 13 & 44    & 46 \\
    A2163 & 0.203 & 104   & 1.7   & 20-80 & 1.5   & 3.23E-02 & 13.29 & 14, 15 & 42    & 46 \\
    A2219 & 0.2256 & 40    &       &       &       & 9.84E-03 & 8.69  & 9, 18 &       & 45 \\
    A2255 & 0.0806 & 79    & 2.73  & 20-80 & 2     & 1.39E-02 & 6.87  & 19, 20 & 43    & 46 \\
    A2256 & 0.058 & 116   & 2.41  & 50-100 & 2     & 4.32E-02 & 7.5   & 21, 22, 23 & 44    & 46 \\
    A2319 & 0.0557 & 198   & 0.67  & 50-100 & 2     & 9.81E-02 & 9.2   & 24, 25    & 40    & 46 \\
    A2744 & 0.308 & 38    & 0.4   & 0.6-7 & 1.66  & 4.40E-03 & 7.24  & 18, 26 & 39    & 45 \\
    A3562 & 0.049 & 50    & 5.52  & 20-80 & 2     & 7.67E-03 & 5.16  & 27, 28 & 43    & 46 \\
    CL0217+70\tablefootmark{a,b} & 0.0655 & 78    &       &       &       & 5.70E-03 & 3.57092783 & 29    &       & 29 \\
    MACSJ0717.5+3745 & 0.5458 & 12    &       &       &       & 9.12E-03 & 11.6  & 30, 31 &       & 47 \\
    MACSJ1752.0+4440\tablefootmark{a,c} & 0.366 & 23    &       &       &       & 1.93E-03 & 6.7   & 32, 33 &       & 32, 33 \\
    PLCK G171.9-40.7\tablefootmark{a} & 0.27  & 8     &       &       &       & 4.48E-03 & 10.65 & 34    &       & 34 \\
    RXCJ1514.9-1523\tablefootmark{a,b} & 0.22  & 40    &       &       &       & 3.44E-03 & 6.42159902 & 35    &       & 48 \\
    RXCJ2003.5-2323\tablefootmark{a,b} & 0.3171 & 20    &       &       &       & 2.06E-03 & 6.78892418 & 36, 37 &       & 48 \\
\hline
\end{tabular}
}
\end{center}
\tablefoot{We estimate the size (in arcmin$^2$) of the halo in the sky in order to scale the ASTRO-H sensitivity correspondingly. In case of approximately spherical sources, this is conservatively estimated to be $\pi D_{LLS}/4$, with $D_{LLS}$ the largest linear size. For elongated sources, this is approximated by a box (or ellipse) whose longest extent (or major axis) is $D_{LLS}$. $F_\mathrm{X,~UL}$ is the current upper limit on the non-thermal hard X-ray flux in units of $10^{-12} \mathrm{\,erg\, cm^{-2}\, s^{-1}}$, with $\Delta E_\mathrm{X,~UL}$ the corresponding energy band. $\Gamma$ is the literature spectral index used to derive $F_\mathrm{X,~UL}$ and then used to calculate the flux density for our analysis. The flux is defined as $f_x = k_x \int_{\nu_\mathrm{min}}^{\nu_\mathrm{max}} \nu_x^{-\Gamma}d\nu_x$, where $k_x$ is a constant. The flux density is $s_{x}(\nu) = k_{x}\nu_x^{-\Gamma}$, with $\Gamma$ the spectral index of the IC component. The ICM temperature, $kT$, in keV is also shown. The three columns of references are for the radio data, the non-thermal hard X-ray upper limit and for the input parameters of the APEC model, respectively.~\tablefoottext{a}{Gas density from \cite{zandanel2014a}.}~\tablefoottext{b}{Temperature from $T-M_{500}$ relation of \cite{mantz2010}.}~\tablefoottext{c}{$M_{500}$ from $L_{X,\mathrm{ROSAT}}-M_{500}$ relation of \cite{mantz2010}.}}
\tablebib{(1)~\cite{liang2000}; (2) \cite{shimwell2014}; (3) \cite{vacca2014}; (4) \cite{brunetti2008}; (5) \cite{giovannini2009}; (6) \cite{macario2013}; (7) \cite{feretti2004}; (8) \cite{weeren2011}; (9) \cite{bacchi2003}; (10) \cite{kassim2001}; (11) \cite{giacintucci2011}; (12) \cite{thierbach2003}; (13) \cite{komissarov1994}; (14) \cite{feretti2001}; (15) \cite{feretti2004}; (16) \cite{kempner2001}; (17) \cite{giovannini2000}; (18) \cite{orru2007}; (19) \cite{feretti1997a}; (20) \cite{govoni2005}; (21) \cite{clarke2006}; (22) \cite{brentjens2008}; (23) \cite{weeren2012}; (24) \cite{feretti1997}; (25) \cite{storm2015}; (26) \cite{govoni2001}; (27) \cite{venturi2003}; (28) \cite{giacintucci2005}; (29) \cite{brown2011}; (30) \cite{bonafede2009}; (31) \cite{pandey-pommier2013};  (32) \cite{bonafede2012}; (33) \cite{weeren2012b}; (34) \cite{giacintucci2013}; (35) \cite{giacintucci2011a}; (36) \cite{venturi2009}; (37) \cite{giacintucci2009}; (38) \cite{wik2014}; (39) \cite{million2009}; (40) \cite{ajello2009}; (41) \cite{wik2011}; (42) \cite{ota2013}; (43) \cite{wik2012}; (44) \cite{ajello2010}; (45) \cite{fukazawa2004}; (46)~\cite{chen2007, pinzke2011}; (47) \cite{ma2008}; (48) \cite{piffaretti2011}.}
\end{table*}
 
\begin{table*}[hbt!]
\begin{center}
\caption{\label{tab:sample_relics} Radio relic sample}
\resizebox{0.99\textwidth}{!}{
\begin{tabular}{l  
	l S[round-mode = places, round-precision=2] l S[round-mode = places, round-precision=1] l 
	S[round-mode = places, round-precision=1, scientific-notation = false, round-integer-to-decimal] 
	S[table-format=1.1e1, scientific-notation = true, round-mode = places, round-precision = 1]  
	S[round-mode = places, round-precision=1, scientific-notation = false, round-integer-to-decimal] 
	rrr}
\hline\hline
\phantom{\Big|}
Cluster  & Source & $z$ & Size  & $F_\mathrm{X,~UL}$ & $\Delta E_\mathrm{X,~UL}$ & $\Gamma$ & \multicolumn{1}{c}{APEC Normalisation}& \multicolumn{1}{c}{$kT$}  & Ref.~1 & Ref.~2 & Ref.~3\\
\hline\\[-0.5em] 
   Toothbrush\tablefootmark{a,c} &       & 0.225 & 18    &       &       &       & 6.40E-04 & 7.8   & 1     &       & 49 \\
    A0013 &       & 0.0943 & 5     &       &       &       & 2.98E-03 & 6     & 2     &       & 50 \\
    A0085 &       & 0.0551 & 26    & 2.51  & 50-100 & 2     & 0.0051 & 6.1   & 2     & 43    & 51 \\
    A0521 &       & 0.2533 & 9     &       &       &       & 5.94E-05 & 4.9   & 3     &       & 52 \\
    A0610\tablefootmark{a,b,e} &       & 0.0954 & 8     &       &       &       & 1.30E-04 & 2.43715373 & 4     &       & 54 \\
    AS753\tablefootmark{a} &       & 0.014 & 318   &       &       &       & 4.20E-04 & 2.5   & 5     &       & 5 \\
    A0754 &       & 0.0542 & 100     &       &       &       & 9.00E-03 & 9     & 6, 7  &       & 6 \\
    A0781\tablefootmark{a,b} &       & 0.3004 & 6     &       &       &       & 3.40E-05 & 7.6   & 8     &       & 55 \\
    A1240 & N     & 0.159 & 12    &       &       &       & 3.70E-06 & 6     & 9     &       & 56 \\
          & S     & 0.159 & 24    &       &       &       & 1.50E-06 &       & 9     &       &  \\
    A1300 &       & 0.3072 & 6     &       &       &       & 1.80E-05 & 6.27  & 10, 11 &       & 52 \\
    A1367 &       & 0.022 & 52    & 8.24  & 20-80 & 2     & 2.70E-04 & 3.55  & 12    & 63    & 51 \\
    A1612\tablefootmark{a,b} &       & 0.179 & 8     &       &       &       & 3.30E-06 & 4.9   & 13    &       & 57 \\
    Coma  &       & 0.0231 & 400   & 0.32  & 0.3-10 & 2.2   & 0.00085 & 8.38  & 14    & 45    & 51 \\
    A1664 &       & 0.1283 & 51    &       &       &       & 6.00E-04 & 6.8   & 8     &       & 27 \\
    A2048\tablefootmark{a,b,d} &       & 0.0972 & 7     &       &       &       & 4.00E-04 & 4.20695596 & 15    &       & 58 \\
    A2061 &       & 0.0784 & 24    &       &       &       & 6.60E-04 & 4.52  & 13, 16 &       & 59 \\
    A2063 &       & 0.0349 & 3     & 7.58  & 20-80 & 2     & 3.60E-03 & 3.68  & 17    & 44    & 51 \\
    A2163 &       & 0.203 & 5     &       &       &       & 1.61E-05 & 13.29 & 18, 19 &       & 51 \\
    A2255 &       & 0.0806 & 32    &       &       &       & 3.20E-04 & 6.87  & 20    &       & 51 \\
    A2256 &       & 0.058 & 152   &       &       &       & 0.023 & 7.5   & 21, 22, 23 &       & 51 \\
    A2345\tablefootmark{a,b} & E     & 0.1765 & 38    &       &       &       & 2.60E-04 & 6.51190837 & 24    &       & 53 \\
          & W     &       & 35    &       &       &       & 8.50E-04 &       & 24    &       &  \\
    A2433\tablefootmark{a,b,e} &       & 0.108 & 11    &       &       &       & 2.70E-05 & 1.79957964 & 25    &       & 60 \\
    A2744 &       & 0.308 & 19    & 0.4   & 0.6-7 & 1.66  & 6.62E-05 & 7.24  & 26, 27 & 46    & 52 \\
    A3376 & E     & 0.0456 & 324   & 3.5   & 8-Apr & 2     & 5.40E-03 & 4.3   & 28    & 47    & 51 \\
          & W     &       & 122   & 1.1   & 8-Apr & 2     & 3.10E-05 &       & 28    & 47    &  \\
    A3411\tablefootmark{a} &       & 0.1687 & 62    &       &       &       & 3.20E-03 & 6.4   & 29    &       & 29 \\
    A3667 & NW     & 0.0556 & 250   & 0.62  & Oct-40 & 1.8   & 6.60E-04 & 7     & 30, 31, 32 & 48    & 51 \\
          & SE    &       & 175   &       &       &       & 1.90E-03 &       & 31, 32 &       &  \\
    A4038 &       & 0.03  & 12    & 7.43  & 20-80 & 2     & 2.80E-04 & 3.15  & 2, 8  & 44    & 51 \\
    El-Gordo\tablefootmark{a} & E     & 0.87  & 3     &       &       &       & 2.70E-02 & 14.5  & 33    &       & 61 \\
          & NW    &       & 3     &       &       &       & 2.70E-02 &       & 33    &       &  \\
          & SE    &       & 3     &       &       &       & 2.70E-02 &       & 33    &       &  \\
    Sausage\tablefootmark{a,b,c} &       & 0.1921 & 18    &       &       &       & 1.10E-05 & 5.55286496 & 34    &       & 57 \\
    MACSJ1149.5+2223\tablefootmark{a} & E     & 0.544 & 2     &       &       &       & 2.49E-07 & 14.5  & 35    &       & 35 \\
          & W     &       & 2     &       &       &       & 4.80E-07 &       & 35    &       &  \\
    MACSJ1752.0+4440\tablefootmark{a,c} & NE    & 0.366 & 8     &       &       &       & 6.30E-06 & 6.7   & 35, 36 &       & 35, 36 \\
          & SW    &       & 3     &       &       &       & 1.20E-05 &       & 35, 36 &       &  \\
    PLCK G287.0+32.9\tablefootmark{a,c} & N     & 0.39  & 7     &       &       &       & 7.40E-05 & 12.86 & 37    &       & 57 \\
          & S     & 0.39  & 6     &       &       &       & 5.20E-07 & 12.86 & 37    &       &  \\
    RXCJ1314.4-2515 & E     & 0.2474 & 8     &       &       &       & 3.50E-06 & 7.2   & 38, 39 &       & 57 \\
          & W     &       & 15    &       &       &       & 2.30E-04 &       & 38, 39 &       &  \\
    ZwCl0008.8-5215\tablefootmark{a,b} & E     & 0.1032 & 36    &       &       &       & 1.30E-04 & 4.98143833 & 40    &       & 53 \\
    ZwCl2341.1+0000\tablefootmark{a} & N     & 0.27  & 4     &       &       &       & 8.70E-05 & 9.3   & 41, 42 &       & 62 \\
          & S     &       & 20    &       &       &       & 1.90E-04 &       & 41, 42 &       &  \\		
\hline
\end{tabular}
}
\end{center}
\tablefoot{Columns are as in Table~\ref{tab:sample_halos}, with the addition of the second column which indicates the specific relic to which we are referring to.~\tablefoottext{a}{Gas density from \cite{zandanel2014a}.}~\tablefoottext{b}{Temperature from $T-M_{500}$ relation of \cite{mantz2010}.}~\tablefoottext{c}{$M_{500}$ from $L_{X,\mathrm{ROSAT}}-M_{500}$ relation of \cite{mantz2010}.}~\tablefoottext{d}{$M_{500}$ from $L_{X,\mathrm{bolometric}}-M_{500}$ of \cite{mantz2010}.}~\tablefoottext{e}{$M_{500}$ scaled from $M_{200}$.}}
\tablebib{(1)~\cite{weeren2012a}; (2) \cite{slee2001}; (3) \cite{macario2013}; (4) \cite{giovannini2000}; (5) \cite{subrahmanyan2003}; (6) \cite{bacchi2003}; (7) \cite{kassim2001}; (8) \cite{kale2012a}; (9) \cite{bonafede2009a}; (10) \cite{reid1999}; (11) \cite{giacintucci2011}; (12) \cite{gavazzi1983}; (13) \cite{weeren2011}; (14) \cite{thierbach2003}; (15) \cite{weeren2011a};  (16) \cite{kempner2001}; (17) \cite{komissarov1994} (18) \cite{feretti2001}; (19) \cite{feretti2004}; (20) \cite{feretti1997a}; (21) \cite{clarke2006}; (22) \cite{brentjens2008}; (23) \cite{weeren2012}; (24) \cite{bonafede2009a}; (25) \cite{cohen2011};  (26) \cite{orru2007}; (27) \cite{govoni2001}; (28) \cite{kale2012}; (29) \cite{weeren2013}; (30) \cite{rottgering1997}; (31) \cite{hindson2014}; (32) \cite{johnston-hollitt2004}; (33) \cite{lindner2014}; (34) \cite{stroe2014}; (35) \cite{bonafede2012}; (36) \cite{weeren2012b}; (37) \cite{bagchi2011}; (38) \cite{feretti2005}; (39) \cite{venturi2007}; (40) \cite{weeren2011b}; (41) \cite{weeren2009a}; (42) \cite{bagchi2002}; (43) \cite{ajello2010}; (44) \cite{wik2012}; (45) \cite{feretti2006}; (46) \cite{million2009}; (47) \cite{kawano2009}; (48) \cite{nakazawa2009}; (49) \cite{ogrean2013}; (50)\cite{juett2008}; (51) \cite{chen2007, pinzke2011}; (52) \cite{fukazawa2004} (53) \cite{planck2013a}; (54) \cite{yoon2008}; (55) \cite{sehgal2008}; (56) \cite{barrena2009, cavagnolo2009};  (57) \cite{piffaretti2011}; (58) \cite{shen2008}; (59) \cite{marini2004}; (60) \cite{popesso2007}; (61) \cite{menanteau2012}; (62) \cite{boschin2013}; (63) \cite{henriksen2001}.}
\end{table*}

\section{The electron population distribution}
\label{sec:electrons}
As anticipated above, we take a phenomenological approach regarding the radio-emitting electrons. We assume a power-law distribution of electrons,
\begin{equation}
	N(\gamma, \theta) = K_0 \gamma^{-\left(2\alpha +1\right)} \frac{\sin \theta}{2},
	\label{eq:N}
\end{equation}
with $K_0$ the normalisation, $\gamma$ the Lorentz factor and
$\alpha$ the photon spectral index, where we consider an isotropic
distribution of pitch angles $\theta$. The electron distribution is then
integrated between $\gamma_\mathrm{min}$ and $\gamma_\mathrm{max}$. In
some of the analysed cases, where spectral steepening is clearly
observed, $\gamma_\mathrm{min}$ and $\gamma_\mathrm{max}$ can be
determined from the radio data. However, more often there is no evidence
for such feature, in which case the cutoffs are fixed to theoretically
motivated values. 

The low energy cutoff is estimated from the lifetime of relativistic
electrons in the ICM. At low energies ($\gamma \lesssim10^2$) Coulomb
cooling dominates, whereas at high energies IC cooling dominates
\citep{rephaeli1979, sarazin1999}. For typical values in the center of
clusters, i.e., a thermal electron number density of
$n_\mathrm{e}\sim10^{-3}\mathrm{\,cm^{-3}}$ and $B=1\mathrm{\,\mu G}$,
this yields a maximum lifetime of $\gamma \approx 300$. However,
outskirts of the cluster are characterised by lower ICM densities which
implies $\gamma \approx 200$. Therefore, the low energy cutoff is fixed
to $\gamma_\mathrm{min} =300$ for halos, and to $200$ for relics.
The high-energy cutoff $\gamma_\mathrm{max}$ is more arbitrary given
that we do not make assumptions on the injection and acceleration history 
of the radio-emitting electrons. If no spectral steepening at high radio frequencies 
is observed, it is fixed to $\gamma_\mathrm{max}=2\times10^5$, corresponding to 
electron energies of 100~GeV. Anyhow, as we will also show in Sect.~\ref{sec:results}, 
our conclusions are not affected by changing $\gamma_\mathrm{min}$ and/or $\gamma_\mathrm{max}$,
unless, of course, spectral steepening is observed.

\section{Radiative processes}
\label{sec:processes}
In this work, we assume that the magnetic field is tangled on scales much smaller than the observed emitting volume, therefore, adopting an isotropic distribution of magnetic field orientations. Such a distribution is well motivated by Faraday rotation (FR) measurements in clusters, which indicate that the coherence scales of magnetic fields are of the order $10\mathrm{\,kpc}$, much smaller than the typical Mpc-size of radio halos and relics \citep{carilli2002, murgia2004,murgia2010}. The synchrotron flux density is then given by
\begin{equation}
	 S_\mathrm{SYNC} =\frac{V}{4 \pi D_L^2}  \int_{\gamma_\mathrm{min}}^{\gamma_\mathrm{max}}\int_0^\pi  P(\nu, \gamma, \theta) N(\gamma, \theta) \frac{\sin\theta}{2} \mathrm{d}\gamma \mathrm{d}\theta\, ,
	 \label{eq:Ssyn}
\end{equation}
where $P(\nu, \gamma, \theta)$ is the synchrotron spectrum emitted by a single electron \citep{blumenthal1970}, $V$ is the emitting volume and $D_{L}$ is the luminosity distance. Note that the factor $\sin\theta/2$ is due to our assumption on the magnetic field distribution. The presence of this factor leads to slightly higher estimates of the magnetic field with respect to assuming coherent magnetic fields on large scales (e.g., \citealp{wik2009}, \citealp{ajello2009} and \citealp{ajello2010}). This assumption of tangled magnetic fields releases part of the tension between the typically high FR measurements and equipartition estimates of the magnetic field on the one hand, and claimed IC detections on the other. Indeed, there are several uncertainties that affect magnetic field estimates from IC emission, and, therefore, the comparison with results from FR measurements. While some of these sources of uncertainties will be discussed in more detail in the next sections, we remind the reader to, e.g., \cite{petrosian2001}, \cite{brunetti2003} and  \cite{bonafede2010} for extensive discussions.

Low-energy photons scattering off of relativistic electrons can produce X-rays through IC emission. We assume that all incoming photons are from the CMB.\footnote{Note that synchrotron-self Comptonization is also possible (e.g., \citealp{ensslin2001}).} While starlight at the very centre of galaxy clusters can be as important as the CMB, this is not true when considering larger areas, as the ones corresponding to radio halos, or the clusters' outskirts where radio relics lie (see, e.g., \citealp{pinzke2011}). The IC flux density is then given by
\begin{equation}
	 S_{\mathrm{IC}} =\frac{V}{4 \pi D_L^2} \int_0^\infty \int_{\gamma_\mathrm{min}}^{\gamma_\mathrm{max}} N(\gamma) F_{IC}(\epsilon, \gamma, \epsilon_1)\mathrm{d}\gamma \mathrm{d}\epsilon\, ,
	 \label{eq:IC}
\end{equation}
where $\epsilon$ is the incoming photon energy, $\epsilon_1$ is he outgoing photon energy, and $F_{IC}$ is the IC kernel function \citep{blumenthal1970} for the Thomson limit, valid for the energy range of interest here. Note that here $N(\gamma)=\int_0^\pi N(\gamma,\theta) \mathrm{d}\theta$.

We conclude this section with an estimate of which HXR IC energies correspond to electrons emitting synchrotron radiation at a given frequency. This depends on the magnetic field strength. The average energy of an IC up-scattered photon is $\left< \epsilon_1 \right>= (4/3) \, \gamma^2 \left< \epsilon\right> \approx 8.6\times10^{-7} \gamma^2 \mathrm{\,keV}$, where we assumed the incoming CMB photon to be at redshift 0. We can use the monochromatic approximation for the synchrotron emission, which yields $(\nu/1\mathrm{\,GHz}) = 1.05\times10^{-9} \, \gamma^2 \, (B/1\mathrm{\,\mu G})$. In case of $B = 1\mathrm{\,\mu G}$, synchrotron radiation at $\sim100\,(20)\mathrm{\,MHz}$ is produced by the electrons that generate IC emission at $80\,(20)\mathrm{\,keV}$, while for  $B = 5\mathrm{\,\mu G}$ one finds that $\sim480\,(120)\mathrm{\,MHz}$ radio waves correspond to $80\,(20)\mathrm{\,keV}$ X-rays. Therefore, low-frequency radio observations are crucial in order to precisely predict HXR IC emission, as will be discussed in Sect.~\ref{sec:results}, and to eventually clarify whether our assumption of Sect.~\ref{sec:electrons} regarding the electron distribution holds down to low energies.

\section{Background emission}
\label{sec:backs}
In this section, we discuss the main background emission for IC searches in clusters. These are the thermal ICM bremsstrahlung and the cosmic X-ray background (CXB) emission. In Section~\ref{sec:detectability}, we will also introduce the instrumental background.

\subsection{Thermal bremsstrahlung from ICM}
\label{sec:brem}
In order to model the thermal bremsstrahlung emission from the ICM, we use the APEC code \citep{smith2001} as provided in the XSPEC software \citep{arnaud1996}. The input parameters are the ICM temperature (taken, for most nearby clusters, from \citealp{chen2007}), the redshift $z$, the metal abundance, which we fix to $0.3$ times the solar abundance for simplicity \citep[e.g., ][]{bohringer2010}, and the normalisation. The latter contains information about the ICM gas distribution that, for our purposes, can be approximated with a beta model (e.g., \citealp{cavaliere1976}),
\begin{equation}
n_\mathrm{e}(r) = \left[ \sum_i n_{0,i}^2 \left(1 + \frac{r^2}{r_{c,i}^2}\right)^{-3 \beta_i}\right]^{1/2}\,,
\label{eq:ne}
\end{equation}
where $n_0$ is the central ICM electron density, and $r_c$ is the core radius. Most clusters are well described by a single beta model ($i = 1$), whereas others are better described by a double beta model ($i=1,2$). The values of $n_0$, $r_c$ and $\beta$ are derived from X-ray observations. For most nearby clusters, these values are taken from \cite{pinzke2011}.

In the few cases where we cannot find a gas density model in the literature, we adopt the phenomenological ICM model of \cite{zandanel2014a} which provides a gas density once the mass $M_{500}$ of the cluster is known. In the few cases where no mass estimate is present in the literature, we adopt the X-ray luminosity-mass relation from \cite{mantz2010}. If there is no ICM temperature available in the literature, we also use the mass-temperature relation of \cite{mantz2010}. Obviously, such approximations are a source of uncertainty when used in particular cases and, therefore, the modelling of those clusters should be taken with caution. This is also true in general for relics where the gas density profiles have been inferred from X-ray observations of the clusters' center.

\begin{figure}[t!]
\centering
\includegraphics[width=0.4\textwidth]{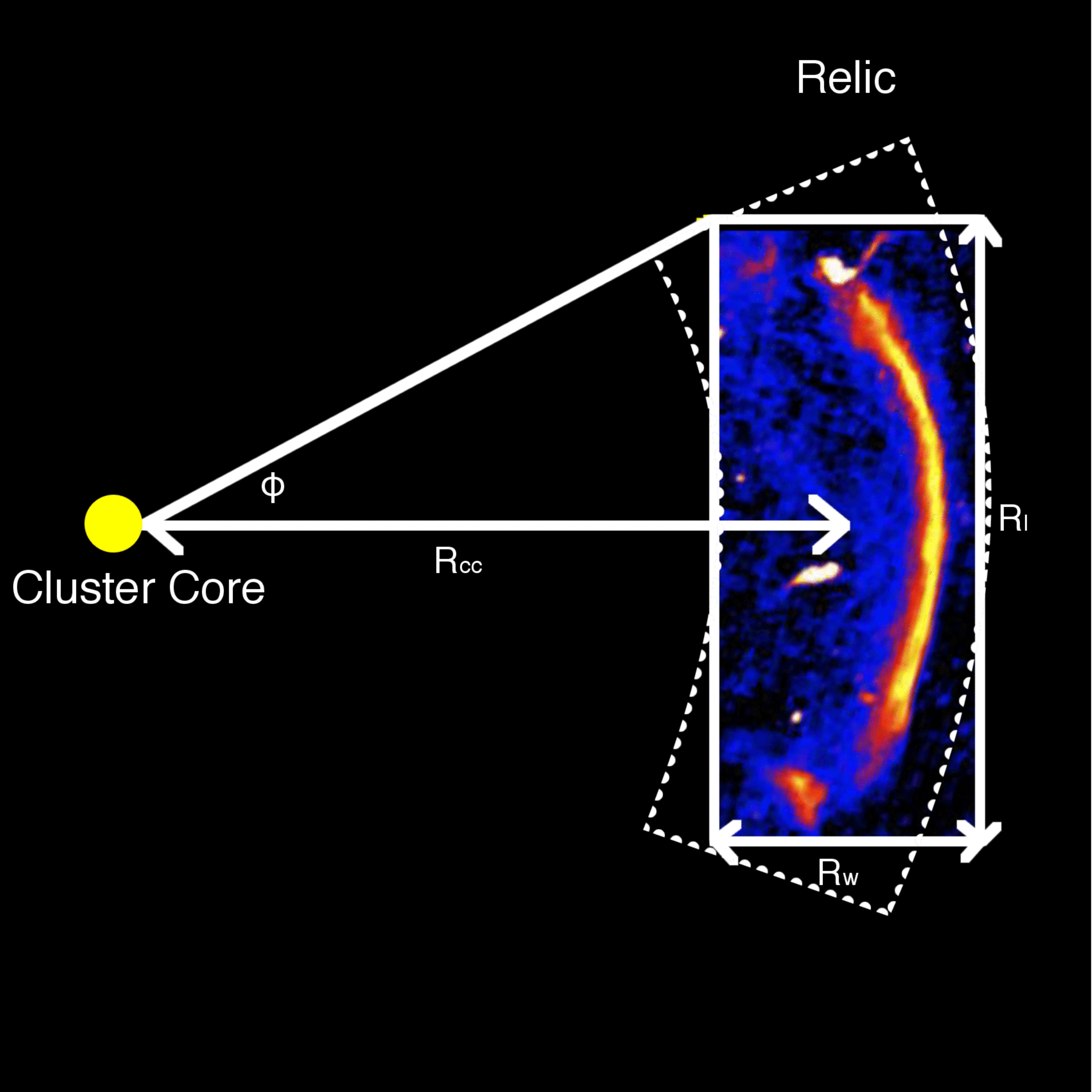}
\caption{\footnotesize{The geometry used in determining the normalisation for the APEC modelling of the radio relic regions (Eq.~\ref{eq:relicnorm}). $R_{cc}$ corresponds to the distance between the relic and the cluster center as projected on the sky. $R_{l}$ is the relic largest linear size and $R_w$ is the relic width. For relics that are classified as roundish by \cite{feretti2012} we use $R_w = R_l$. $\phi$ is given by $\tan^{-1}\left(\frac {0.5 R_{l}}{R_{cc} - 0.5 R_w}\right)$. In result, the dashed curve is the solid angle over which we integrate. The relic image, corresponding to the so-called Sausage, is adapted from \cite{rottgering2013}.}}
\label{fig:relicgeom}
\end{figure}

In the case of radio halos, the bremsstrahlung luminosity, $L_{\mathrm{B}}$, is obtained by integrating over the full extent of the halo, where we approximate the halo to be spherical and use the largest linear size as given in \cite{feretti2012} as halo diameter. In the case of radio relics in the outskirts of clusters, we only integrate over the relic region as:
\begin{equation}
L_\mathrm{B} \sim (1 - \cos \phi) \int_{ R_{cc} - 0.5 R_w}^{ R_{cc} + 0.5 R_w} n_\mathrm{e}(r)^2 \, r^2  \, \mathrm{d}r \,, 
\label{eq:relicnorm}
\end{equation} 
where $R_{cc}$ is the distance to the cluster center, $R_w$ is the approximate width of the relic, and $\phi = \tan^{-1}[0.5 R_{l}/(R_{cc} - 0.5 R_w)]$ with $R_l$ the largest linear size of the relic (see Fig.~\ref{fig:relicgeom}). For relics that are classified as roundish by \cite{feretti2012}, we use $R_w = R_l$.

The parameters that enter the APEC model are shown in Tables~\ref{tab:sample_halos} and \ref{tab:sample_relics} for all the clusters in our sample.

\subsection{Cosmic X-ray background}
\label{sec:cxb}
The CXB \citep[e.g.,][]{gilli2007} is another important background for the case of radio relics in the outskirts of clusters, where it often dominates over the thermal ICM emission. We assume the CXB to be isotropic across the sky and model it in the $0.5-100\mathrm{\,keV}$ regime as follows \citep{gruber1992, comastri1995}:
\begin{subequations}
	\begin{gather}
	13.6 \, E^{-1} \,\, \mathrm{for} \, \,  0.5 \leq E < 2 \mathrm{\,keV}, \nonumber
	\end{gather}
	\begin{gather}
	7.877 \, E^{-0.29} \, e^{-E/(41.13\mathrm{\,keV})} \, \, \mathrm{for} \, \, 2 \leq E < 60 \mathrm{\,keV},  \nonumber
	\end{gather}
	\begin{gather}
	1652 \, E^{-2} + 1.754 \, E^{-0.7} \, \, \mathrm{for} \, \, 60 \leq E \leq 100 \mathrm{\,keV}, \nonumber
	\end{gather}
\end{subequations}
in units of $\mathrm{\,keV\,cm^{-2}\,s^{-1}\,sr^{-1}\,keV^{-1}}$ and
with $E$ the photon energy in units of keV. Note that below $0.5\mathrm{\,keV}$ the galactic emission becomes an important source of background, but these low energies are not considered in this work.

\section{The ASTRO-H and NuSTAR satellites}
\label{sec:astroh}
In this section, we describe the characteristics of ASTRO-H that are relevant for this work. While we provide detailed predictions for the IC-emission detection by ASTRO-H, we also discuss in the same context the recently launched NuSTAR satellite.

\subsection{ASTRO-H}
ASTRO-H is a next generation X-ray satellite that is scheduled for launch in 2015 \citep{takahashi2012,takahashi2014}. The instruments that are of interest for our purposes are the Hard X-ray Imager (HXI) and the Soft X-ray Imager (SXI). In Table~\ref{tab:astroh}, we summarise the main characteristics of these instruments.

\begin{table}[hbt!]
    \caption{\label{tab:astroh}Properties of the SXI and HXI detectors that will be on-board of ASTRO-H.}
\centering
\begin{tabular}{lccc}
\hline\hline
\phantom{\Big|}
& SXI & HXI \\
\hline\\[-0.5em] 
    Energy range [$\mathrm{keV}$]             & 0.4--12.0 & 5--80 \\
    Angular resolution [$\mathrm{arcmin}$] & 1.3   & 1.7 \\
    Field of view (FoV) [$\mathrm{arcmin^2}$]      & $38\times38$ & $9\times9$ \\
    Energy resolution [$\mathrm{eV}$]        & 150 @$6\mathrm{\,keV}$    &  $< 2000$ @$60\mathrm{\,keV}$ \\
    Effective area [$\mathrm{cm^2}$]          & 360 @$6\mathrm{\,keV}$    &  300 @$30\mathrm{\,keV}$ \\
\hline
\end{tabular}
\tablefoot{
Values adopted from the ASTRO-H Quick Reference (http://astro-h.isas.jaxa.jp/ahqr.pdf).
}
\end{table}

To estimate the detectability of the non-thermal X-ray component in galaxy clusters, we make use of the sensitivity curves as published by the ASTRO-H collaboration for both point and extended sources \citep[ASTRO-H Quick Reference,\footnote{http://astro-h.isas.jaxa.jp/ahqr.pdf}][]{takahashi2010, takahashi2012}. We adopt the sensitivity curve for $1\mathrm{\,Ms}$ of observation, and scale it to other observation times as $\sqrt{\text{time}}$. Note that the above mentioned ASTRO-H sensitivity curves are $3\sigma$ sensitivities, and that the sensitivity curve for extended sources assumes a uniform source of $1\mathrm{\, deg^2}$.

We assume, for simplicity, that halos and relics are uniform sources. In order to scale the sensitivity to the size, $\Omega \mathrm{\, arcmin^2}$, of a given object, we proceed as follows. For sources that are smaller than the HXI field of view, we scale the point-source sensitivity as $\sqrt{\Omega/2\mathrm{\,arcmin^2}}$, where $2\mathrm{\,arcmin^2}$ is roughly the pixel size. For sources that are larger, we apply a linear scaling to the $1\mathrm{\, deg^2}$ diffuse source sensitivity as $\Omega/3600\mathrm{\,arcmin^2}$ (Hiroki Akamatsu, Madoka Kawaharada, private communication).

\subsection{NuSTAR}
The NuSTAR satellite was launched on June 13, 2012. Its performances in HXR should be comparable with ASTRO-H \citep{harrison2013}, with the exception that ASTRO-H will be equipped with a pre-collimator designed to mitigate stray light \citep{mori2012}. 

Recently, \cite{wik2014} pointed out that a possible IC signal has to be extracted from both the thermal and instrumental background, with the former dominating at low and the latter at high energies. In fact, the instrumental background is likely to dominate the count rate in the HXR. Since the instrumental background is modelled by a power-law, any claim of IC detection is extremely sensitive to the applied model. While the HXI detector onboard ASTRO-H will not improve much upon NuSTAR - in fact, our predictions are approximately valid in both cases - the presence of the SXI detector will greatly improve the modelling of the thermal component in clusters over the NuSTAR results. 

\section{Multi-band modelling and HXR detectability}
\label{sec:detectability}
For any given object in Tables~\ref{tab:sample_halos} and \ref{tab:sample_relics}, we proceed as follows. If there exist an upper limit on the non-thermal hard X-ray flux $F_\mathrm{X,UL}$, we calculate the corresponding lower limit on the volume-average magnetic field $B_\mathrm{V}$. First, we fix $K_0$ by fitting the IC spectrum of Eq.~\ref{eq:IC} to $F_\mathrm{X,UL}$. Next, we fit the synchrotron emission of Eq.~\ref{eq:Ssyn} to the radio data\footnote{Note that we neglect the possible effect at high frequencies of thermal Sunyaev-Zeldovich (SZ) decrement (e.g., \citealp{ensslin2002}) that, however, should have a small impact (e.g., \citealp{brunetti2013}). Anyhow, ignoring the SZ decrement implies more conservative results.}, with $B_\mathrm{V}$ as free parameter. Depending on whether spectral steepening at low and/or high frequencies is observed, $\gamma_\mathrm{min}$ and/or $\gamma_\mathrm{max}$ are also determined from the data (see Sect.~\ref{sec:electrons}). The spectral index $\alpha$ is taken from the literature, unless it returns a poor fit, in which case we determine a new spectral index. The resulting broadband spectrum provides a lower limit on the magnetic field.

We then compare the background spectra (thermal bremsstrahlung and the CXB) to the properly scaled ASTRO-H sensitivity. We use the intersection between the ASTRO-H sensitivity curve and the dominant background component, either thermal ICM or CXB, to fix a new normalisation of the electron distribution $K_0$ using Eq.~\ref{eq:IC}. This represents the lower limit on the normalisation detectable by ASTRO-H. If no upper limit on the hard X-ray flux, $F_\mathrm{X,UL}$, exist for a given object, we start from this step.
Finally, we fit again the synchrotron emission of Eq.~\ref{eq:Ssyn} to the radio data. The resulting magnetic field value is a cap on the volume-average magnetic field below which HXR IC emission would be detectable by ASTRO-H, and we call it $B_\mathrm{dt}$. To be clearer, HXR IC emission would be detectable by ASTRO-H if the cluster's magnetic field is lower or equal to $B_\mathrm{dt}$, always meaning a volume-average. We underline that this simple approach is meant to be just a first-order approximation to what can really be achieved. In fact, e.g., the CXB background is well know and can be easily subtracted. We provide the result in such a case, along with the rest, in Tables~\ref{tab:halos} and \ref{tab:relics}.

The above approach leads to a rough estimate which we complement with one based on the total number of counts in the $20-80\mathrm{\,keV}$ band as follows.
We require the signal-to-noise ratio $\mathrm{s/n} = N_\mathrm{S}/\sqrt{N_\mathrm{B} + N_\mathrm{S}} \ge 5$, where $N_S$ is the total number of source counts and $\sqrt{N_B + N_S}$ is the corresponding noise, with $N_\mathrm{B} = N_\mathrm{ICM} + N_\mathrm{CXB} + N_\mathrm{instr}$ including all background counts (ICM, CXB and instrumental). 

We approximate the instrumental background by a power-law corresponding to $6\times10^{-3}\mathrm{\,counts/s/keV/FoV}$ at $10\mathrm{\,keV}$ and $2\times10^{-4}\mathrm{\,counts/s/keV/FoV}$ at $50\mathrm{\,keV}$, values adopted from the ASTRO-H Quick Reference (table~2). For our purpose, a simple power-law approximation of the instrumental noise suffices, since uncertainties will be dominated by the modelling of the thermal and IC emission.\footnote{Additionally, comparing our power-law estimate to the ASTRO-H response files (http://astro-h.isas.jaxa.jp/researchers/sim/response.html) in the $20-80\mathrm{\,keV}$ band shows that we overestimate the predicted instrumental background by just a few percent.}
We convert the thermal bremsstrahlung and CXB fluxes into count rates by multiplying by the HXI effective area taken from the ASTRO-H Quick Reference (figure~1). $N_S$ is then related to the HXR flux density, $S_\mathrm{x}(E) = k_\mathrm{x} E^{-\alpha}$, through 
\begin{equation}
N_\mathrm{S} = T_\mathrm{obs}\int^{80\mathrm{\,keV}}_{20\mathrm{\,keV}} \frac{S_\mathrm{x}(E) A_\mathrm{eff}(E)}{E}\mathrm{d}E\,,
\label{eq:Ns}
\end{equation}
where $T_\mathrm{obs}$ is the observation time and $A_\mathrm{eff}(E)$ is the effective area. Using the $N_S$ for which $\mathrm{s/n} \ge 5$, we solve for $k_\mathrm{x}$ and, finally, calculate the corresponding flux as $F_\mathrm{x} = k_\mathrm{x} \int^{80\mathrm{\,keV}}_{20\mathrm{\,keV}} E^{-\alpha}\mathrm{d}E$. The spectral index used here is the same as above. This flux is then converted into a magnetic field, $B_\mathrm{dt}^\mathrm{s/n}$, using
\begin{eqnarray}
& &\frac{F	_{x} \nu_r^{-\alpha} }{s_r \int_{\nu_\mathrm{min}}^{\nu_\mathrm{max}} \nu_{x}^{-\alpha} \mathrm{d}\nu_x} = \nonumber \\
& &\quad \frac{2.46 \times 10^{-19} T_\mathrm{CMB}^3}{B_\mathrm{dt}^\mathrm{s/n}} \frac{b(p)}{a(p)} \left(\frac{4.96 \times 10^3 T_\mathrm{CMB}}{B_\mathrm{dt}^\mathrm{s/n}}\right)^\alpha,
\label{eq:Banalytic1}
\end{eqnarray}
where $T_\mathrm{CMB} = T_0\left(1+z\right)$ is the CMB temperature, $F_x$ is the X-ray flux, $s_r$ is the radio flux density at $\nu_r$, $p = 2 \alpha + 1$, and
\begin{subequations}
	\begin{align}
	a(p) &= \frac{2^{(p-1)/2} \sqrt 3 \Gamma\left(\frac{3p -1}{12}\right)\Gamma\left(\frac{3p +19}{12}\right) \Gamma\left(\frac{p +7}{4}\right)} {16 \pi^{1/2} \Gamma\left(\frac{p + 9}{4}\right)},
	\label{eq:anew1} \\
	b(p) &= 2^{p+3} \frac{(p^2 + 4p + 11) \Gamma\left(\frac{p + 5}{2}\right) \zeta\left(\frac{p+5}{2}\right)}{(p+3)^2(p+1)(p+5)},
	\end{align}
\end{subequations}
Here $\Gamma$ and $\zeta$ are the gamma and Riemann zeta functions, respectively \citep{pacholczyk1970, blumenthal1970, sarazin1988, longair2011}.\footnote{Note that the arguments of the gamma function ($\Gamma$) are slightly different with respect to the values when a coherent magnetic field is assumed.} In this approach, for objects whose size in the sky ($\Omega$~arcmin$^2$) is larger than the HXI FoV ($9\times9$~arcmin$^2$), the radio flux density $s_\mathrm{r}$ is roughly scaled down as $9\times9/\Omega$. This implies slightly conservative results as halos and relics are not uniform sources. Eventually, the resulting magnetic field value, $B_\mathrm{dt}^\mathrm{s/n}$, is again a cap on the volume-average magnetic field below which HXR IC emission is detectable by ASTRO-H.

The analysis based on the signal-to-noise ratio is done assuming a straight power-law in the 20--80~keV energy band. That is, we ignore any possible cutoff effect in that range, otherwise Eq.~\ref{eq:Banalytic1} would not be valid anymore. In most cases, the cutoffs are at much lower and higher energies. However, e.g., for A0085 and A1914 (see Fig.~\ref{modeling_1}), and for A2063 and A0013 (see Appendix~\ref{sec:app_comments} for the corresponding figures), this method could give slightly optimistic results.

Note also that in some cases (only radio relics, see Appendix~\ref{sec:app_comments} for the figures), the CXB dominates over the thermal ICM emission, and the intersection point between the ASTRO-H sensitivity and the CXB emission is at an energy lower than the rise in the CXB spectrum. The corresponding results for $B_\mathrm{dt}$ should be taken with caution and, in fact, in these cases, the signal-to-noise approach gives more conservative results. We underline again that our simple approach is meant to be only a first-approximation, and that, in the following, we base all our conclusions only on the more robust signal-to-noise approach. Nevertheless, we also note that the two approaches turn out to give similar results.

We eventually decided to exclude the A2218 halo and the ZwCl0008.8-5215 western relic from the final modelling. This is due to their low radio flux, and to the apparent presence of a quite high $\gamma_\mathrm{min}$, or break in the radio spectra, that would render rather impossible the IC search in these objects. 

\begin{figure}[t!]
\includegraphics[width=0.5\textwidth]{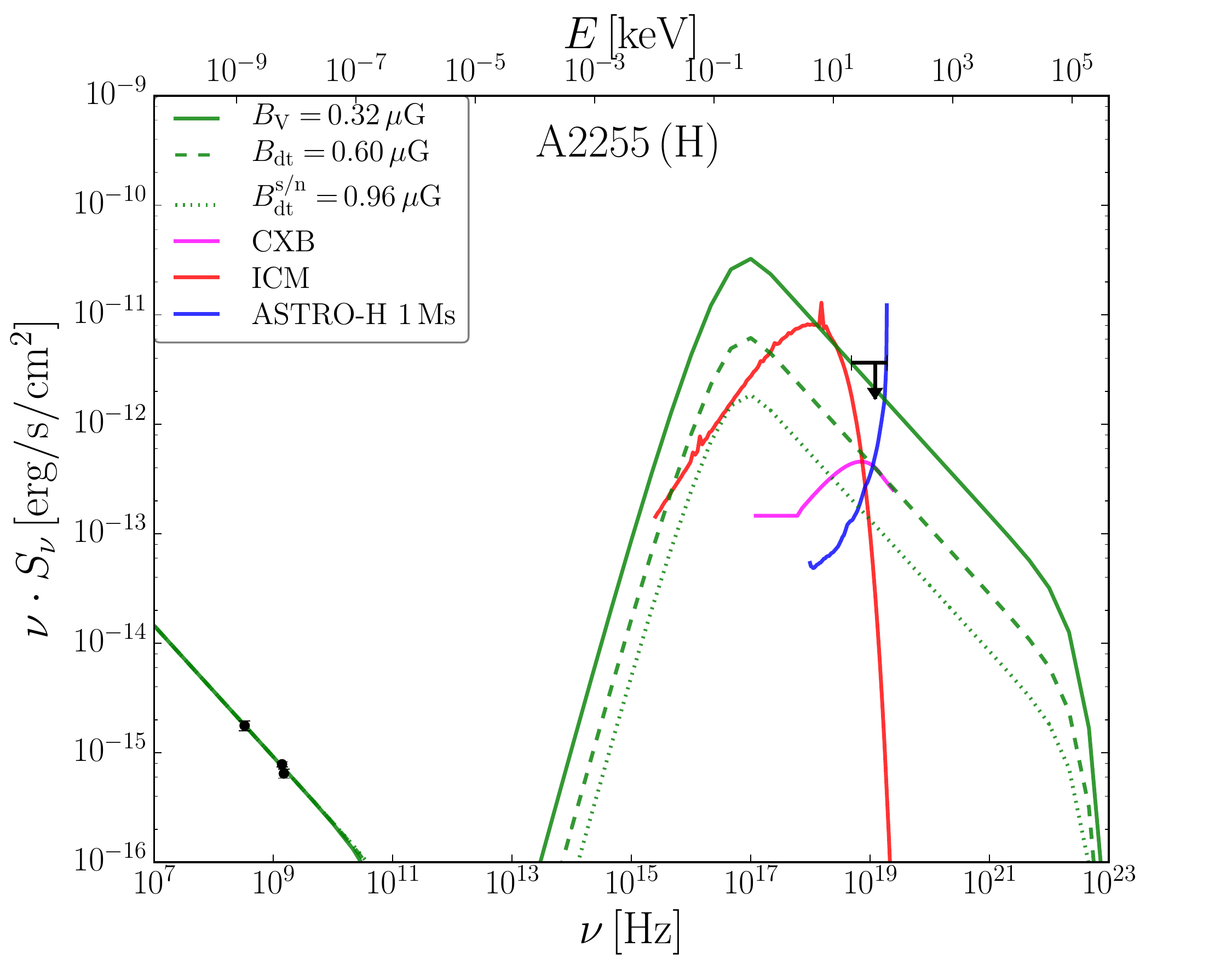}
\caption{\footnotesize{Synchrotron and IC modelling of the radio halo (H) in A2255. The black points and arrow are the currently available radio data and HXR upper-limit, respectively, from Table~\ref{tab:sample_halos}. We show the thermal ICM and CXB emissions, together with the ASTRO-H HXI sensitivity for 1~Ms of observations scaled to the halo angular size. The green lines show the models for the current volume-average lower limit $B_\mathrm{V}$ (solid), and for $B_\mathrm{dt}$ (dashed) and $B_\mathrm{dt}^\mathrm{s/n}$ (dotted) for 1~Ms of observations.}}
\label{A2255Halo}
\end{figure}

We show an example of synchrotron and IC modelling in Fig.~\ref{A2255Halo} for the case of the radio halo hosted by A2255, and we present and discuss all results in the next sections.

\section{Results}
\label{sec:results}
The results of our analysis are given in Tables~\ref{tab:halos} and \ref{tab:relics} for radio halos and relics, respectively. The third column contains the current lower limit on the magnetic field for objects where an HXR flux upper limit is available in literature. Note that our lower limits on the magnetic field are slightly higher than those in the original works from which the HXR upper limits fluxes are taken because of our assumption on the magnetic field distribution (see Sect.~\ref{sec:processes}). The sixth column contains $B_\mathrm{dt}$, corresponding to the cap value of the volume-averaged magnetic field up to which ASTRO-H can still detect non-thermal IC for $1\mathrm{\,Ms}$ of observation time. The eighth column shows an optimistic value for the detection sensitivity on the volume-average magnetic field, $B_\mathrm{dt}^\mathrm{opt}$, obtained assuming that the CXB can be perfectly subtracted and, therefore, considering always the intersection of the ASTRO-H sensitivity curve with the thermal ICM emission as estimation point for $B_\mathrm{dt}^\mathrm{opt}$ (see Sect.~\ref{sec:detectability} for details). 

In Tables~\ref{tab:sn_halos} and \ref{tab:sn_relics}, we show the estimate of the ASTRO-H detection sensitivity of the volume-average magnetic field,  $B_\mathrm{dt}^\mathrm{s/n}$, obtained with the signal-to-noise approach in the 20--80\,keV band, for halos and relics, respectively. We show the results for $100\mathrm{\,ks}$, $500\mathrm{\,ks}$ and $1\mathrm{\,Ms}$ of observation time. 

Comparing the results in Tables~\ref{tab:halos}$-$\ref{tab:sn_relics},
it is clear that the former results based on the intersection point
between the background emission and the ASTRO-H sensitivity are more
pessimistic. In particular, for sources larger than the HXI FoV (e.g.,
the halos of Coma and A2319, and the relics of Coma and A3667), our
initial analysis shows little room for improvement on the current
constraints while the signal-to-noise analysis shows differently. The
simplified assumption of a uniform source in our initial analysis is the
most obvious reason for the pessimistic perspective for sources that
cover a large area on the sky. This is also the origin for the
$B_\mathrm{dt}$ values of the Coma and A2319 radio halos and the Coma
and A3667-NW relic to be at the same level of current upper
limits. Additionally, as a general remark, assuming that the IC emission
is detectable only at energies where it dominates over the background is
a conservative approach, giving that backgrounds are known and can be
modelled in detail in a real analysis. Indeed, as already stressed 
earlier, this simple method is meant to be just a first-order approximation,
and we take our signal-to-noise approach as reference when drawing
conclusions.

Note that the soft X-ray (SXR) and extreme ultra-violet (EUV) bands could also be promising for the detection of IC emission, since the flux would be much higher at those lower frequencies \citep[e.g.,][]{sarazin1997}. However, this is advantageous only if the background can be well constrained, and crucially depends on the low energy cutoff of the relativistic electron population, $\gamma_\mathrm{min}$, which is, in most cases, unconstrained by current radio data. Indeed, assuming an electron spectrum cutoff of $\gamma_\mathrm{min}\sim10^2$, as discussed in Section~\ref{sec:electrons}, implies IC spectra peaking in the EUV band and often dominating over the X-ray background in SXRs, in particular in case of very steep radio spectra. However, such steep spectra down to very low frequencies seem unlikely. The radio spectrum depends on acceleration history of relativistic electrons and, e.g., steady state injection models feature a spectral break and would imply a low-frequency flattening \citep[e.g.,][]{sarazin1999}. In the near future, low-frequency observations with LOFAR \citep{rottgering2012} will provide crucial information and certainly shed more light on the spectrum of cosmic-ray electrons in clusters. This will be vital in order to asses the possible IC contribution in the SXR and EUV regimes.

\begin{table*}[hbt!]
\sisetup{round-mode = places}
\caption{Results for halos}
\centering
\begin{tabular}{l
	S[round-precision=2, table-omit-exponent, scientific-notation = false]
	S[round-precision=2,table-omit-exponent, scientific-notation = false]
	S[round-precision=2, table-omit-exponent, scientific-notation = false]
	S[round-precision=2,table-omit-exponent, scientific-notation = false]
	S[round-precision=0, table-omit-exponent]
	S[round-precision=2,table-omit-exponent, fixed-exponent = -12]}
	\hline\hline
\multicolumn{2}{l}{} \phantom{\Big|}  &\multicolumn{1}{|c}{Current Limits} &\multicolumn{1}{|c|}{ASTRO-H $1~\mathrm{Ms}$} &\multicolumn{1}{c|}{} \\ 
\multicolumn{1}{l}{Cluster} & \multicolumn{1}{c}{$\alpha$} & \multicolumn{1}{|c}{$B_\mathrm{V}$} & \multicolumn{1}{|c|}{$B_\mathrm{dt}$} & \multicolumn{1}{c|}{$B_\mathrm{dt}^\mathrm{opt}$} &  \multicolumn{1}{c}{$E$} & \multicolumn{1}{c}{$S_x\times E$} \phantom{\Big|} \\
& &\multicolumn{1}{|c}{[$\mathrm{\mu G}$]}&   \multicolumn{1}{|c|}{[$\mathrm{\mu G}$]} &  \multicolumn{1}{c|}{[$\mathrm{\mu G}$]} & \multicolumn{1}{c}{[$\mathrm{keV}$]} \phantom{\Big|} \\       
\hline\\[-0.5em] 
    Bullet & 1.5   & 0.37861042 & 1.00124787 & 1.00124787 & 45    & 1.95E-13 \\
    A0520 & 1.12  &       & 0.36644326 & 0.56672889 & \multicolumn{1}{c}{33~(16)} & \multicolumn{1}{c}{0.14~(0.06)} \\
    A0521 & 1.81  &       & 0.80121416 & 1.15779557 & \multicolumn{1}{c}{31~(18)} & \multicolumn{1}{c}{0.11~(0.06)} \\
    A0665 & 1.04  & 0.11014153 & 0.28160707 & 0.4125128 & \multicolumn{1}{c}{46~(29)} & \multicolumn{1}{c}{0.42~(0.20)} \\
    A0697 & 1.52  &       & 0.32835093 & 0.32835093 & 52.5  & 2.87E-13 \\
    A0754  & 1.1   & \multicolumn{1}{c}{$0.09^{\gamma_\mathrm{max} = 6.0\times 10^4}$} & \multicolumn{1}{c}{$0.45^{\gamma_\mathrm{max} = 2.7\times 10^4}$} & 0.44647248 & 52    & 4.73E-13 \\
    A1300  & 1.75633064 &       & 1.01661562 & 1.26857941 & \multicolumn{1}{c}{30~(23)} & \multicolumn{1}{c}{0.10~(0.07)} \\
    Coma  & 1.16  & \multicolumn{1}{c}{$0.32^{\gamma_\mathrm{max} = 3.9\times 10^4}$} & \multicolumn{1}{c}{$0.36^{\gamma_\mathrm{max} = 3.7\times 10^4}$} & 0.36124246 & 47.5  & 3.67E-12 \\
    A1758a & 1.5   &       & 0.70805184 & 0.95032008 & 34    & 1.60E-13 \\
    A1914 & 1.9473499 & \multicolumn{1}{c}{$0.59_{\gamma_\mathrm{min} = 2.3\times 10^3}$} & \multicolumn{1}{c}{$1.12_{\gamma_\mathrm{min} = 1.7\times 10^3}$} & 1.11967715 & 51    & 3.28E-13 \\
    A2163 & 1.18  & 0.13208016 & 0.4862356 & 0.4862356 & 56.5  & 8.18E-13 \\
    A2219 & 0.72054679 &       & 0.29218091 & 0.30259662 & \multicolumn{1}{c}{39~(37)} & \multicolumn{1}{c}{0.22~(0.21)} \\
    A2255  & 1.6   & 0.31864914 & 0.60424839 & 0.76802879 & \multicolumn{1}{c}{46~(34)} & \multicolumn{1}{c}{0.42~(0.27)} \\
    A2256  & 1.34543266 & 0.19979998 & 0.48935604 & 0.52651021 & 47.5  & 6.00E-13 \\
    A2319 & 1.65  & 0.66288771 & 0.64910507 & 0.64910507 & 53    & 1.36E-12 \\
    A2744  & 0.91734375 & 0.28639805 & 0.4369414 & 0.58639693 & \multicolumn{1}{c}{39~(26)} & \multicolumn{1}{c}{0.21~(0.12)} \\
    A3562 & 1.3   & \multicolumn{1}{c}{$0.10^{\gamma_\mathrm{max} = 6.8\times 10^4}$} & \multicolumn{1}{c}{$0.37^{\gamma_\mathrm{max} = 3.5\times 10^4}$} & 0.53162899 & \multicolumn{1}{c}{41~(27)} & \multicolumn{1}{c}{0.27~(0.14)} \\
    CL0217+70\tablefootmark{a} & 1.2   &       & 0.35927871 & 0.68532691 & \multicolumn{1}{c}{46~(17)} & \multicolumn{1}{c}{0.41~(0.12)} \\
    MACSJ0717.5+3745 & 1.23288714 &       & 1.2914362 & 1.2914362 & 39.5  & 1.24E-13 \\
    MACSJ1752.0+4440\tablefootmark{a} & 1.6   &       & 0.97283482 & 1.3695487 & \multicolumn{1}{c}{32~(20)} & \multicolumn{1}{c}{0.13~(0.07)} \\
    PLCK G171.9-40.7\tablefootmark{a} & 1.84  &       & 1.14964615 & 1.14964615 & 41.5  & 1.10E-13 \\
    RXCJ1514.9-1523\tablefootmark{a} & 1.6   &       & 0.52029484 & 0.76484548 & \multicolumn{1}{c}{39~(24)} & \multicolumn{1}{c}{0.22~(0.11)} \\
    RXCJ2003.5-2323\tablefootmark{a} & 1.32  &       & 0.92348963 & 1.19073595 & \multicolumn{1}{c}{31~(22)} & \multicolumn{1}{c}{0.11~(0.07)} \\
    \hline
    \end{tabular}%
    \tablefoot{For each object, we show $\alpha$ and report the values
 of the current lower limit on the volume-average magnetic field,
 $B_\mathrm{V}$, where available, and the detection sensitivity on the volume-average magnetic field, $B_\mathrm{dt}$, for which IC emission would be detectable by ASTRO-H in 1~Ms. Additionally, $B_\mathrm{dt}^\mathrm{opt}$ is an optimistic estimate of the latter that corresponds to the intersection of the sensitivity curve and the thermal ICM emission, thus assuming the CXB is fully known and perfectly subtracted. The last two columns shows the energy where non-thermal X-rays start to dominate over the background for $B_\mathrm{dt}$ ($B_\mathrm{dt}^\mathrm{opt}$), and the corresponding non-thermal flux density in units of in $10^{-12} \mathrm{\,erg\, cm^{-2}\, s^{-1}}$, respectively. In the cases where $\gamma_\mathrm{min}$ and/or $\gamma_\mathrm{min}$ are calculated from the radio data and differ from the standard values that we adopt (see Sect.~\ref{sec:electrons}), we show them explicitly.~\tablefoottext{a}{Thermal ICM emission computed using the model by \cite{zandanel2014a} for the gas density.}
    }
  \label{tab:halos}%
\end{table*}

\begin{table*}[hbt!]
\sisetup{round-mode = places}
\caption{Results for relics}
\centering
\begin{tabular}{l l
	S[round-mode = places, round-precision=2, table-omit-exponent, scientific-notation = false]
	S[round-precision=2,table-omit-exponent, scientific-notation = false]
	S[round-precision=2, table-omit-exponent, scientific-notation = false]
	S[round-precision=2,table-omit-exponent, scientific-notation = false]
	S[round-precision=0, table-omit-exponent]
	S[round-precision=2,table-omit-exponent]}
	\hline\hline
	\multicolumn{3}{l}{}  \phantom{\Big|} &\multicolumn{1}{|c}{Current Limits} &\multicolumn{1}{|c|}{ASTRO-H $1\mathrm{Ms}$} &\multicolumn{1}{c|}{}\\ 
    	\multicolumn{1}{l}{Cluster} & \multicolumn{1}{c}{Source}&\multicolumn{1}{c}{$\alpha$} & \multicolumn{1}{|c}{$B_\mathrm{V}$} & \multicolumn{1}{|c|}{$B_\mathrm{dt}$} & \multicolumn{1}{c|}{$B_\mathrm{dt}^\mathrm{opt}$} &  \multicolumn{1}{c}{$E$} & \multicolumn{1}{c}{$S_x\times E$} \phantom{\Big|} \\
        && &\multicolumn{1}{|c}{[$\mathrm{\mu G}$]}&   \multicolumn{1}{|c|}{[$\mathrm{\mu G}$]} &  \multicolumn{1}{c|}{[$\mathrm{\mu G}$]} & \multicolumn{1}{c}{[$\mathrm{keV}$]} \phantom{\Big|} \\
\hline\\[-0.5em] 
    Toothbrush\tablefootmark{a}  &       & 1.1   &       & 1.65  & 2.13  & \multicolumn{1}{c}{31~(22)} & \multicolumn{1}{c}{0.11~(0.07)} \\
    A0013 &       & 2.3   &       & \multicolumn{1}{c}{$1.98_{\gamma_{min} = 2.6\times 10^3}$} & 1.98  & \multicolumn{1}{c}{31~(31)} & \multicolumn{1}{c}{0.06~(0.06)} \\
    A0085 &       & 1.61282313 & \multicolumn{1}{c}{$0.43_{\gamma_{min} = 3.1\times 10^3}^{\gamma_{max} = 1.6\times 10^4}$} & \multicolumn{1}{c}{$1.79_{\gamma_{min} = 1.5\times 10^3}^{\gamma_{max} = 7.7\times 10^3}$} & 1.96  & \multicolumn{1}{c}{34~(31)} & \multicolumn{1}{c}{0.15~(0.13)} \\
    A0521 &       & 1.45  &       & 1.00  & 2.00  & \multicolumn{1}{c}{24~(6)} & \multicolumn{1}{c}{0.05~(0.02)} \\
    A0610\tablefootmark{a}  &       & 1.4   &       & 0.86  & 1.51  & \multicolumn{1}{c}{23~(8)} & \multicolumn{1}{c}{0.05~(0.02)} \\
    AS753\tablefootmark{a}  &       & 2.1   &       & \multicolumn{1}{c}{$1.08^{\gamma_{max} = 2.0\times 10^4}$} & 4.82  & \multicolumn{1}{c}{49~(5)} & \multicolumn{1}{c}{1.75~(0.23)} \\
    A0754  &       & 1.1   &       & 0.28  & 0.32  & \multicolumn{1}{c}{48(40)} & \multicolumn{1}{c}{0.52(0.39)} \\
    A0781\tablefootmark{a} &       &1.23  &       & 1.058223082   & 1.76  & \multicolumn{1}{c}{17(5)} &	\multicolumn{1}{c}{0.03(0.01)}\\
    A1240 & N     & 0.85782044 &       & 0.16  & 0.28  & \multicolumn{1}{c}{28~(5)} & \multicolumn{1}{c}{0.07~(0.02)} \\
          & S     & 0.71033074 &       & 0.10  & 0.18  & \multicolumn{1}{c}{34~(5)} & \multicolumn{1}{c}{0.15~(0.03)} \\
    A1300  &       & 0.9   &       & 0.60  & 0.79  & 17    & 0.03 \\
    A1367 &       & 1.9   & 0.89294364 & 0.76  & 2.16  & \multicolumn{1}{c}{43~(10)} & \multicolumn{1}{c}{0.29~(0.05)} \\
    A1612\tablefootmark{a}  &       & 1.4   &       & 1.64  & 3.39  & \multicolumn{1}{c}{23~(5)} & \multicolumn{1}{c}{0.05~(0.02)} \\
    Coma  &       & 1.18  & 0.87  & 0.28  & 0.65  & \multicolumn{1}{c}{49~(13)} & \multicolumn{1}{c}{2.20~(0.46)} \\
    A1664 &       & 1.1   &       & 0.52  & 0.90  & \multicolumn{1}{c}{43~(18)} & \multicolumn{1}{c}{0.29~(0.10)} \\
    A2048\tablefootmark{a}  &       & 2.29  &       & 2.43  & 2.96  & \multicolumn{1}{c}{20~(15)} & \multicolumn{1}{c}{0.04~(0.03)} \\
    A2061 &       & 1.03  &       & 0.31  & 0.48  & \multicolumn{1}{c}{34~(16)} & \multicolumn{1}{c}{0.15~(0.06)} \\
    A2063 &       & 1.20592569 & \multicolumn{1}{c}{$0.42^{\gamma_{max} = 1.4\times 10^4}$} & \multicolumn{1}{c}{$5.51^{\gamma_{max} = 4.0\times 10^3}$} & 5.51  & 24    & 0.03 \\
    A2163  &       & 1.02  &       & 0.75  & 0.99  & 16    & 0.03 \\
    A2255  &       & 1.4   &       & 0.38  & 0.63  & \multicolumn{1}{c}{37~(17)} & \multicolumn{1}{c}{0.19~(0.08)} \\
    A2256   &       & 0.81  &       & 0.34  & 0.41  & \multicolumn{1}{c}{49~(39)} & \multicolumn{1}{c}{0.84~(0.57)} \\
    A2345\tablefootmark{a}  & E     & 1.3   &       & 0.49  & 0.99  & \multicolumn{1}{c}{40~(14)} & \multicolumn{1}{c}{0.22~(0.06)} \\
          & W     & 1.55582187 &       & 0.73  & 1.20  & \multicolumn{1}{c}{39~(19)} & \multicolumn{1}{c}{0.20~(0.09)} \\
    A2433\tablefootmark{a}  &       & 1.74  &       & \multicolumn{1}{c}{$1.38^{\gamma_{max} = 9.3\times 10^3}$} & 2.52  & \multicolumn{1}{c}{27~(5)} & \multicolumn{1}{c}{0.07~(0.02)} \\
    A2744  &       & 1.1   & 0.35  & 0.51  & 1.09  & \multicolumn{1}{c}{31~(6)} & \multicolumn{1}{c}{0.11~(0.03)} \\
    A3376 & E     & 0.88178006 & \multicolumn{1}{c}{$0.13^{\gamma_{max} = 3.9\times 10^4}$} & \multicolumn{1}{c}{$0.30^{\gamma_{max} = 2.5\times 10^4}$} & 0.59  & \multicolumn{1}{c}{49~(16)} & \multicolumn{1}{c}{1.78~(0.45)} \\
          & W     & 1     & \multicolumn{1}{c}{$0.26^{\gamma_{max} = 3.0\times 10^4}$ }& \multicolumn{1}{c}{$0.48^{\gamma_{max} = 2.2\times 10^4}$}& 1.32  & \multicolumn{1}{c}{49~(5)} & \multicolumn{1}{c}{0.67~(0.09)} \\
    A3411\tablefootmark{a}  &       & 1     &       & 0.38  & 0.59  & \multicolumn{1}{c}{45~(25)} & \multicolumn{1}{c}{0.34~(0.14)} \\
    A3667 & NW     & 1.02  & 1.29  & 0.94  & 2.12  & \multicolumn{1}{c}{49~(12)} & \multicolumn{1}{c}{1.37~(0.27)} \\
          & SE    & 0.95  &       & 0.31  & 0.53  & \multicolumn{1}{c}{49~(21)} & \multicolumn{1}{c}{0.96~(0.32)} \\
    A4038 &       & 1.18028581 & \multicolumn{1}{c}{$0.20^{\gamma_{max} = 2.5\times 10^4}$} & \multicolumn{1}{c}{$1.88^{\gamma_{max} = 8.2\times 10^3}$} & 3.11  & \multicolumn{1}{c}{28~(11)} & \multicolumn{1}{c}{0.07~(0.03)} \\
    El-Gordo\tablefootmark{a}  & E     & 0.9   &       & 0.12  & 0.12  & 51    & 0.10 \\
          & NW    & 1.19  &       & 0.77  & 0.77  & 51    & 0.10 \\
          & SE    & 1.4   &       & 0.41  & 0.41  & 51    & 0.10 \\
    Sausage\tablefootmark{a}  &       & 1.06  &       & \multicolumn{1}{c}{$0.73^{\gamma_{max} = 5.4\times 10^4}$} & 1.62  & \multicolumn{1}{c}{31~(5)} & \multicolumn{1}{c}{0.11~(0.02)} \\
    MACSJ1149.5+2223\tablefootmark{a}  & E     & 1.2   &       & 1.63  & 1.63  & 5     & 0.01 \\
          & W     & 0.8   &       & 0.66  & 0.66  & 5     & 0.01 \\
    MACSJ1752.0+4440\tablefootmark{a}  & NE    & 1.16  &       & 1.69  & 3.17  & \multicolumn{1}{c}{23~(5)} & \multicolumn{1}{c}{0.05~(0.02)} \\
          & SW    & 1.1   &       & 2.45  & 2.45  & 5     & 0.01 \\
    PLCK G287.0+32.9\tablefootmark{a}  & N     & 1.26  &       & 1.57  & 1.91  & \multicolumn{1}{c}{20~(14)} & \multicolumn{1}{c}{0.04~(0.03)} \\
          & S     & 1.54  &       & 2.25714321  & 4.14  &\multicolumn{1}{c}{17(5)}& \multicolumn{1}{c}{0.03(0.01)} \\
    RXCJ1314.4-2515 & E     & 1.2   &       & 0.62  & 1.18  & \multicolumn{1}{c}{23~(5)} & \multicolumn{1}{c}{0.05~(0.02)} \\
          & W     & 1.4   &       & 0.83  & 1.28  & \multicolumn{1}{c}{30~(15)} & \multicolumn{1}{c}{0.09~(0.04)} \\
    ZwCl0008.8-5215\tablefootmark{a}  & E     & 1.59  &       & 0.85  & 2.14  & \multicolumn{1}{c}{39~(9)} & \multicolumn{1}{c}{0.21~(0.04)} \\
    ZwCl2341.1+0000\tablefootmark{a}  & N     & 0.49  &       & 0.13  & 0.13  & 16    & 0.02 \\
          & S     & 0.76  &       & 0.19  & 0.29  & \multicolumn{1}{c}{32~(16)} & \multicolumn{1}{c}{0.12~(0.05)} \\
    \hline
    \end{tabular}%
    \tablefoot{Columns are as in Table~\ref{tab:halos}, with the addition of the second column which indicates the specific relic to which we are referring to.~\tablefoottext{a}{Thermal ICM emission computed using the model by \cite{zandanel2014a} for the gas density.}~\tablefoottext{b}{No improvement upon current constraints are possible according to this analysis (see main text for details).}
    }
  \label{tab:relics}%
\end{table*}

\begin{table*}[hbt!]
\sisetup{round-mode = places}
\caption{S/N results for halo}
\centering
   \begin{tabular}{l 
       	S[round-precision=0, table-omit-exponent, scientific-notation = false]
	S[round-precision=0,table-omit-exponent, scientific-notation = false]
    	S[round-precision=2, table-omit-exponent, scientific-notation = false]
	S[round-precision=2,table-omit-exponent, scientific-notation = false]
	S[round-precision=2,table-omit-exponent, scientific-notation = false]
	S[round-precision=2, table-omit-exponent, scientific-notation = false]   
	S[round-precision=2, table-omit-exponent, scientific-notation = false] 
	S[round-precision=2,table-omit-exponent, scientific-notation = false] 
	}
    \hline\hline
    \multicolumn{3}{l}{} \phantom{\Big|}  &\multicolumn{2}{|c}{ASTRO-H $100\mathrm{\,ks}$} &\multicolumn{2}{|c}{$500\mathrm{\,ks}$} &\multicolumn{2}{|c}{$1\mathrm{\,Ms}$} \\ 
    Cluster & \multicolumn{1}{c}{$s_{r}$} & $\nu_{r}$ &\multicolumn{1}{|c}{$B_\mathrm{dt}^\mathrm{s/n}$} & \multicolumn{1}{c}{$F_{20-80\mathrm{\,keV}}$\tablefootmark{a}} & \multicolumn{1}{|c}{$B_\mathrm{dt}^\mathrm{s/n}$} & \multicolumn{1}{c}{$F_{20-80\mathrm{\,keV}}$\tablefootmark{a}} & \multicolumn{1}{|c}{$B_\mathrm{dt}^\mathrm{s/n}$} & \multicolumn{1}{c}{$F_{20-80\mathrm{\,keV}}$\tablefootmark{a}} \phantom{\Big|} 
    \\
      &\multicolumn{1}{c}{[$\mathrm{mJy}$]}  & \multicolumn{1}{c}{[$\mathrm{MHz}$]}  & \multicolumn{1}{|c}{[$\mathrm{\mu G}$]} &  &\multicolumn{1}{|c}{[$\mathrm{\mu G}$]}  & & \multicolumn{1}{|c}{[$\mathrm{\mu G}$]} \phantom{\Big|} \\ 
\hline\\[-0.5em] 
    Bullet & 51.65 & 1340  & 0.84246925 & 0.42223745 & 1.18347438 & 0.18052807 & 1.36526871 & 0.12629802 \\
    A0520 & 85    & 325   & 0.25674214 & 0.40136475 & 0.38512572 & 0.16990097 & 0.45632279 & 0.11858131 \\
    A0521 & 328   & 153   & 0.61036093 & 0.28386061 & 0.83066019 & 0.11940543 & 0.94457344 & 0.08321305 \\
    A0665 & 197   & 327   & 0.2535752 & 0.72494478 & 0.38176663 & 0.3146407 & 0.45403707 & 0.22091056 \\
    A0697 & 135   & 153   & 0.25350152 & 0.85062378 & 0.35182673 & 0.37240874 & 0.40450642 & 0.26200916 \\
    A0754  & 86    & 1365  & 0.27242868 & 0.97001719 & 0.40377615 & 0.42453644 & 0.4773898 & 0.29865981 \\
    A1300  & 130   & 325   & 0.74263486 & 0.28362588 & 1.01732336 & 0.11912432 & 1.15988689 & 0.08298682 \\
    Coma* & 640   & 1400  & 0.19930699 & 1.8758301 & 0.29072503 & 0.82994478 & 0.34172555 & 0.58536947 \\
    \hspace{0.78cm}** & "     & "     & 0.54401148 & "     & 0.79353075 & "     & 0.93273399 & " \\
    \hspace{0.78cm}*** & "     & "     & 0.25371109 & 1.1116303 & 0.37138801 & 0.48809615 & 0.43690324 & 0.34363819 \\
    A1758 & 146   & 325   & 0.55443245 & 0.38235398 & 0.78044986 & 0.162639 & 0.90077535 & 0.11364386 \\
    A1914 & 64    & 1465  & 1.18390008 & 0.53479813 & 1.57096393 & 0.23232958 & 1.77111885 & 0.16315566 \\
    A2163* & 861   & 325   & 0.45820493 & 1.08205055 & 0.6685599 & 0.47492569 & 0.78537408 & 0.33433523 \\
    \hspace{0.88cm}** & "     & "     & 0.51181799 & "     & 0.74672774 & "     & 0.87718361 & " \\
    A2219 & 232   & 325   & 0.18702454 & 0.65997084 & 0.30509768 & 0.28434081 & 0.37510514 & 0.19928833 \\
    A2255  & 536   & 330   & 0.6077641 & 0.63840536 & 0.83714917 & 0.27766441 & 0.95894955 & 0.19504672 \\
    A2256* & 6600  & 63    & 0.41019157 & 0.87350835 & 0.58361931 & 0.38212765 & 0.67808441 & 0.26879688 \\
    \hspace{0.88cm}* & "     & "     & 0.47547529 & "     & 0.67641991 & "     & 0.78588239 & " \\
    A2319* & 1450  & 408   & 0.5647961 & 1.19931796 & 0.76932829 & 0.52878775 & 0.87793851 & 0.37265304 \\
    \hspace{0.88cm}** & "     & "     & 0.78994679 & "     & 1.07598569 & "     & 1.22788079 & " \\
    A2744  & 218   & 325   & 0.31706629 & 0.54717387 & 0.49318684 & 0.23456385 & 0.59400131 & 0.16420481 \\
    A3562 & 20    & 1400  & 0.26418174 & 0.54635642 & 0.38088202 & 0.23552358 & 0.44450552 & 0.16509541 \\
    CL0217+70\tablefootmark{a}  & 326   & 325   & 0.32870904 & 0.67950577 & 0.48043816 & 0.29483313 & 0.56424621 & 0.20698923 \\
    MACSJ0717.5+3745 & 492.5 & 235   & 0.91406511 & 0.3666468 & 1.34525443 & 0.15470647 & 1.58087912 & 0.10789367 \\
    MACSJ1752.0+4440\tablefootmark{a} & 164   & 323   & 0.73493696 & 0.33849525 & 1.02297022 & 0.14327125 & 1.17472625 & 0.09999251 \\
    PLCK G171.9-40.7\tablefootmark{a} & 483   & 235   & 0.98194619 & 0.26319796 & 1.33413737 & 0.11021533 & 1.51561227 & 0.07672578 \\
    RXCJ1514.9-1523\tablefootmark{a} & 102   & 327   & 0.45694266 & 0.43919064 & 0.63270566 & 0.18843824 & 0.72566192 & 0.13194236 \\
    RXCJ2003.5-2323\tablefootmark{a} & 360   & 240   & 0.64257178 & 0.35343671 & 0.93223636 & 0.14906973 & 1.08894795 & 0.10395212 \\
        \hline
    \end{tabular}%
          \tablefoot{For each halo, we show the radio data values $s_{r}$ and $\nu_{r}$ used in Eq.~\ref{eq:Banalytic1} to calculate $B_\mathrm{dt}^\mathrm{s/n}$. We show the latter for 100~ks, 500~ks and 1~Ms of ASTRO-H observations, together with the corresponding flux in the $20-80\mathrm{\,keV}$ energy band in units of $10^{-12}\mathrm{\,erg\, cm^{-2}\, s^{-1}}$. Within a few percent error the fluxes and magnetic fields scale with observation time according to Eq.~\ref{eq:BvsT}.~\tablefoottext{*}{In this cases, the source is larger than the HXI FoV of $9\times9=81$~arcmin$^2$. In our work, for simplicity, we assumed a uniform surface brightness and, therefore, $s_{r}$ was scaled by a factor $81/x$ where $x\mathrm{\,arcmin^2}$ is the size that a source covers in the sky (see Tables~\ref{tab:sample_halos} and \ref{tab:sample_relics}). Note that this estimate results in slightly conservative predictions since halos and relics are non-uniform sources.}~\tablefoottext{**}{This gives results assuming all radio emission is coming from an area equal to the HXI FoV and, therefore, $s_\mathrm{r}$ was not scaled. This can be considered an optimistic estimate for the detectable magnetic field.}~\tablefoottext{***}{Additional analysis for Coma. Since the Coma radio halo is extremely large in the sky, we estimated the thermal ICM emission only in the HXI FoV, integrating Eq.~\ref{eq:ne} out to $\sim180 \mathrm{\,kpc}$, roughly corresponding to the HXI FoV, rather than over the full halo size.}~\tablefoottext{a}{Thermal ICM emission computed using the model by \cite{zandanel2014a} for the gas density.}
     }
\label{tab:sn_halos}%
\end{table*}
    
\begin{table*}[hbt!]
\sisetup{round-mode = places}
\caption{S/N results for relics}
\centering
   \begin{tabular}{l l
       	S[round-precision=0, table-omit-exponent, scientific-notation = false]
	S[round-precision=0,table-omit-exponent, scientific-notation = false]
    	S[round-precision=2, table-omit-exponent, scientific-notation = false]
	S[round-precision=2,table-omit-exponent, scientific-notation = false]
	S[round-precision=2,table-omit-exponent, scientific-notation = false]
	S[round-precision=2, table-omit-exponent, scientific-notation = false]   
	S[round-precision=2, table-omit-exponent, scientific-notation = false] 
	S[round-precision=2,table-omit-exponent, scientific-notation = false] 
	}
    \hline\hline
    \multicolumn{4}{l}{} \phantom{\Big|}  &\multicolumn{2}{|c}{ASTRO-H $100\mathrm{\,ks}$} &\multicolumn{2}{|c}{$500\mathrm{\,ks}$} &\multicolumn{2}{|c}{$1\mathrm{\,Ms}$} \\ 
    Cluster &Source& \multicolumn{1}{c}{$s_{r}$} & $\nu_{r}$ &\multicolumn{1}{|c}{$B_\mathrm{dt}^\mathrm{s/n}$} & \multicolumn{1}{c}{$F_{20-80\mathrm{\,keV}}$\tablefootmark{a}} & \multicolumn{1}{|c}{$B_\mathrm{dt}^\mathrm{s/n}$} & \multicolumn{1}{c}{$F_{20-80\mathrm{\,keV}}$\tablefootmark{a}} & \multicolumn{1}{|c}{$B_\mathrm{dt}^\mathrm{s/n}$} & \multicolumn{1}{c}{$F_{20-80\mathrm{\,keV}}$\tablefootmark{a}} \phantom{\Big|} 
    \\
      &&\multicolumn{1}{c}{[$\mathrm{mJy}$]}  & \multicolumn{1}{c}{[$\mathrm{MHz}$]}  & \multicolumn{1}{|c}{[$\mathrm{\mu G}$]} &  &\multicolumn{1}{|c}{[$\mathrm{\mu G}$]}  & & \multicolumn{1}{|c}{[$\mathrm{\mu G}$]} \phantom{\Big|} \\ 
\hline\\[-0.5em] 
    Toothbrush\tablefootmark{a}  &       & 7600  & 74    & 1.08457717 & 0.35315156 & 1.64014863 & 0.14816764 & 1.9484692 & 0.10319334 \\
    A0013 &       & 6000  & 80    & 1.10423783 & 0.14038969 & 1.4543923 & 0.05657085 & 1.62766585 & 0.03901903 \\
    A0085 &       & 8330  & 160   & 1.32874398 & 0.36638457 & 1.84307671 & 0.15582945 & 2.11412293 & 0.1088829 \\
    A0521 &       & 297   & 153   & 0.54038939 & 0.22938127 & 0.77893602 & 0.09365054 & 0.90528191 & 0.06479914 \\
    A0610\tablefootmark{a}  &       & 59    & 600   & 0.47074363 & 0.22293735 & 0.68518987 & 0.09055643 & 0.7992371 & 0.06258122 \\
    AS753*\tablefootmark{a}  &       & 460   & 1398  & 1.30138923 & 0.4765311 & 1.70371734 & 0.20673843 & 1.90967389 & 0.14513764 \\
    \hspace{0.88cm}** &       & "     & "     & 2.02302585 & "     & 2.64844995 & "     & 2.96861198 & " \\
    A0754* &       & 1450  & 74    & 0.22465425 & 0.79742918 & 0.33370778 & 0.3472949 & 0.39475483 & 0.24403667 \\
    \hspace{0.88cm}** &       &       &       & 0.24940389 & "     & 0.37051432 & "     & 0.43830659 & " \\
    A0781\tablefootmark{a} &       & 93.3  & 325   & 0.47641747 & 0.2121494 & 0.71789193 & 0.08502421 & 0.84850479 & 0.05856731 \\
    A1240 & N     & 21    & 325   & 0.08525916 & 0.3226037 & 0.13710894 & 0.13346121 & 0.16688403 & 0.09263858 \\
          & S     & 28.5  & 325   & 0.05871424 & 0.46231905 & 0.09706831 & 0.19566739 & 0.11978545 & 0.13655882 \\
    A1300  &       & 75    & 325   & 0.27249124 & 0.23893564 & 0.44079106 & 0.09580948 & 0.53629987 & 0.0660049 \\
    A1367 &       & 180   & 610   & 0.78091327 & 0.4301486 & 1.04402395 & 0.18531318 & 1.1801557 & 0.12988011 \\
    A1612\tablefootmark{a}  &       & 775.8 & 241   & 0.91989221 & 0.22633277 & 1.33799301 & 0.09209305 & 1.56042846 & 0.06366941 \\
    Coma* &       & 3300  & 151   & 0.27073036 & 0.70075282 & 0.39701015 & 0.30428082 & 0.46693454 & 0.21366016 \\
    \hspace{0.78cm}** &       & "     & "     & 0.56007878 & "     & 0.82117803 & "     & 0.96576968 & " \\
    A1664 &       & 1250  & 150   & 0.44864173 & 0.57576273 & 0.66997404 & 0.24803394 & 0.79353624 & 0.17383717 \\
    A2048\tablefootmark{a}  &       & 559   & 325   & 1.4144506 & 0.14591479 & 1.86229136 & 0.0590305 & 2.08426752 & 0.04075456 \\
    A2061 &       & 104   & 327   & 0.19978195 & 0.41736057 & 0.3050484 & 0.17675514 & 0.36415021 & 0.12337895 \\
    A2063 &       & 13200 & 178   & 2.25126027 & 0.17135063 & 3.4623632 & 0.06629713 & 4.11582144 & 0.0452757 \\
    A2163   &       & 82.2  & 325   & 0.30115925 & 0.21560594 & 0.47542142 & 0.08572944 & 0.57231343 & 0.05893978 \\
    A2255  &       & 103   & 330   & 0.22038292 & 0.42010701 & 0.31426583 & 0.17925603 & 0.36477823 & 0.12534805 \\
    A2256*\tablefootmark{a}   &       & 5600  & 63    & 0.26191186 & 0.94219051 & 0.41423771 & 0.41105744 & 0.5033053 & 0.2889608 \\
    \hspace{0.88cm}** &       & "     & "     & 0.36858263 & "     & 0.58285482 & "     & 0.70815106 & " \\
    A2345\tablefootmark{a}  & E     & 188   & 325   & 0.40565679 & 0.46979439 & 0.58650056 & 0.20121278 & 0.68492906 & 0.1408277 \\
          & W     & 291   & 325   & 0.63311957 & 0.41532312 & 0.88275052 & 0.17760186 & 1.01516114 & 0.12425583 \\
    A2433\tablefootmark{a}  &       & 5310  & 74    & 0.90045259 & 0.23038292 & 1.24366136 & 0.09510189 & 1.42119871 & 0.06597796 \\
    A2744  &       & 98    & 325   & 0.33060506 & 0.36772336 & 0.49928471 & 0.15471763 & 0.59295149 & 0.10782785 \\
    A3376* & E     & 122   & 1400  & 0.11064754 & 0.77761 & 0.17241571 & 0.33765345 & 0.20806628 & 0.23709374 \\
    \hspace{0.88cm}** &       & "     & "     & 0.22929241 & "     & 0.35720307 & "     & 0.43103649 & " \\
    \hspace{0.88cm}* & W     & 113   & 1400  & 0.23835318 & 0.74241678 & 0.36181876 & 0.32232375 & 0.43181353 & 0.22632151 \\
    \hspace{0.88cm}** &       & "     & "     & 0.29055306 & "     & 0.44096404 & "     & 0.52624311 & " \\
    A3411\tablefootmark{a}  &       & 2074  & 74    & 0.37838351 & 0.66285438 & 0.57534757 & 0.28669623 & 0.68692337 & 0.20112495 \\
    A3667* & NW     & 81000 & 85.5  & 1.12160849 & 0.74084647 & 1.69539985 & 0.32169081 & 2.01979039 & 0.22588522 \\
    \hspace{0.88cm}** &       & "     & "     & 1.94806319 & "     & 2.94411187 & "     & 3.50727304 & " \\
    \hspace{0.88cm}* & SE     & 3200  & 120   & 0.29888475 & 0.76589471 & 0.45846131 & 0.3326574 & 0.54960282 & 0.23360077 \\
    \hspace{0.88cm}** &       & "     & "     & 0.4413838 & "     & 0.67693281 & "     & 0.81147522 & " \\
    A4038 &       & 380   & 606   & 0.66699485 & 0.29244277 & 0.99925204 & 0.1211396 & 1.18124332 & 0.08411184 \\
    Sausage\tablefootmark{a}  &       & 96    & 1400  & 0.5520268 & 0.36397161 & 0.84106105 & 0.15288354 & 1.00238003 & 0.10650713 \\
    El-Gordo\tablefootmark{a}  & E     & 1.2   & 610   & 0.0630405 & 0.43938335 & 0.09906693 & 0.18614714 & 0.11969715 & 0.12994562 \\
          & NW    & 19    & 610   & 0.4686601 & 0.39711602 & 0.6937004 & 0.16824022 & 0.81742537 & 0.11744513 \\
          & SE    & 3     & 610   & 0.2968055 & 0.36917742 & 0.42450579 & 0.15640373 & 0.49308657 & 0.10918229 \\
    MACSJ1149.5+2223\tablefootmark{a}  & E     & 23    & 323   & 0.40320491 & 0.14418144 & 0.6317179 & 0.05369269 & 0.75456363 & 0.03631911 \\
          & W     & 17    & 323   & 0.17667527 & 0.16584155 & 0.30585128 & 0.06175771 & 0.38004375 & 0.04177431 \\
    MACSJ1752.0+4440\tablefootmark{a}  & NE    & 410   & 323   & 0.8970468 & 0.2477492 & 1.35973074 & 0.10088686 & 1.61296805 & 0.06976237 \\
          & SW    & 163   & 323   & 0.65594638 & 0.17215371 & 1.03400781 & 0.06619724 & 1.2408276 & 0.04513837 \\
    PLCK G287.0+32.9\tablefootmark{a}  & N     & 550   & 150   & 0.76718182 & 0.22834563 & 1.14439679 & 0.09248678 & 1.34808379 & 0.06387085 \\
          & S     & 780   & 150   & 1.05286647 & 0.19267677 & 1.50810683 & 0.07734638 & 1.74622218 & 0.0532997 \\
    RXCJ1314.4-2515 & E     & 28    & 610   & 0.3267129 & 0.24411178 & 0.49151896 & 0.09939533 & 0.5812587 & 0.06872927 \\
          & W     & 64.8  & 610   & 0.55498502 & 0.2989983 & 0.79853002 & 0.12486616 & 0.92886177 & 0.08686826 \\
    ZwCl0008.8-5215\tablefootmark{a}  & E     & 820   & 241   & 0.72006669 & 0.41263489 & 0.99947583 & 0.17650258 & 1.14724709 & 0.12349511 \\
    ZwCl2341.1+0000\tablefootmark{a}  & N     & 36    & 157   & 0.05384757 & 0.2403794 & 0.10084052 & 0.09438838 & 0.12993831 & 0.06469435 \\
          & S     & 70    & 157   & 0.10307667 & 0.42474066 & 0.16841496 & 0.17900128 & 0.20671874 & 0.12480086 \\
      \hline
      \end{tabular}
          \tablefoot{Columns, and footnotes, are as in Table~\ref{tab:sn_halos}, with the addition of the second column which indicates the specific relic to which we are referring to.
}
\label{tab:sn_relics}%
\end{table*}

\begin{table}[ht!]
\centering
\caption{\label{tab:Beq} Magnetic field equipartition estimates}
\sisetup{round-mode = places,
round-precision = 3, scientific-notation = true}
\begin{tabular}{ll
	S[round-precision=1, table-omit-exponent]
	S[round-precision=1, table-omit-exponent]  
	r}
    \hline\hline
    \phantom{\Big|}
    Cluster& &$B_\mathrm{eq} \mathrm{\,\left(\mu G\right)}$ & $B_\mathrm{eq}^\mathrm{rev} \mathrm{\,\left(\mu G\right)}$ & Reference \\
    \hline
    \multicolumn{5}{c}{\textbf{Halos}}  \phantom{\Big|}  \\
    \hline\\[-0.5em]
    Bullet &&       & 1.2   & 1 \\
    A0520 && 1.43  &       & 2\\
    A0665 && 0.61  &       &3 \\
    A0754 && 0.66  &       & 4\\
    Coma && 0.45  & \multicolumn{1}{c}{0.7-1.9}   & 1, 5, 6, 21 \\
    A1914 && 0.55  & 1.3   & 1, 4 \\
    A2163 && 0.70  & 0.97  & 1, 3 \\
    A2219 && 0.4   & 0.7   & 7  \\
    A2255 && 0.53  &       & 8 \\
    A2256 && >1    & 1.1   & 1, 9\\
    A2319 && 0.53  &       & 10 \\
    A2744 && 0.5   & 1.0   & 1, 7\\
    A3562 && 0.4   &       & 11 \\
    MACSJ0717.5+3745 && 3.43  & 6.49  & 12 \\
    RXCJ2003.5-2323 &&       & 1.7   & 13\\
    \hline
    \multicolumn{5}{c}{\textbf{Relics}} \phantom{\Big|} \\
    \hline\\[-0.5em]
   Toothbrush &&       & \multicolumn{1}{c}{7.4-9.2}   &14 \\
    A0085 && \multicolumn{1}{c}{1.1-2.4}   &       &  15, 20\\
    A0754 && 0.3   &       & 15 \\
    AS753 && 1.3   &       & 15 \\
    A1240 &N& 1     & 2.4   & 16\\
          &S& 1     & 2.5   & 16 \\
    A1367 && 1     &       &15 \\
    Coma && 0.55  &\multicolumn{1}{c}{0.7-1.7}      & 5, 6, 21\\
    A1664 && 0.88  &       & 2, 17\\
    A2163 && 2.2  &       & 22 \\
    A2255 && 0.53  &       & 8 \\
    A2256 && >1    &       & 9 \\
    A2345 &E& 0.8   & 2.2   & 16\\
          &W& 1     & 2.9   & 16\\
    A2744 && 0.6   & 1.3   & 7\\
    A3667 &NW& \multicolumn{1}{c}{1.5-2.5} &       & 6, 18\\
    A4038 && 3     &       & 17  \\
    ZwCl0008.8-5215 &E& 2.5   & 6.6   & 19 \\
    ZwCl2341.1+0000 	& N & 0.59 & 0.64 & 23 \\
          				& E & 0.54 & 0.66 &23 \\
    \hline
    \end{tabular}
  \tablefoot{Magnetic field estimates based on the equipartition theorem. $B_\mathrm{eq}$ is the equipartition magnetic field, while $B_\mathrm{eq}^\mathrm{rev}$ is the so-called revised equipartition magnetic field.
  }
  \tablebib{\\
  (1) \cite{petrosian2006}; (2) \cite{govoni2001}; (3) \cite{feretti2004}; (4) \cite{bacchi2003}; (5) \cite{giovannini1991}; (6) \cite{govoni2004a}; (7) \cite{orru2007}; (8) \cite{feretti1997a}; (9) \cite{clarke2006}; (10) \cite{feretti1997}; (11) \cite{venturi2003}; (12) \cite{pandey-pommier2013}; (13) \cite{giacintucci2009}; (14) \cite{weeren2012}; (15) \cite{chen2008}; (16) \cite{bonafede2009a}; (17) \cite{kale2012a}; (18) \cite{johnston-hollitt2004}; (19) \cite{weeren2011b}; (20) \cite{ensslin1998}; (21) \cite{thierbach2003}; (22) \cite{feretti2001}.
  }
\end{table}

\subsection{Promising targets}
In order to pinpoint the most promising targets for ASTRO-H HXR observations, we focus on magnetic field estimates (e.g., \citealp{govoni2004a}). FR measurements range from $\sim \mu$G for merging clusters up to 10~$\mu$G for cool-core clusters \citep{carilli2002,clarke2004,vogt2005,bonafede2010,bonafede2013}. Therefore, $\mu$G-level magnetic fields in clusters are nowadays widely acknowledged. 

Analytical estimates can be obtained by the equipartition argument, i.e., minimising the total energy content, sum of the energies of the magnetic field and relativistic particles, of a radio source. While standard equipartition estimates, $B_\mathrm{eq}$, are based on computing the synchrotron radio luminosity between two fixed frequencies, the so-called revised equipartition estimates, $B_\mathrm{eq}^\mathrm{rev}$, are based on integrating over the relativistic electron distribution \citep{brunetti1997, beck2005}. In Table~\ref{tab:Beq}, we summarise all the (revised) equipartition estimates that we could find in literature for the clusters in our sample. 

We stress that the equipartition estimates compiled in Table~\ref{tab:Beq} should be considered only as reference values for comparison purposes. Not only these suffer from several uncertainties such as, e.g., the ratio of electrons to protons in the source and the low-energy cutoff of the electron distribution which are, a priori, unknown (see \citealp{govoni2004a} for a discussion), but also one could question whether the equipartition assumption is motivated as the evolution of relativistic electrons and magnetic fields in clusters may be decoupled given the short lifetime of the electrons compared to that of clusters. We also note that different authors often adopt different assumptions when applying the equipartition argument.

We consider a given halo or relic to be a promising candidate for ASTRO-H if its $B_\mathrm{dt}^\mathrm{s/n}$ in 100~ks of observations is $\ge1$~$\mu$G; while if smaller, we consider it promising if that value is comparable to, or higher than, the corresponding (revised) equipartition estimate (where available). The value of $1$~$\mu$G is just a rough educated guess assuming as prototype magnetic field value that of the Coma cluster for which the volume-average is $\sim2\mathrm{\,\mu G}$ \citep{bonafede2010,bonafede2013}. We summarise all objects that meet these criteria in Table~\ref{tab:summary}.

There is a crucial point to be considered when looking at our summary list of promising targets in Table~\ref{tab:summary}. That is the size in the sky of a given halo; this plays a crucial role for our predictions. Several objects in Table~\ref{tab:summary} have a size well below the ASTRO-H HXI FoV of $9\times9=81$~arcmin$^2$. The smaller the size of a given object with respect to the FoV, the higher is the contamination given by the thermal ICM emission which obviously peaks in the centre of clusters. This is particularly important for radio relics in the cluster's outskirts which are the best targets for HXR IC detection, as clear from the results presented in the tables of this work. On the other hand, an object much larger than the ASTRO-H FoV will call for multiple pointings with the corresponding increase in observation time. ASTRO-H observation simulations of these objects, considering their morphology, is needed in order to give a definitive answer on which are the most promising targets. 

Note that we include Coma in our list of promising targets mainly because it is one of the best studied clusters and its observation by ASTRO-H is desirable. In fact, Coma represents one of the best targets where to detect and map for the first time the bulk of the ICM turbulent velocities with the Soft X-ray Spectrometer (SXS) on-board ASTRO-H \citep{kitayama2014}. However, it does not represent the best hopes for the HXR IC detection.

In order to facilitate the interpretation of our results, we provide a simple, approximate, scaling of the ASTRO-H observation time with the magnetic field value $B_\mathrm{dt}^\mathrm{s/n}$ corresponding to the signal-to-noise approach. We note that the number of background counts scales linearly with time $N_\mathrm{B} \propto T_\mathrm{obs}$, and, under the assumption that $N_\mathrm{B} \gg N_\mathrm{S}$, one finds that the number of source counts necessary to reach a fixed signal-to-noise ratio scales as $N_\mathrm{S} \sim \sqrt{T_\mathrm{obs}}$. From Eq.~\ref{eq:Ns}, $F_x \propto N_\mathrm{S}/T_\mathrm{obs} \propto T_\mathrm{obs}^{-1/2}$, and, eventually, from Eq.~\ref{eq:Banalytic1}, we obtain that, for a given $\alpha$,
\begin{equation}
B_\mathrm{dt}^\mathrm{s/n} \propto T_\mathrm{obs}^{\frac{1}{2\left(1 + \alpha\right)}}\,.
\label{eq:BvsT}
\end{equation}

Analogously, it can be shown that for a fixed observation time, i.e., fixed $N_B$, the detectable magnetic field depends on the signal-to-noise ratio as follows:
\begin{equation}
B_\mathrm{dt}^\mathrm{s/n} \propto \left(\mathrm{s/n}\right)^{-\frac{1}{\left(1 + \alpha\right)}}.
\label{eq:BvsSN}
\end{equation}

\subsection{Comments on individual clusters}
\label{sec:comments}
\begin{figure*}[hbt!]
\includegraphics[width=0.5\textwidth]{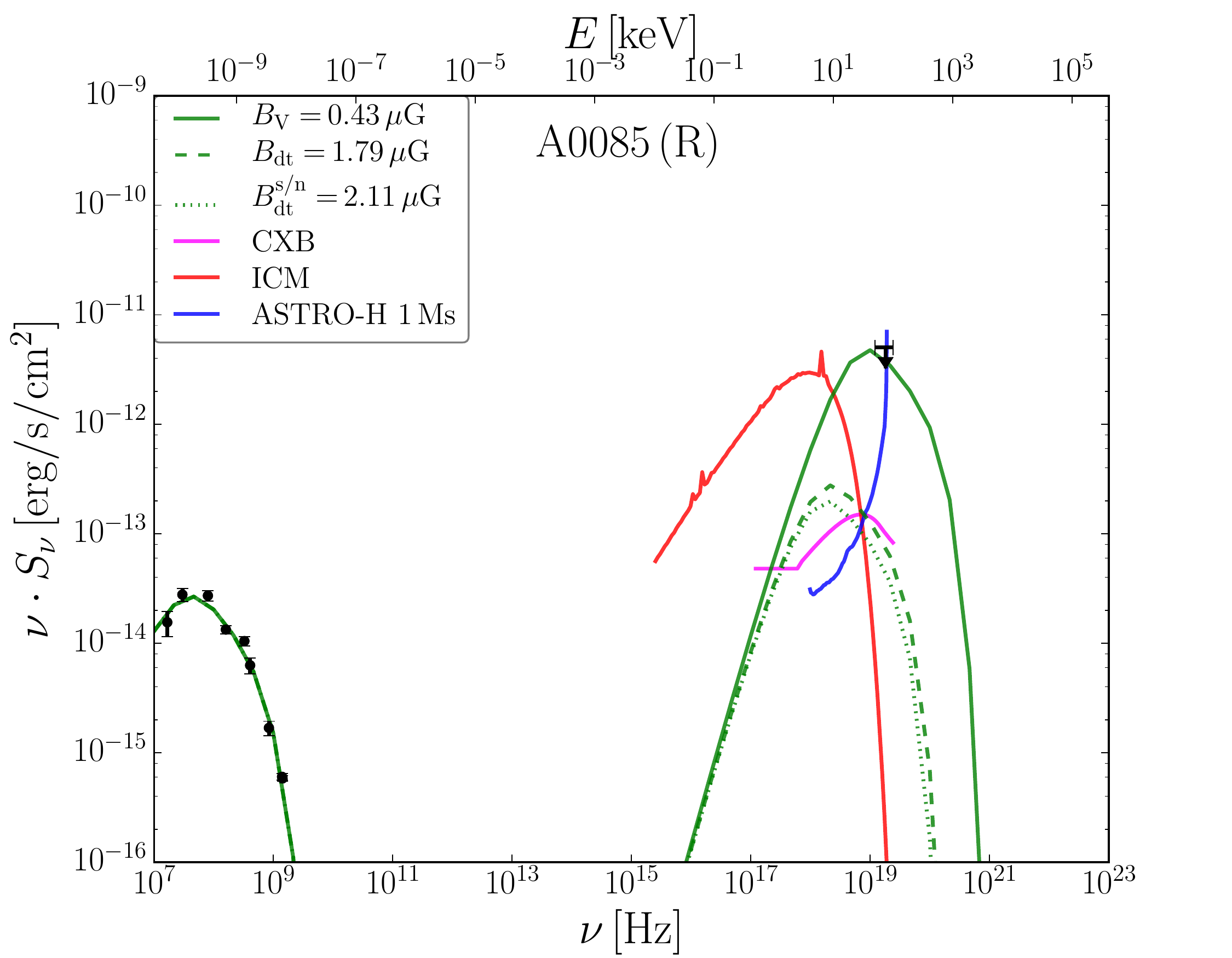}
\includegraphics[width=0.5\textwidth]{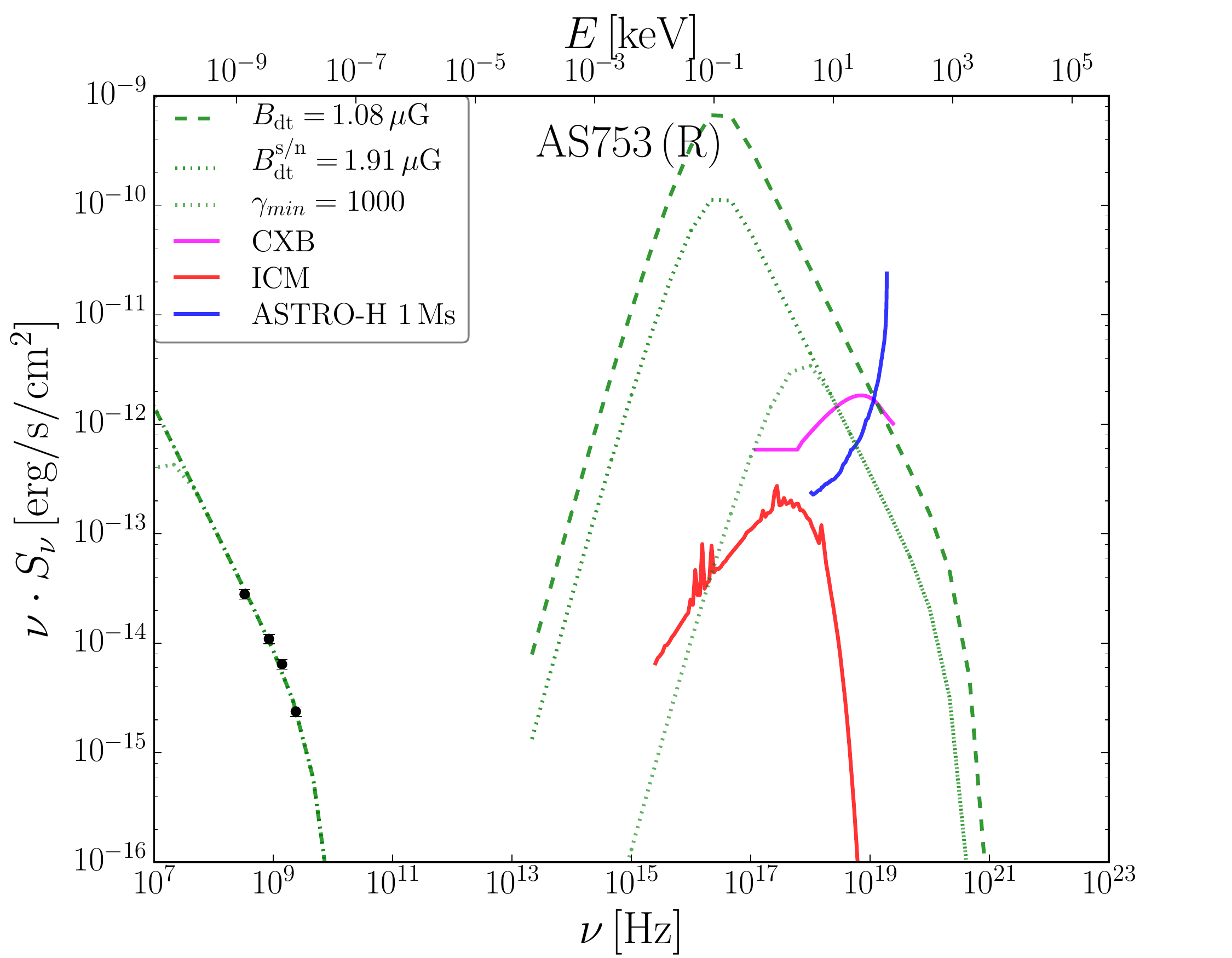}
\includegraphics[width=0.5\textwidth]{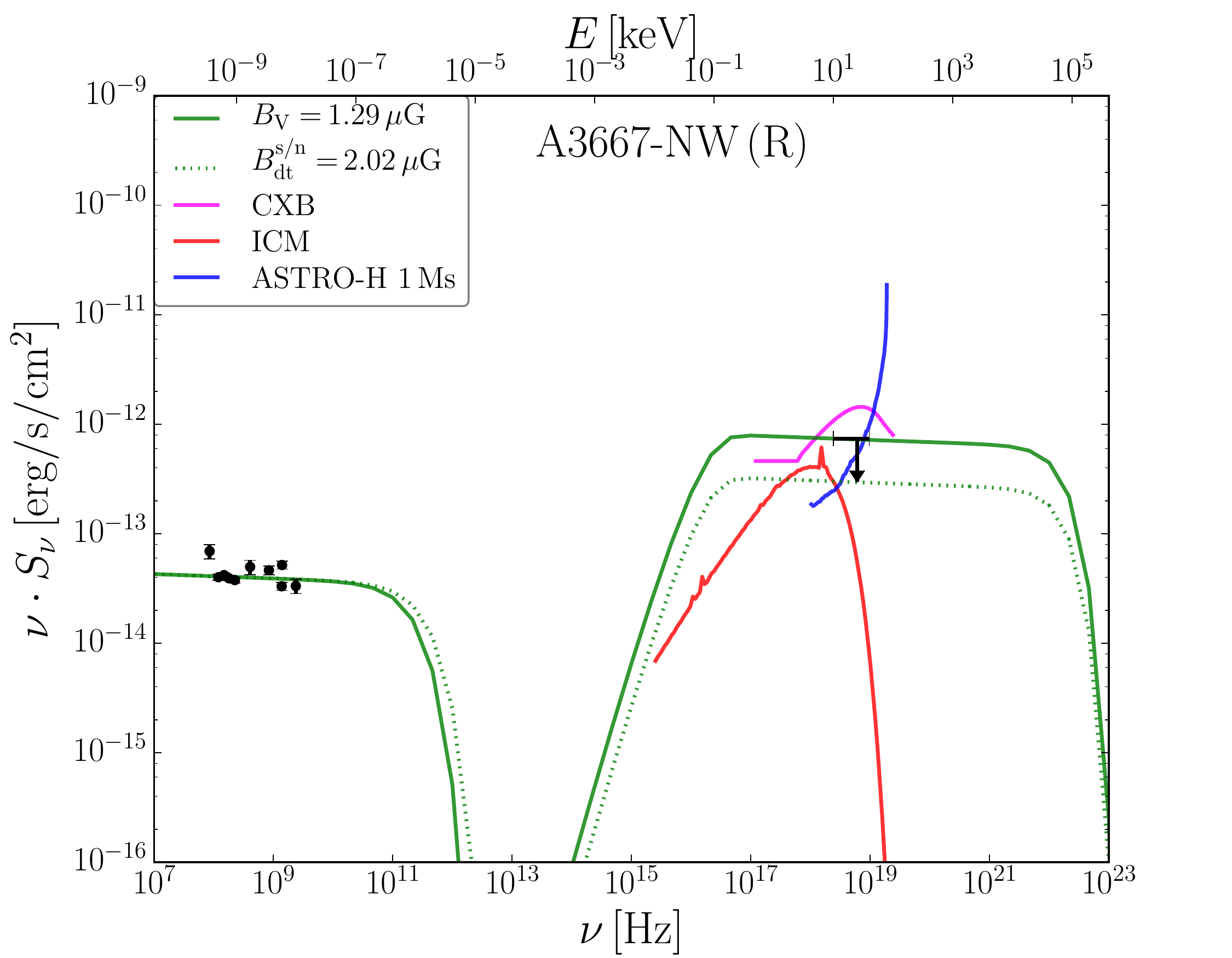}
\includegraphics[width=0.5\textwidth]{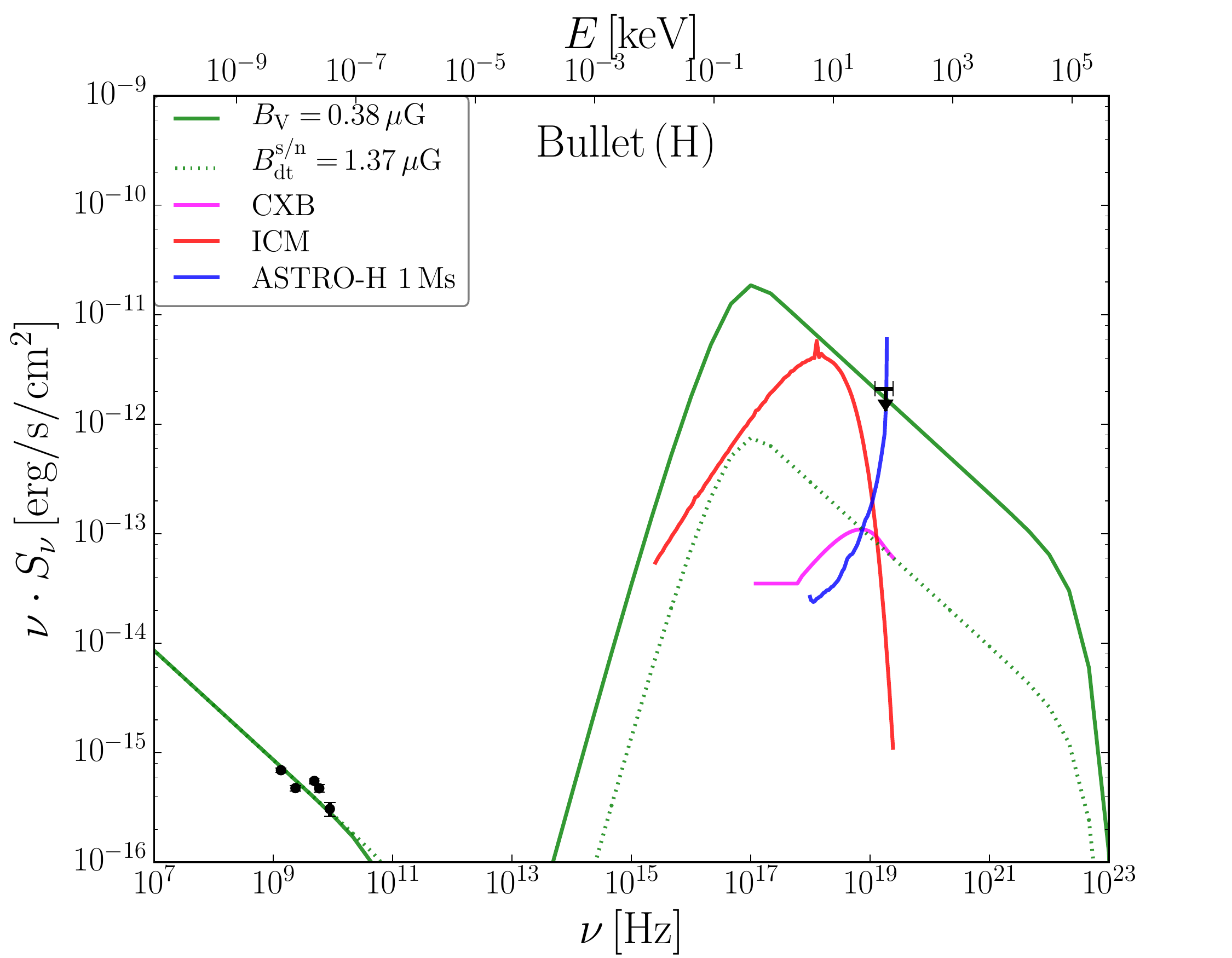}
\includegraphics[width=0.5\textwidth]{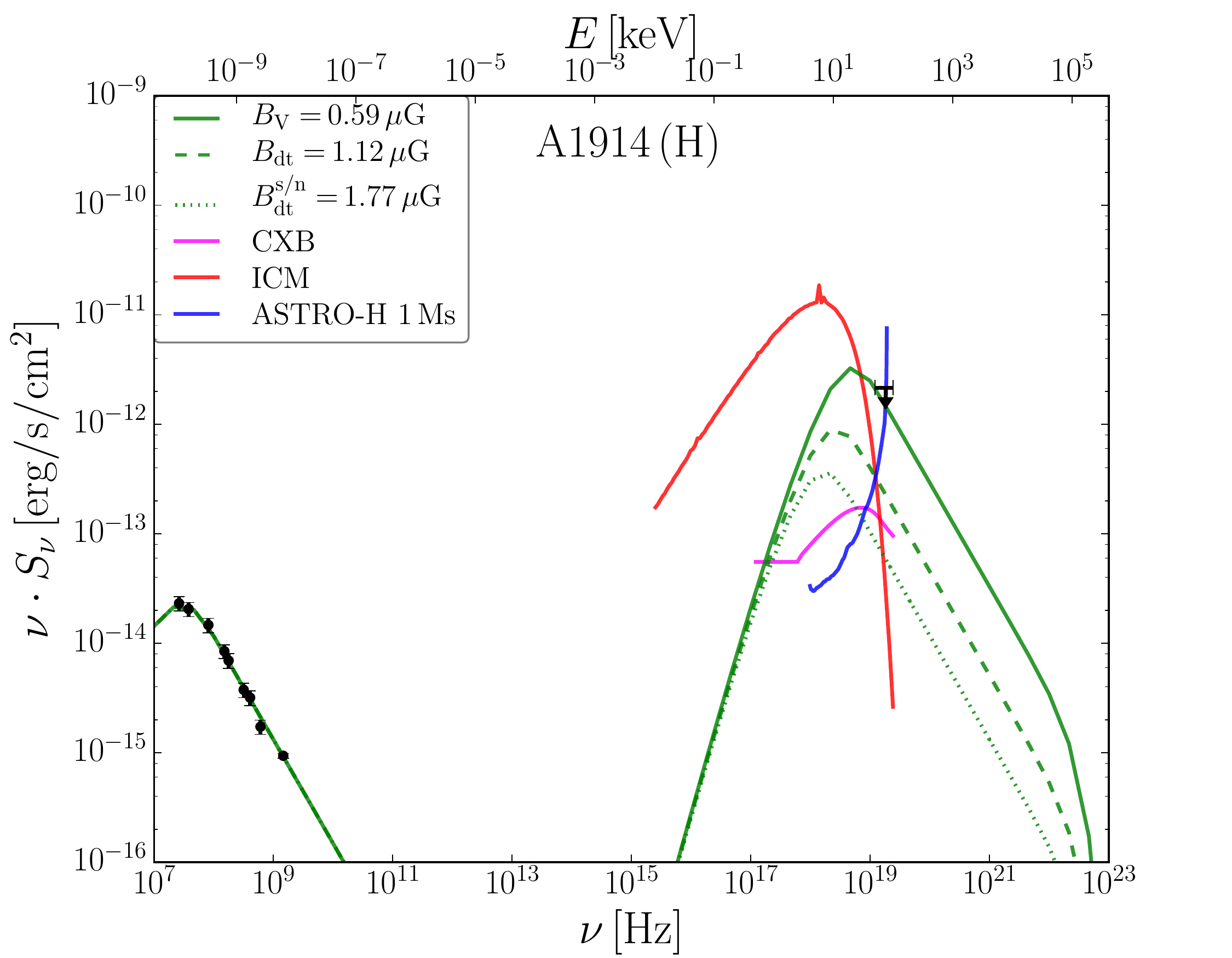}
\includegraphics[width=0.5\textwidth]{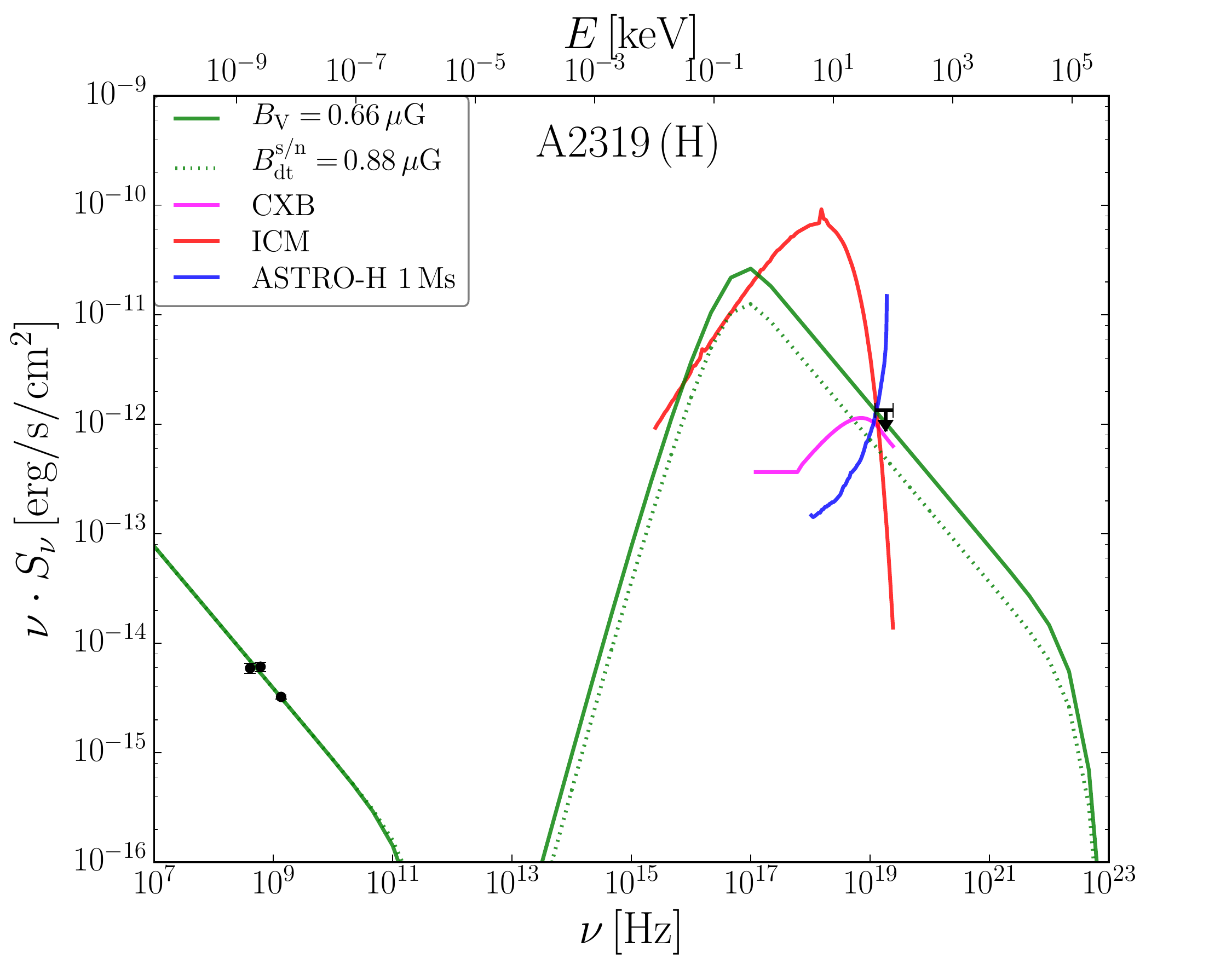}
\caption{\footnotesize{Synchrotron and IC modelling of the A0085, AS753, and A3667-NW radio relics (R), and of the Bullet, A1914 and A2319 radio halos (H). Caption as in Fig.~\ref{A2255Halo}.} For A3667-NW and A2319, the lines corresponding to $B_\mathrm{dt}$ are not shown as they correspond to that of $B_\mathrm{V}$.}
\label{modeling_1}
\end{figure*}

\begin{figure*}[hbt!]
\includegraphics[width=0.5\textwidth]{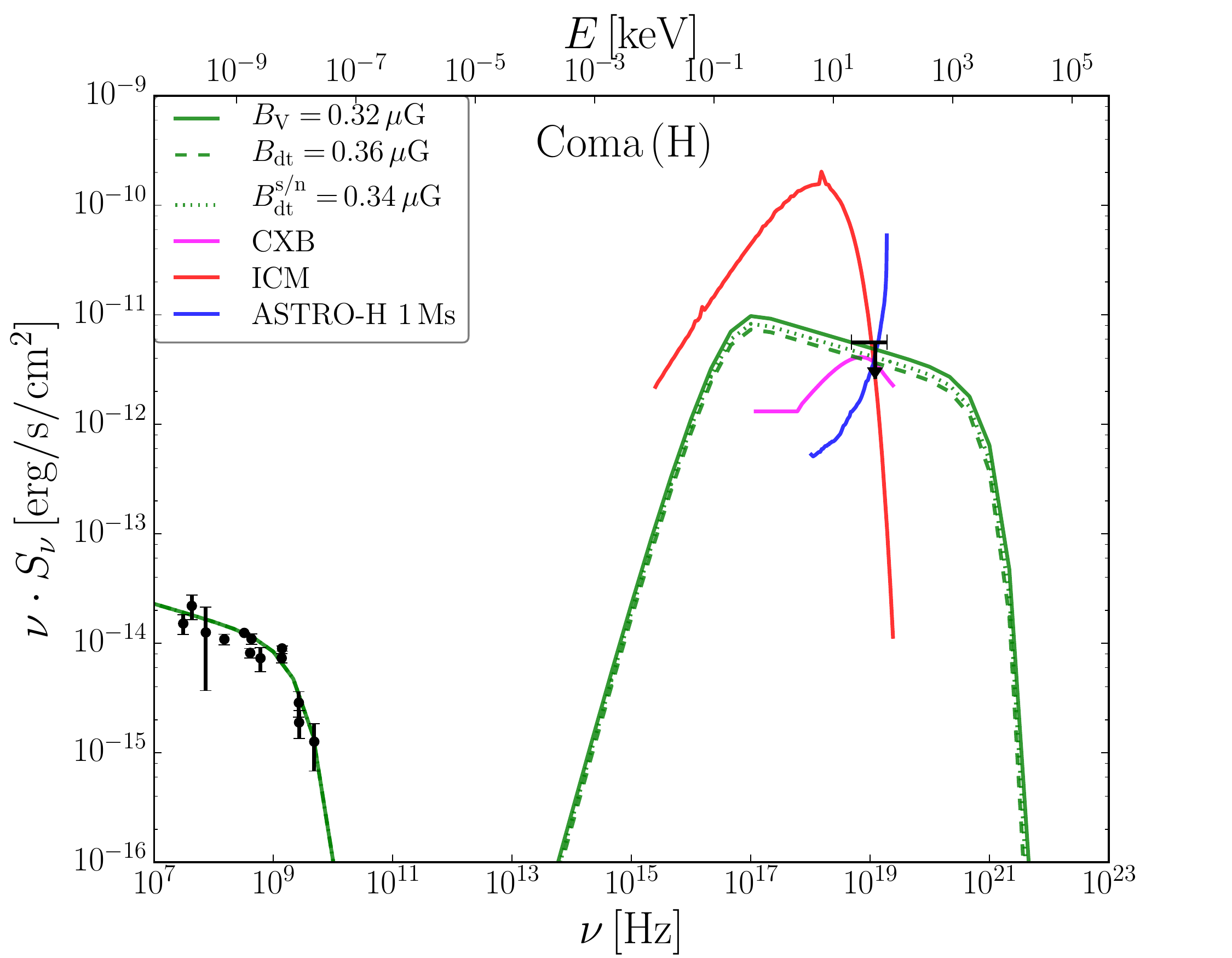}
\includegraphics[width=0.5\textwidth]{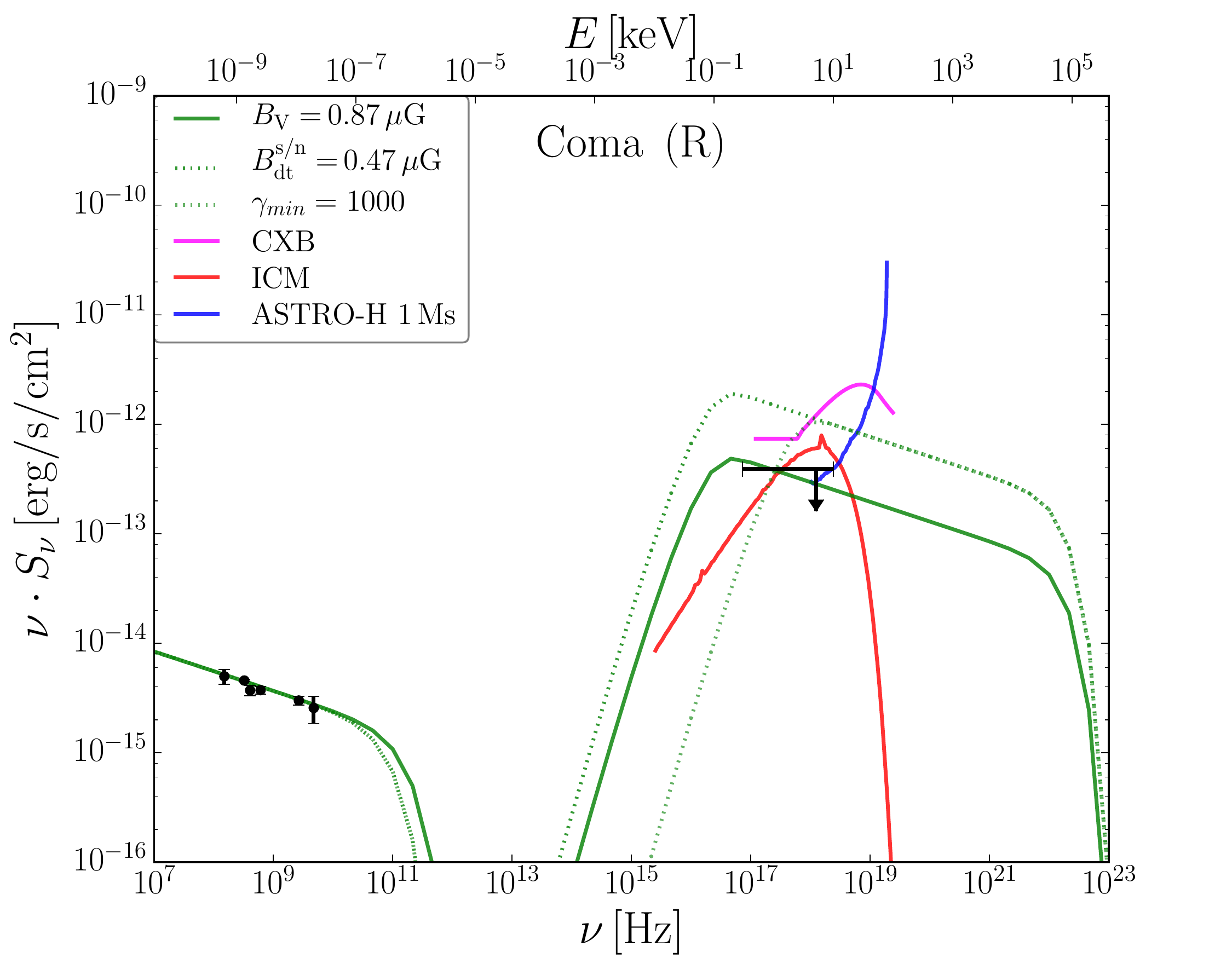}
\caption{\footnotesize{Synchrotron and IC modelling of the radio halo (H) and relic (R) hosted by Coma. Caption as in Fig.~\ref{A2255Halo}. In the case of the relic, we do not show the line corresponding to $B_\mathrm{dt}$ because it gives worst estimates than that of $B_\mathrm{V}$. This is due to our assumption of uniform radio sources which is particularly inappropriate in case of objects with a large size in the sky, like Coma. See main text for details.}}
\label{modeling_2}
\end{figure*}

We report in the following some comments on the individual clusters in our final list of most promising objects - A0085,  AS753, A3667, Bullet, A1914, A2255, A2319,  and Coma - while we leave the rest for Appendix~\ref{sec:app_comments}. We show the corresponding synchrotron and IC modelling in Figs.~\ref{A2255Halo},  \ref{modeling_1}, and \ref{modeling_2}. 

\vspace{0.1cm}
{\bf A0085.} As can be seen from the top-left panel of Fig.~\ref{modeling_1}, the radio relic of A0085 appears to have a low-energy cutoff in the radio spectrum. While this cutoff seems to not impact our conclusion for the ASTRO-H detectability, would the magnetic field be much larger than $B_\mathrm{s/n}$, the low-energy turnover of the spectrum would move to higher energies and could impact the detectability estimates. Note also that the current HXR upper limit for this cluster \citep{ajello2010} refers to a much larger area than that of the relic. We remind the reader that the result of signal-to-noise approach should be taken with caution in this case as its assumes a straight power-law in the 20--80~keV range, while some spectral features could be expected.

\vspace{0.1cm}
{\bf AS753.} The top-right panel of Figure~\ref{modeling_1} shows the synchrotron and IC modelling for the radio relic hosted by this cluster. There are two things that should be noted. First, the very steep radio spectrum which implies a very-large IC flux in SXRs if we stick to the theoretically motivated $\gamma_\mathrm{min} = 200$. For this reason, we show in the figure also the case with $\gamma_\mathrm{min} = 1000$ and see that it does not impact much on our predictions for the HXR regime. Additionally, we must note that, lacking literature information on the gas density in this cluster, we used the phenomenological ICM model of \cite{zandanel2014a}. This is, obviously, a source of a uncertainty and likely resulted in an underestimation of the gas density at the relic location.

\vspace{0.1cm}
{\bf A3667.} This cluster hosts the prototype double relic system \citep{rottgering1997}. The $1.9\mathrm{\,Mpc}$ northwestern relic contains most of the diffuse emission. HXR emission from the A3667-NW relic has been studied by \cite{ajello2009} and \cite{nakazawa2009}. We use the latter upper limit on the HXR non-thermal emission since it specifically refers to the relic region. In addition, a detailed X-ray study of the ICM can be found in \cite{finoguenov2010} and \cite{akamatsu2012a}. We show the corresponding synchrotron and IC modelling in the central-left panel of Figure~\ref{modeling_1}. The current HXR upper limit is at the same level of our $B_\mathrm{dt}$ prediction due to our assumption of uniform radio sources. We try to asses this issue in Table~\ref{tab:sn_relics} providing an alternative estimate by assuming that all the radio emission comes from a region within the ASTRO-H HXI FoV.

Recently, \cite{kitayama2014} pointed out that the A3667-NW relic is one of the more promising targets for ASTRO-H to detect non-thermal HXR emission. They predict that HXI can constrain the magnetic field to $B\geq 4\mathrm{\,\mu G}$ in $200\mathrm{\,ks}$ of observations. This is significantly higher, about a factor of two, than our predicted values of Tables \ref{tab:relics} and \ref{tab:sn_relics}. However, their approach is quite different as they estimate the $90$\% confidence-level (CL) upper limit obtainable with ASTRO-H in $200\mathrm{\,ks}$ of observations. Moreover, the discrepancy is mostly due to the fact that they assume that the SXR observed by \cite{finoguenov2010} with XMM-\emph{Newton} is of non-thermal origin, a hypothesis that cannot be excluded, and extrapolate this to HXRs with a power-law spectrum with a photon index of $\alpha = 2$.

\vspace{0.1cm}
{\bf Bullet.} We include the Bullet cluster for the broad interest in it, and because it has been recently observed by NuSTAR, even if it does not fulfil our criteria to be in the promising targets list. \cite{wik2014} recently reported the results of the NuSTAR observations of the Bullet cluster where they do not find any evidence of non-thermal HXR emission from the radio halo. We derive slightly better prospects for ASTRO-H than their derived upper limit as can be seen in the central-right panel of Fig.~\ref{modeling_1}. Although ASTRO-H and NuSTAR have comparable performance in HXR, the former has the advantage of the simultaneous operation of the SXI, allowing for a better determination of the thermal ICM emission.

\vspace{0.1cm}
{\bf A1914.} The radio halo of A1914 has a very steep spectrum \citep{bacchi2003, komissarov1994}. Note that while we modelled the low-energy flattening in radio with a spectral cutoff, the data is insufficient to distinguish it from a broken power-law, as can be seen in the bottom-left panel of Fig.~\ref{modeling_1}. In this sense, therefore, the result of signal-to-noise approach should be taken with caution as its assumes a straight power-law in the 20--80~keV range, while some spectral features could be expected. The X-ray upper limit is from \cite{ajello2010}.

\vspace{0.1cm}
{\bf A2255.} This cluster hosts both a radio halo and a relic \citep{feretti1997a}. Since the relic is at $\sim10'$ from the cluster centre and the XMM-\emph{Newton} FoV is $30'$, current HXR upper limits likely contain contributions from both the halo and the relic \citep{turner2001,wik2012}. The synchrotron and IC emission modelling of the radio halo is shown in Figure~\ref{A2255Halo}. Note that while we use radio observations from \cite{feretti1997a} and \cite{govoni2005}, more recent observations were performed by \cite{pizzo2009}. However, considering the different areas used for the flux extraction, the latter results are comparable with the former for the purpose of our modelling.

\vspace{0.1cm}
{\bf A2319.} We use the HXR upper limit from \cite{ajello2009} and note that we obtain a more constraining lower limit $B_\mathrm{V}>0.7$~$\mu$G due to our assumptions regarding the magnetic field distribution (see Sect.~\ref{sec:processes}). The bottom-right panel of Figure~\ref{modeling_1} shows the corresponding synchrotron and IC modelling. The current HXR upper limit is at the same level of our $B_\mathrm{dt}$ prediction due to our assumption of uniform radio sources, where we try to asses this issue in Table~\ref{tab:sn_halos} providing an alternative estimate by assuming that all the radio emission comes from a region within the ASTRO-H HXI FoV.

\vspace{0.1cm}
{\bf Coma.} We included the Coma cluster more for \emph{historical} reasons than for it being a real promising target. Coma is one of the best studied clusters and hosts the prototype radio halo and relic (e.g., \citealp{thierbach2003,brown2011b}). A HXR excess at the halo location was claimed by \cite{rephaeli2002} and \cite{fusco-femiano2004}, but later never confirmed \citep{wik2009, wik2011, gastaldello2014}. The radio relic (1253+275) is located at $75'$ from the cluster centre. Fig.~\ref{modeling_2} shows the synchrotron and IC modelling for both the radio halo and relic. 

Note that, for the relic, the current upper limit \citep{feretti2006} appears more constraining than what is achievable with ASTRO-H HXI in 1 Ms of observation according to the signal-to-noise analysis. However, this upper limit is determined in the $20-80\mathrm{\,keV}$ band and could be evaded with a low-energy spectral cutoff or flattening. Therefore, as for AS753, we also show the case with $\gamma_\mathrm{min} = 1000$ in the right panel of Fig.~\ref{modeling_2}. Also in the case of the halo, the current HXR upper limit is at the same level of our predictions for $B_\mathrm{dt}$. This is due to our assumption of uniform radio sources which is particularly inappropriate in case of objects with a large size in the sky, like Coma. In fact, in Tables~\ref{tab:sn_halos} and \ref{tab:sn_relics}, we try to asses this issue providing alternative estimates by i) assuming that all the radio emission comes from a region within the ASTRO-H HXI FoV (we show this also for few other objects), and ii) computing the thermal ICM emission for the case of the halo only within the halo boundaries (we normally integrate over the full halos size, but again, due to the large sky size of Coma, here the difference is appreciable). 

\cite{kitayama2014} recently discussed the prospects for $500 \mathrm{\,ks}$ of ASTRO-H observation of the halo region. They estimate a corresponding $90\% \mathrm{CL}$ upper limit of $B\sim 0.4 \mathrm{\, \mu G}$, slightly more optimistic than our estimates, where, however, they assumed a photon index of $\alpha = 2$.

\begin{table*}[hbt!]
\centering
\caption{\label{tab:summary}Promising targets for ASTRO-H observations}
\sisetup{round-mode = places}
\begin{tabular}{lll
	S[round-precision=2]
	S[round-precision=0]
	S[round-precision=1]
	S[round-precision=2]
	S[round-precision=2]
	S[round-precision=2]	
	ll}
    \hline\hline
    \multicolumn{1}{l}{Cluster} & \multicolumn{1}{l}{Source} & \multicolumn{1}{c}{Type} & \multicolumn{1}{c}{$z$}  & \multicolumn{1}{c}{$\mathrm{Size}$}  &\multicolumn{1}{c}{$\alpha$} & \multicolumn{1}{c}{$B_\mathrm{dt}^\mathrm{s/n} (100\mathrm{\,ks})$} & \multicolumn{1}{c}{$B_\mathrm{dt}^\mathrm{s/n} (1\mathrm{\,Ms})$} & \multicolumn{1}{c}{$B_\mathrm{dt} (1\mathrm{\,Ms})$} & \multicolumn{1}{c}{$B_\mathrm{eq}$}& \multicolumn{1}{c}{$B_\mathrm{eq}^\mathrm{rev}$}  \phantom{\Big|} \\
    \hline\\[-0.5em]
    Bullet &       & H     & 0.296 & 19    & 1.5   & 0.84246925 & 1.36526871 & 1.00124787 &       & 1.2 \\
    Toothbrush\tablefootmark{a}  &       & R     & 0.225 & 18    & 1.1   & 1.08457717 & 1.9484692 & 1.65  &       & 7.4-9.2 \\
    A0013\tablefootmark{a }  &       & R     & 0.0943 & 5     & 2.3   & 1.10423783 & 1.62766585 & 1.98  &       &  \\
    A0085 &       & R     & 0.0551 & 26    & 1.61282313 & 1.32874398 & 2.11412293 & 1.79  & 1.1-2.4 &  \\
    AS753\tablefootmark{a}  &       & R     & 0.014 & 318   & 2.1   & 1.30138923 & 1.90967389 & 1.08  & 1.3   &  \\
    Coma  &       & H     & 0.0231 & 710   & 1.16  & 0.19930699 & 0.34172555 & 0.36  & 0.5   & 0.7-1.9 \\
          &       & R     &       & 400   & 1.18  & 0.27073036 & 0.46693454 & 0.28  & 0.6   & 0.7-1.7 \\
    A1914 &       & H     & 0.1712 & 30    & 1.9473499 & 1.18390008 & 1.77111885 & 1.12  &  0.6  & 1.3 \\
    A2048\tablefootmark{a }  &       & R     & 0.0972 & 7     & 2.29  & 1.4144506 & 2.08426752 & 2.43  &       &  \\
    A2063  &       & R     & 0.0349 & 3     & 1.20592569 & 2.25126027 & 4.11582144 & 5.51  &       &  \\
    A2255 &       & H     & 0.0806 & 79    & 1.60023457 & 0.6077641 & 0.95894955 & 0.60424839 & 0.5   &  \\
    A2319 &       & H     & 0.0557 & 198   & 1.65  & 0.5647961 & 0.87793851 & 0.64910507 & 0.5   &  \\
    A3667 & NW     & R     & 0.0556 & 250   & 1.02  & 1.12160849 & 2.01979039 & 0.94  & 1.5-2.5 &  \\
    PLCK G287.0+32.9\tablefootmark{a }  & N     & R     & 0.39  & 7     & 1.26  & 0.76718182 & 1.34808379 & 1.57  &       &  \\
          & S     & R     & 0.39  & 6     & 1.54  & 1.05286647 & 1.74622218 & 4.14  &       &  \\
    \hline
    \end{tabular}
  \tablefoot{We summarise all objects that meet the criteria to be considered promising targets (see main text for details) and the corresponding magnetic field values testable by ASTRO-H in 100~ks and 1~Ms of observations in $\mu$G. We show the most optimistic values obtained adopting the signal-to-noise approach, $B_\mathrm{dt}^\mathrm{s/n}$, and also the more pessimistic estimates based on the intersection point between the background emission and the ASTRO-H sensitivity, $B_\mathrm{dt}$, for 1~Ms of observations. We indicate the type of source, either radio halo (H) or radio relic (R), and, in case more than one relic is present in a given cluster, we explicitly mention to which source we are referring to. We also report the approximate surface on the sky in arcmin$^2$ and the spectral index $\alpha$. Finally, we also show (revised) equipartition estimates, in $\mu$G, where available. \tablefoottext{a}{Thermal ICM emission computed using the model by \cite{zandanel2014a} for the gas density.}
  }
\end{table*}

\section{Conclusions}
\label{sec:conclusions}
In this work, we modelled the synchrotron and IC emission of all known radio halos and relics for which the spectral index can be determined. Our approach is phenomenological towards the generation mechanism of relativistic electrons. We simply assume that the same electrons generating the observed radio synchrotron emission are responsible for HXR IC emission, where we adopt a power-law distribution with low- and high-energy cutoffs that we either fit to the radio data, if steepening is observed, or we fix to theoretically motivated values. 

We provide updated lower limits on the magnetic field values for those objects for which an HXR upper limit is available. Subsequently, by considering the thermal ICM and CXB emissions, and the instrumental background, we provide predictions for the volume-average magnetic filed values up to which HXR IC emission can be detected by the ASTRO-H HXI instrument, adopting different approaches and considering different observations times. 

Our first approach is to estimate the magnetic field testable by ASTRO-H by taking as reference the intersection point between the dominant background emission, either thermal ICM or CXB, and the properly scaled ASTRO-H sensitivity for 1~Ms of observations. We then adopt a more robust approach based on photon counts in the $20-80\mathrm{\,keV}$ regime, requiring the signal-to-noise ratio to be higher than 5, where the noise includes the ICM and CXB emissions, and the instrumental background. With this approach, on which we base our conclusions, we estimate the magnetic field values testable by ASTRO-H in 100~ks, 500~ks and 1~Ms of observations. In this latter case, we find that the observation time needed to test a certain magnetic field value $B$ roughly scales as $B \propto T_\mathrm{obs}^{1/ \left(2 (1 + \alpha)\right)}$.

We identify as promising all those halos and relics for which the magnetic field testable by ASTRO-H in 100~ks is larger than $1$~$\mu$G or than the equipartition estimates, where available. We provide the corresponding list of 15 halos and relics in Table~\ref{tab:summary}. Among the most promising targets, we have the AS753 and A3667-NW relics, and the A2255 and A2319 halos, with an extension in the sky of the same order or larger than the ASTRO-H HXI FoV. Additionally, we can also identify as very promising the A0085, A2048 and A2063 relics, and the A1914 halo, with an extension smaller than the HXI FoV. The size in the sky of the considered object plays a crucial role in such observations, both for evaluating the contribution of the thermal ICM emission to the observed spectrum, and to plan multiple pointings with a corresponding increase in observation time.

We stress that we provide a theoretical expectation suffering from several approximations that we discuss openly in the text. While we estimate that most of these approximations imply conservative results, some of them could go in the other direction. In particular, the uncertainty in the ICM density distribution of some objects, particularly for radio relics in the clusters' outskirts, and the assumption of a power-law distribution of electrons down to low energies that are not tested by current radio observations, but could be tested soon by LOFAR. Therefore, while it is clear that detailed ASTRO-H observation simulations should be performed to select the best target among the ones we propose here, in particular against their angular dimension in the sky, we conclude that HXR IC detection could be at hand for ASTRO-H in several objects. 

Concluding, with the operation of NuSTAR and the upcoming launch of ASTRO-H, we will eventually be able to probe HXR IC emission in clusters for magnetic fields $\geq1$~$\mu$G. Although detailed observation simulation are needed case by case, for a few clusters we might finally break the degeneracy between the magnetic field and the relativistic electrons distribution, and be able to shed new light on the non-thermal phenomena in clusters of galaxies.

\begin{acknowledgements}
We thank the anonymous referee for helping improving our manuscript.
We thank Hiroki Akamatsu, Madoka Kawaharada, Matteo Murgia, Anders
Pinzke, Jacco Vink and Franco Vazza for useful comments and discussions. 
We also profited from the Radio Observation of Galaxy Clusters database,
collected by Klaus Dolag. This work was supported by the Netherlands
Organization for Scientific Research (NWO) through a Veni and a Vidi 
grant (FZ and SA).
\end{acknowledgements}

\bibliographystyle{bibtex/aa} 
\bibliography{library-thesis.bib}


\begin{appendix} 
\section{Other figures and comments on individual clusters}
\label{sec:app_comments}
This appendix contains the figures for the synchrotron and IC modelling of all clusters not shown in the main text. For some of these, the IC emission peaks in SXRs or EUV above the CXB and ICM emissions, and we, therefore, also show the case with $\gamma_\mathrm{min} = 1000$. 

Note that several objects, mainly radio relics, have the peculiarity of a very low thermal ICM emission, probably an underestimation by observations pointed to the cluster's center, or due to the use of the phenomenological model of \cite{zandanel2014a} for the gas density. In some cases (A0781, A2163, MACSJ1149,5-2223-E and W, MACSJ1752.0+4440-SW, PLCK~G287.0+32.9-N and S), the CXB dominates over the thermal ICM emission, and, additionally, the intersection point between the ASTRO-H and CXB emissions is at an energy lower than the rise in the CXB spectrum. The corresponding results for $B_\mathrm{dt}$ should be taken with caution and, in fact, in these cases, the signal-to-noise approach gives more conservative, and reasonable, results. 

We also remind the reader that the signal-to-noise approach is done assuming a straight power-law in the $20-80$~keV energy band (see Sect.~\ref{sec:detectability}). While in most cases the cutoffs are at much lower and higher energies, for A2063 and A0013 this method could give slightly optimistic results. 

All figures can be read as explained in the caption of Fig.~\ref{A2255Halo}.

\begin{figure*}[htb!]
\includegraphics[width=0.5\textwidth]{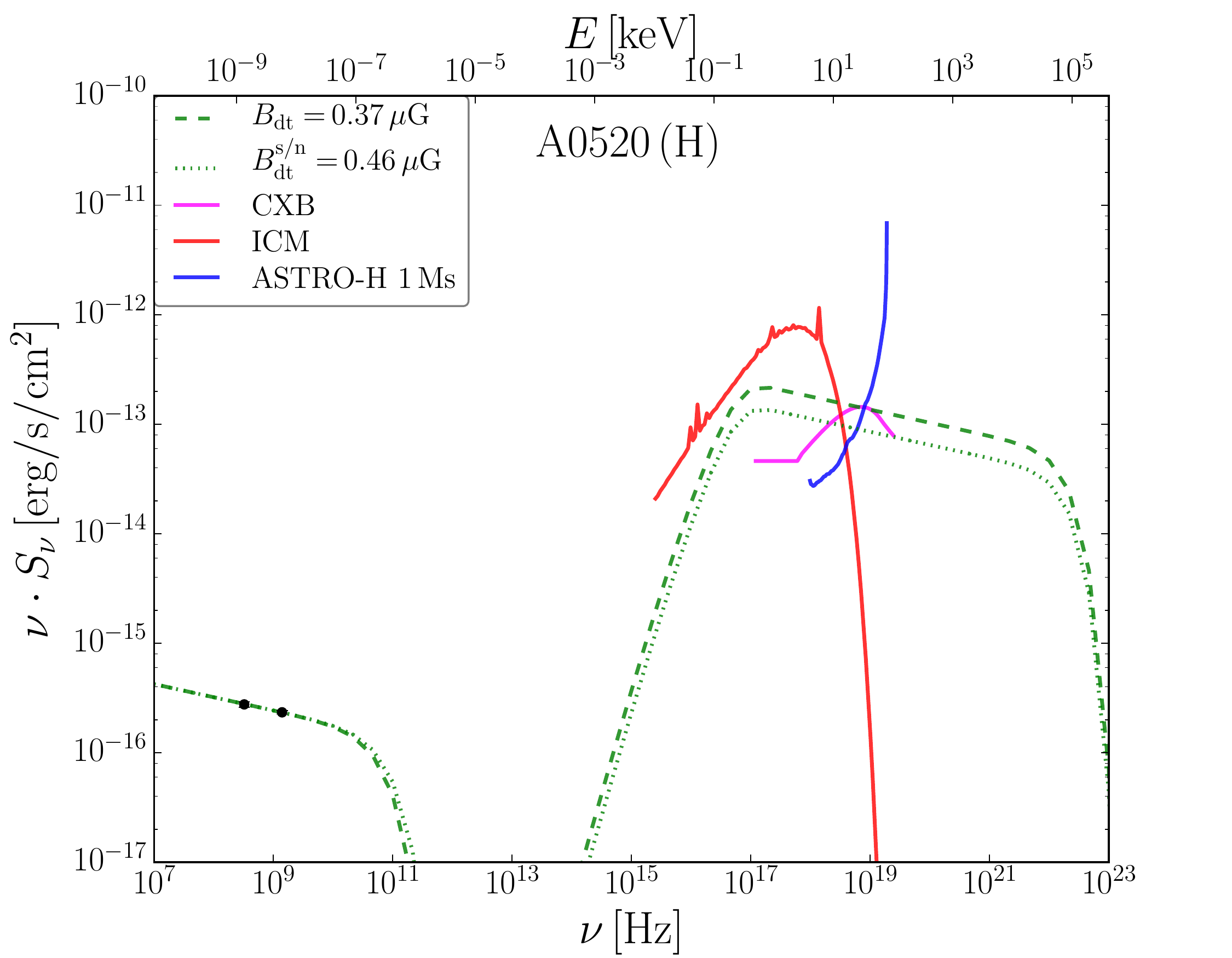}
\includegraphics[width=0.5\textwidth]{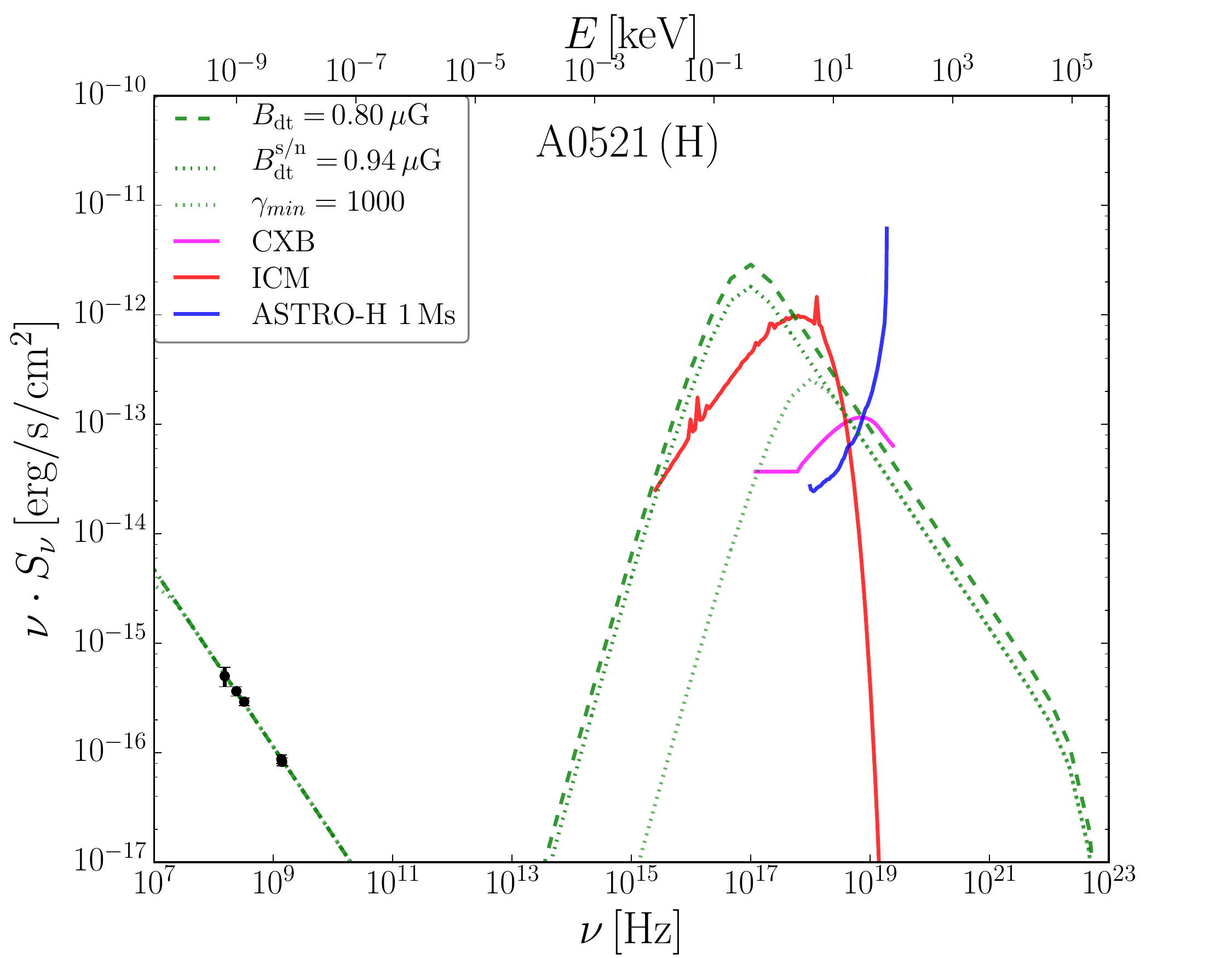}
\includegraphics[width=0.5\textwidth]{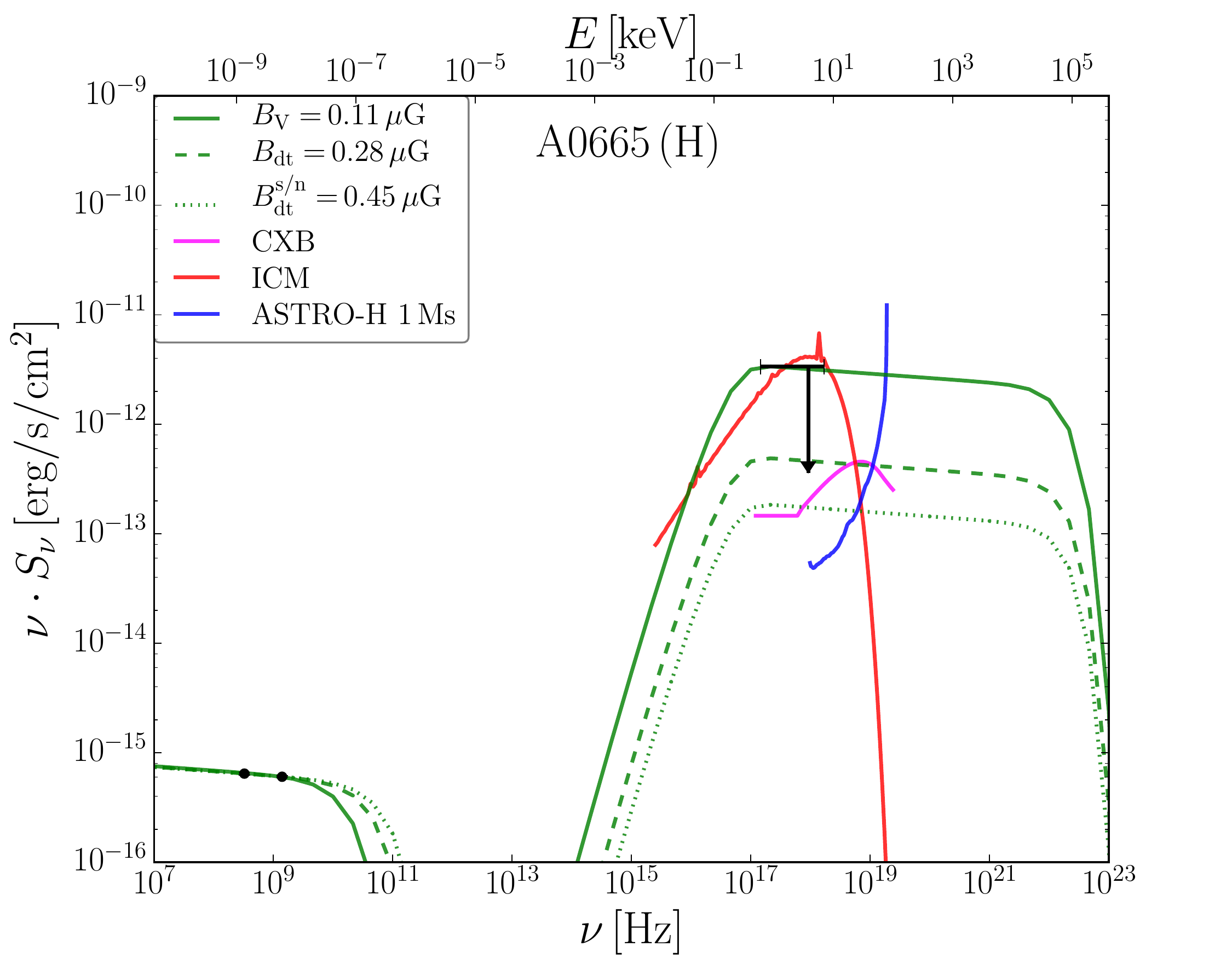}
\includegraphics[width=0.5\textwidth]{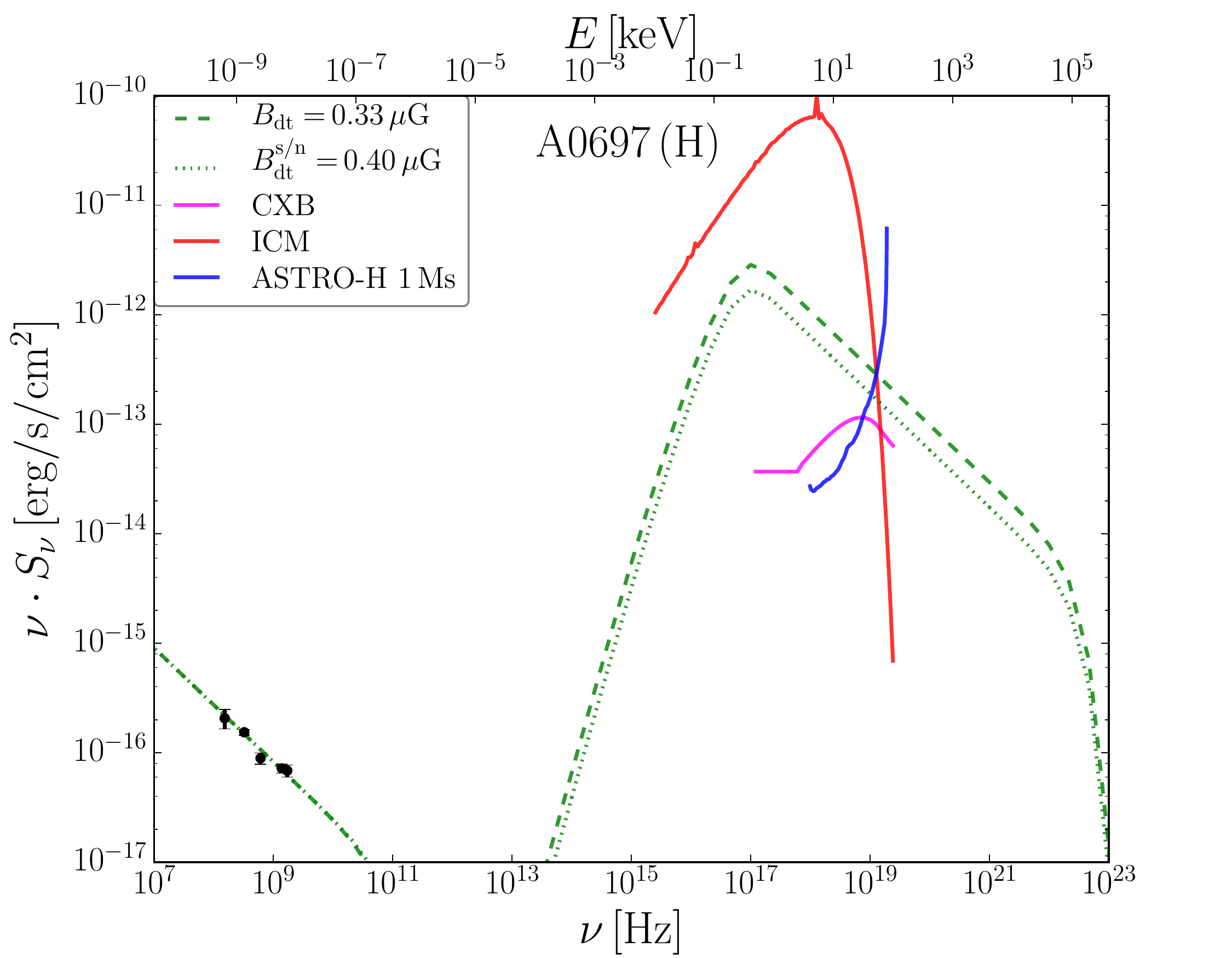}
\includegraphics[width=0.5\textwidth]{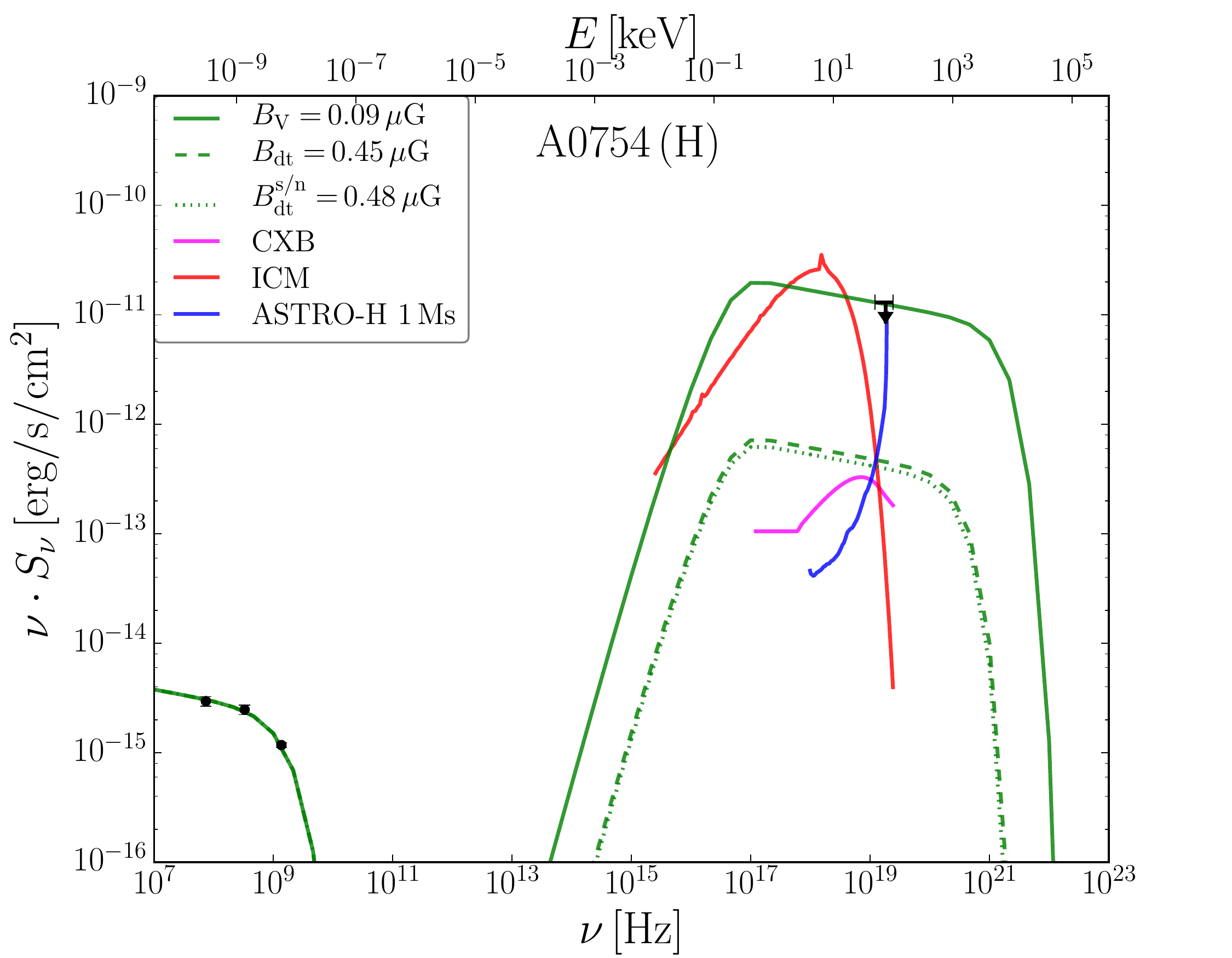}
\includegraphics[width=0.5\textwidth]{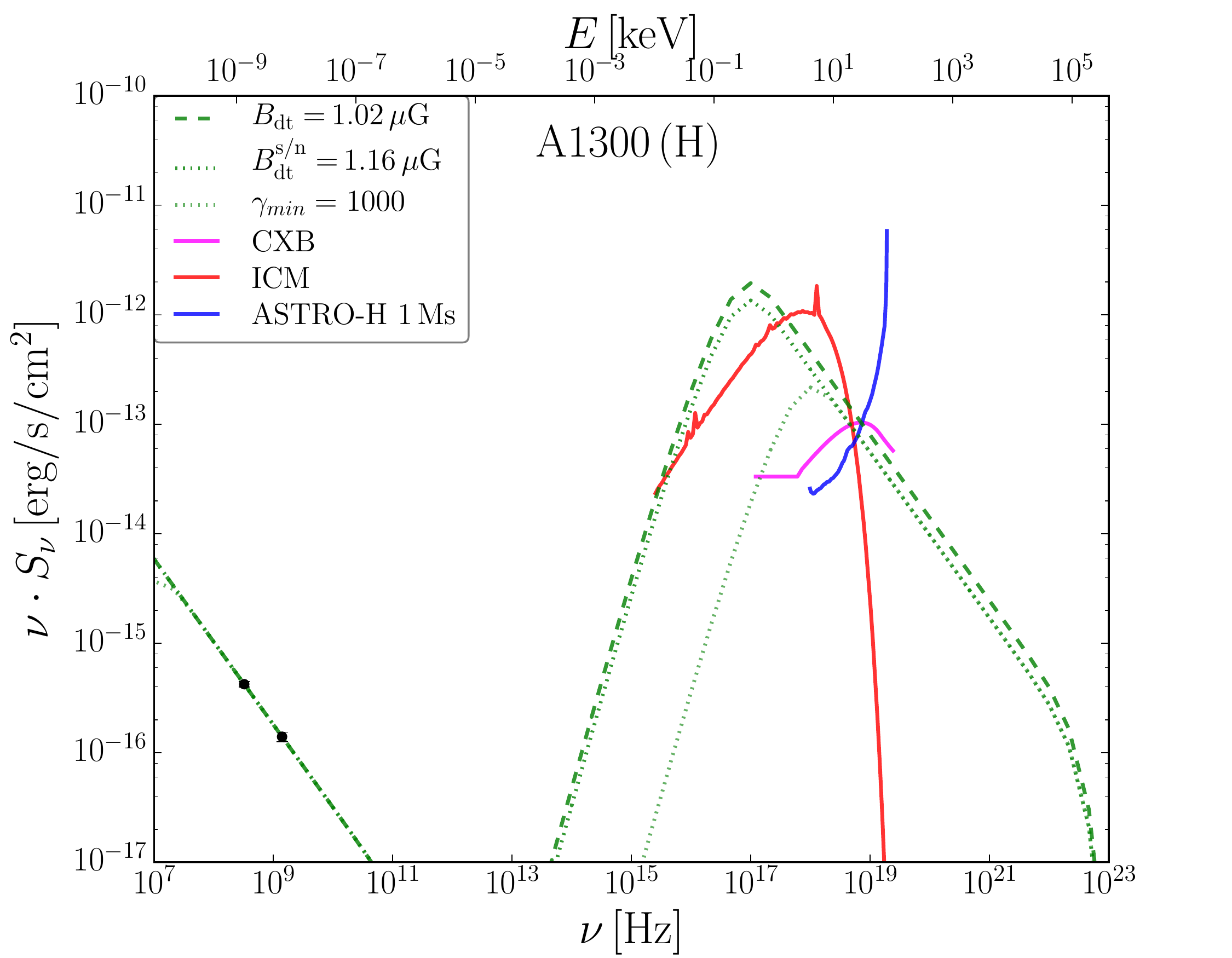}
\end{figure*}
\begin{figure*}[hbt!]
\includegraphics[width=0.5\textwidth]{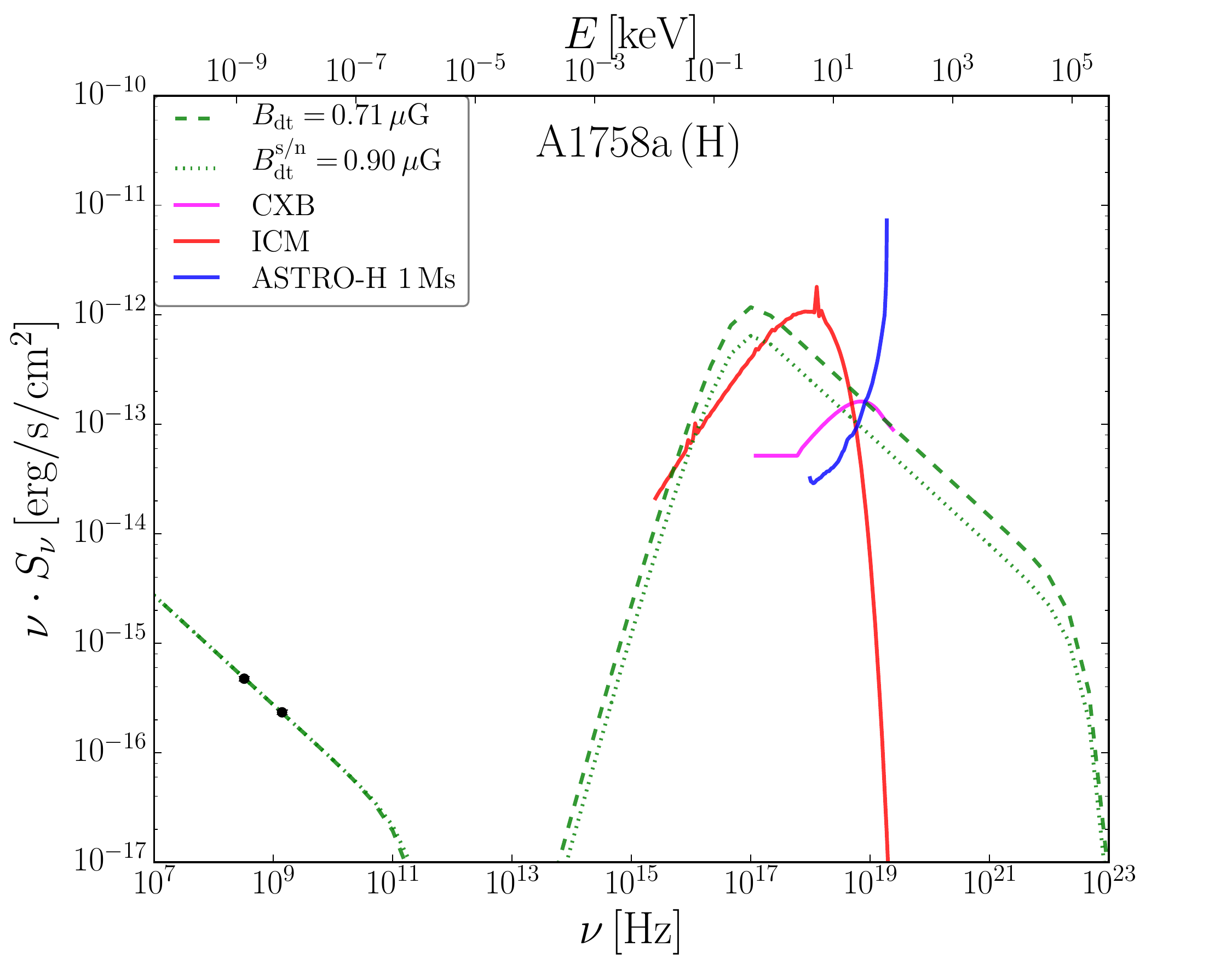}
\includegraphics[width=0.5\textwidth]{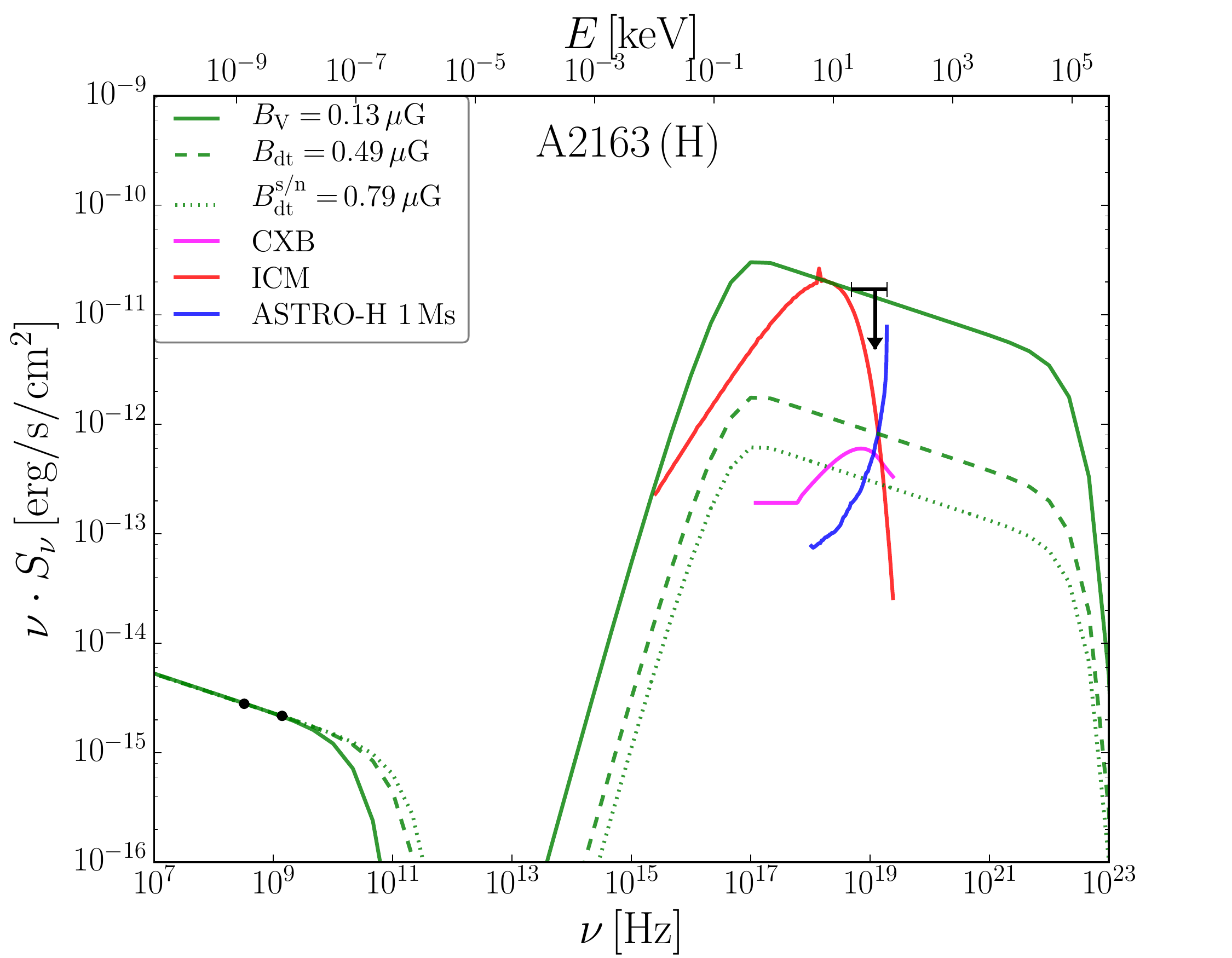}
\includegraphics[width=0.5\textwidth]{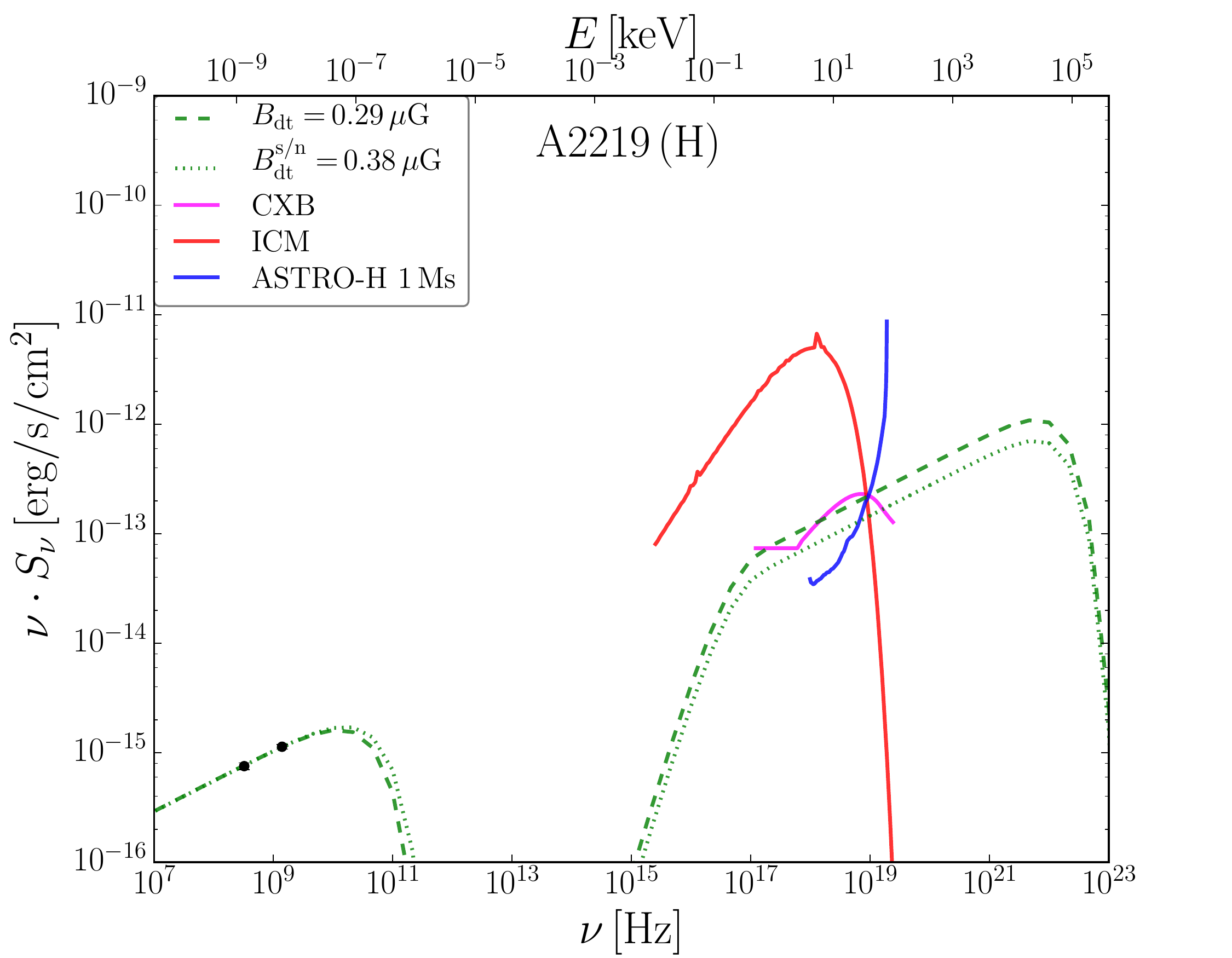}
\includegraphics[width=0.5\textwidth]{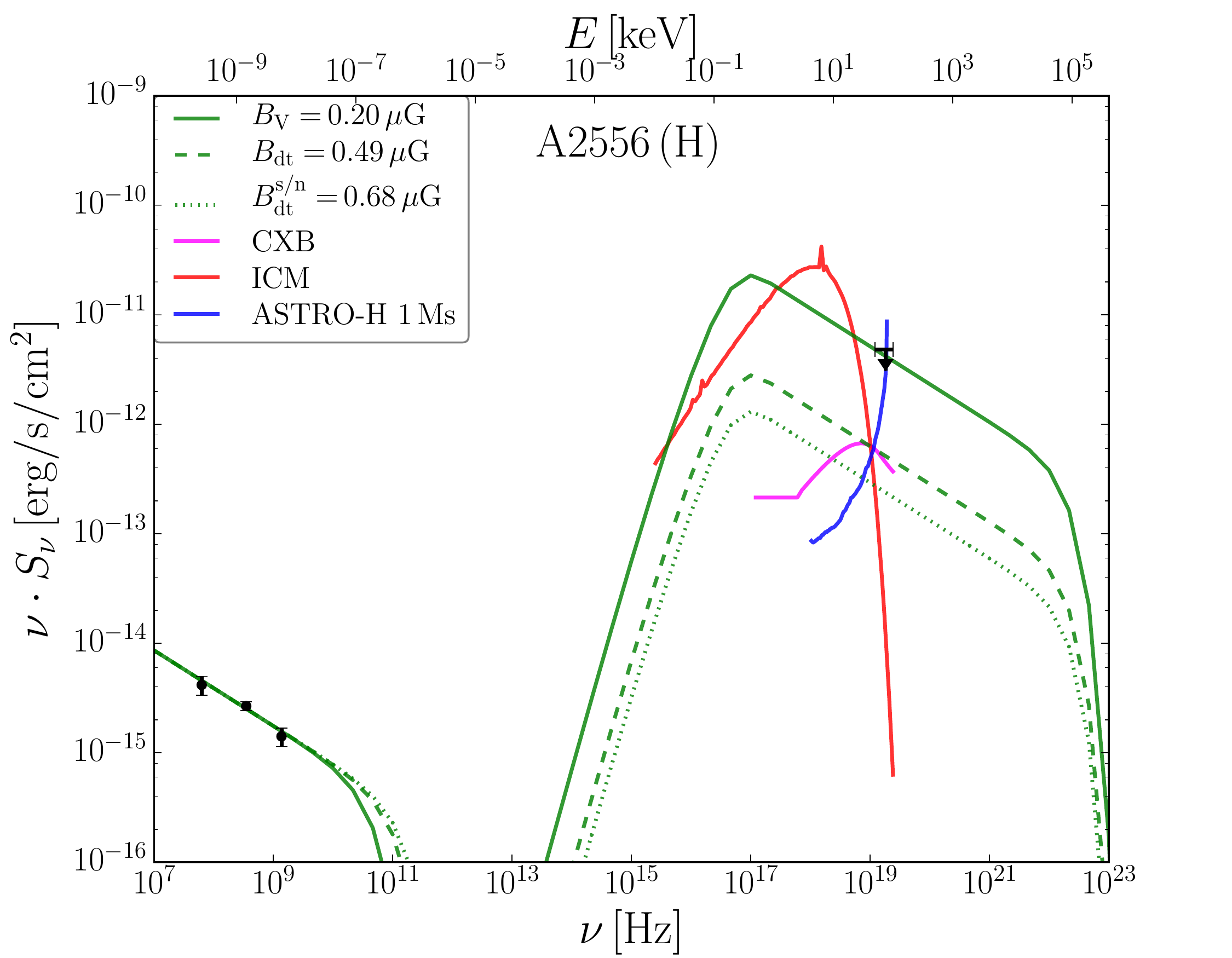}
\includegraphics[width=0.5\textwidth]{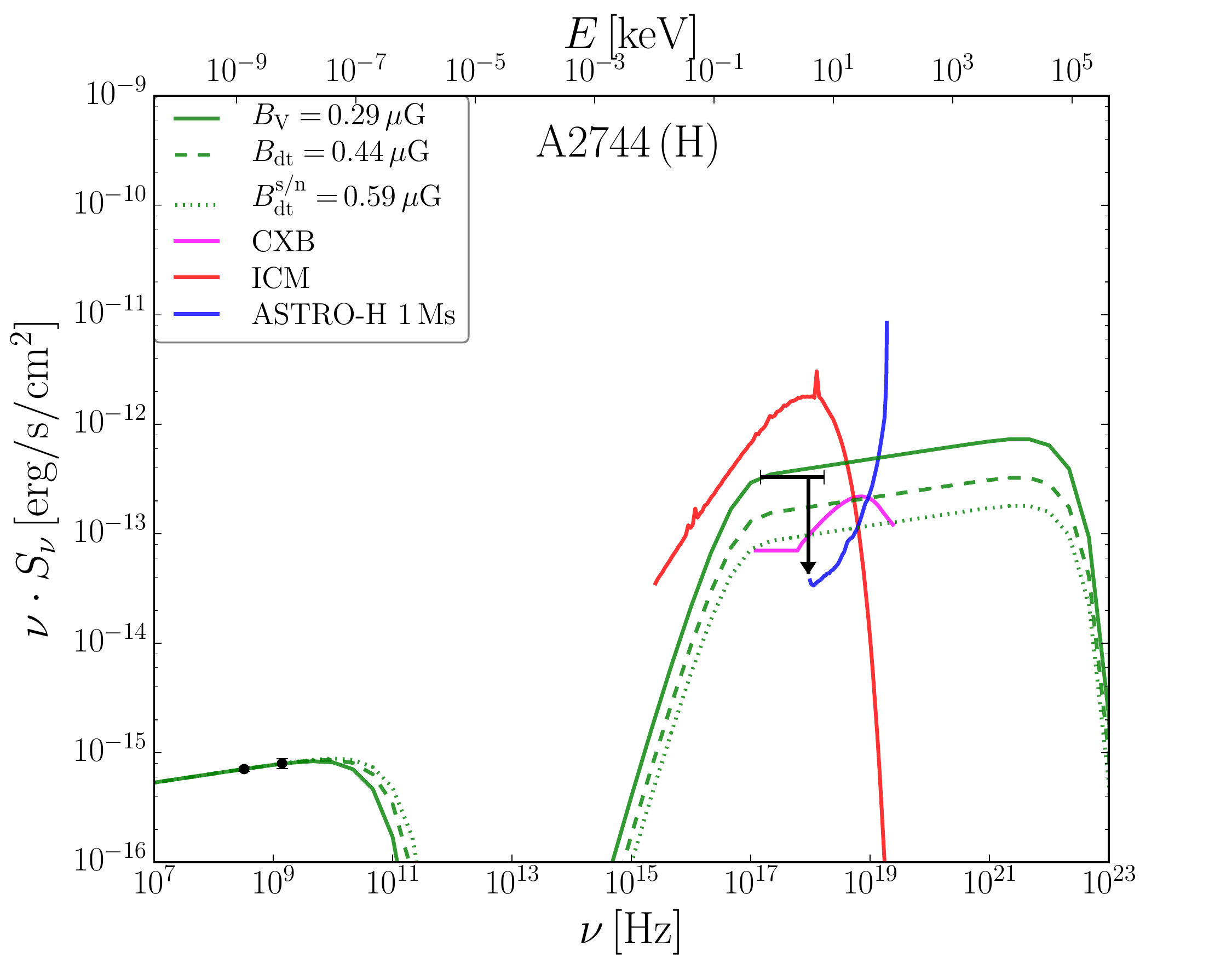}
\includegraphics[width=0.5\textwidth]{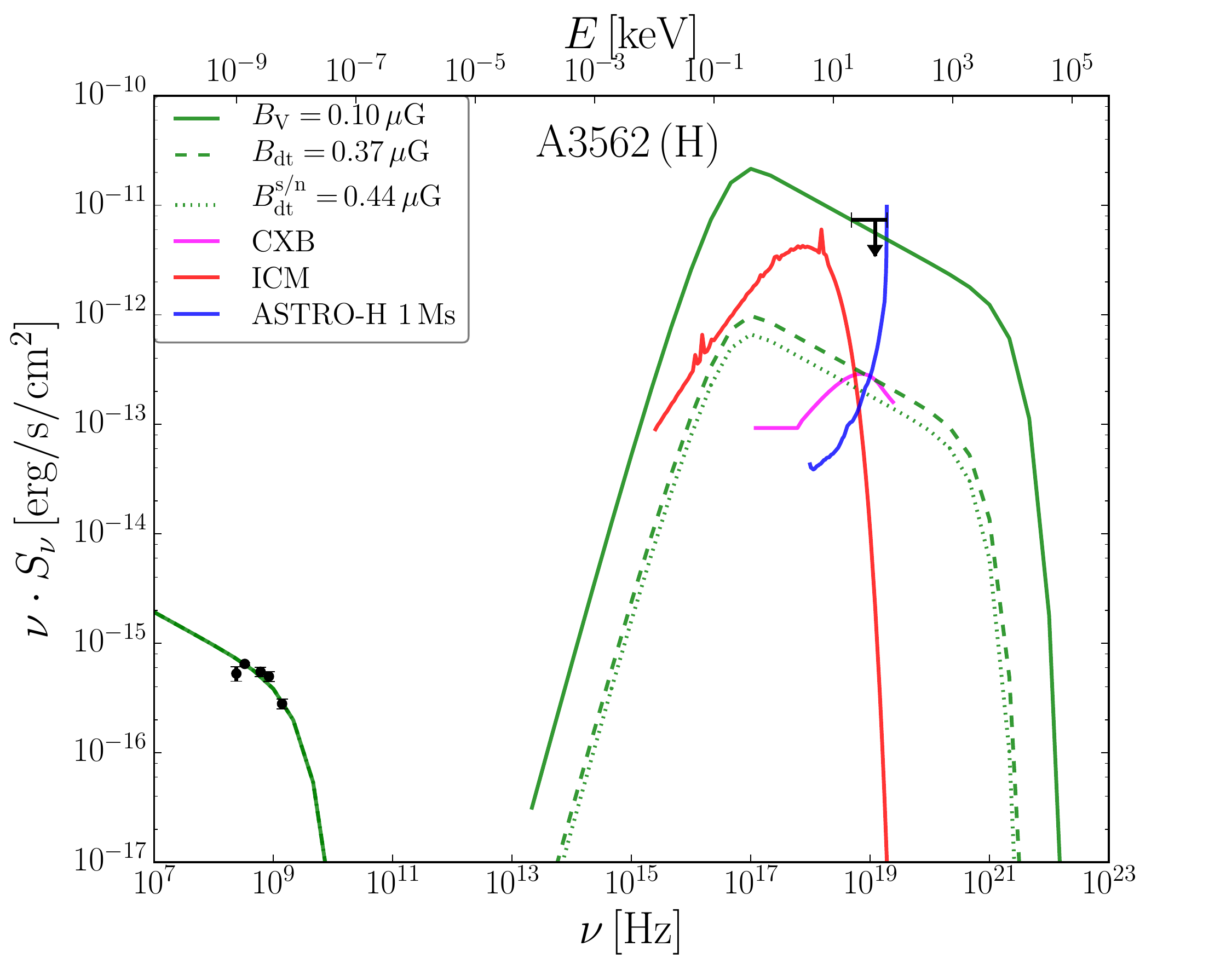}
\end{figure*}
\begin{figure*}[hbt!]
\includegraphics[width=0.5\textwidth]{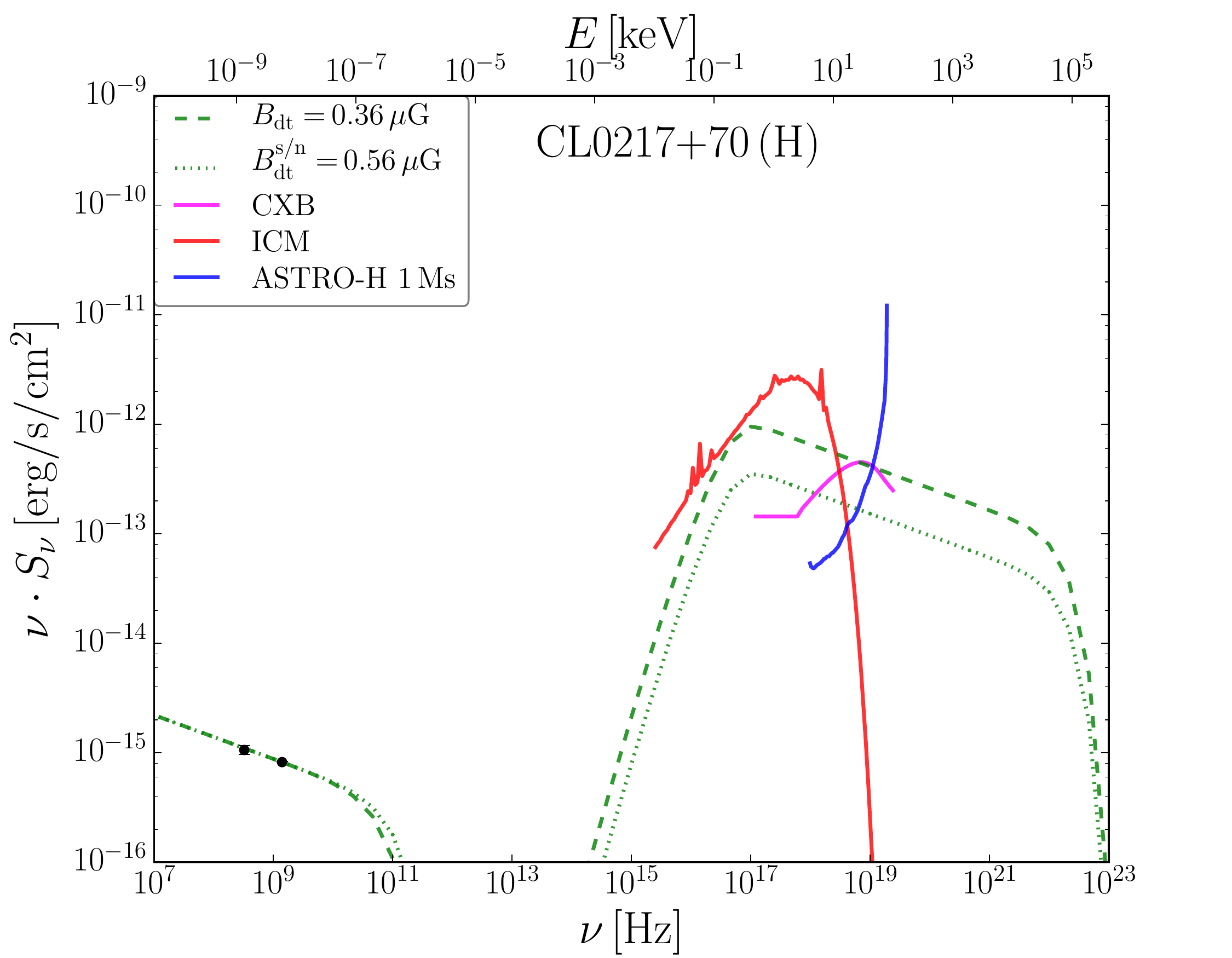}
\includegraphics[width=0.5\textwidth]{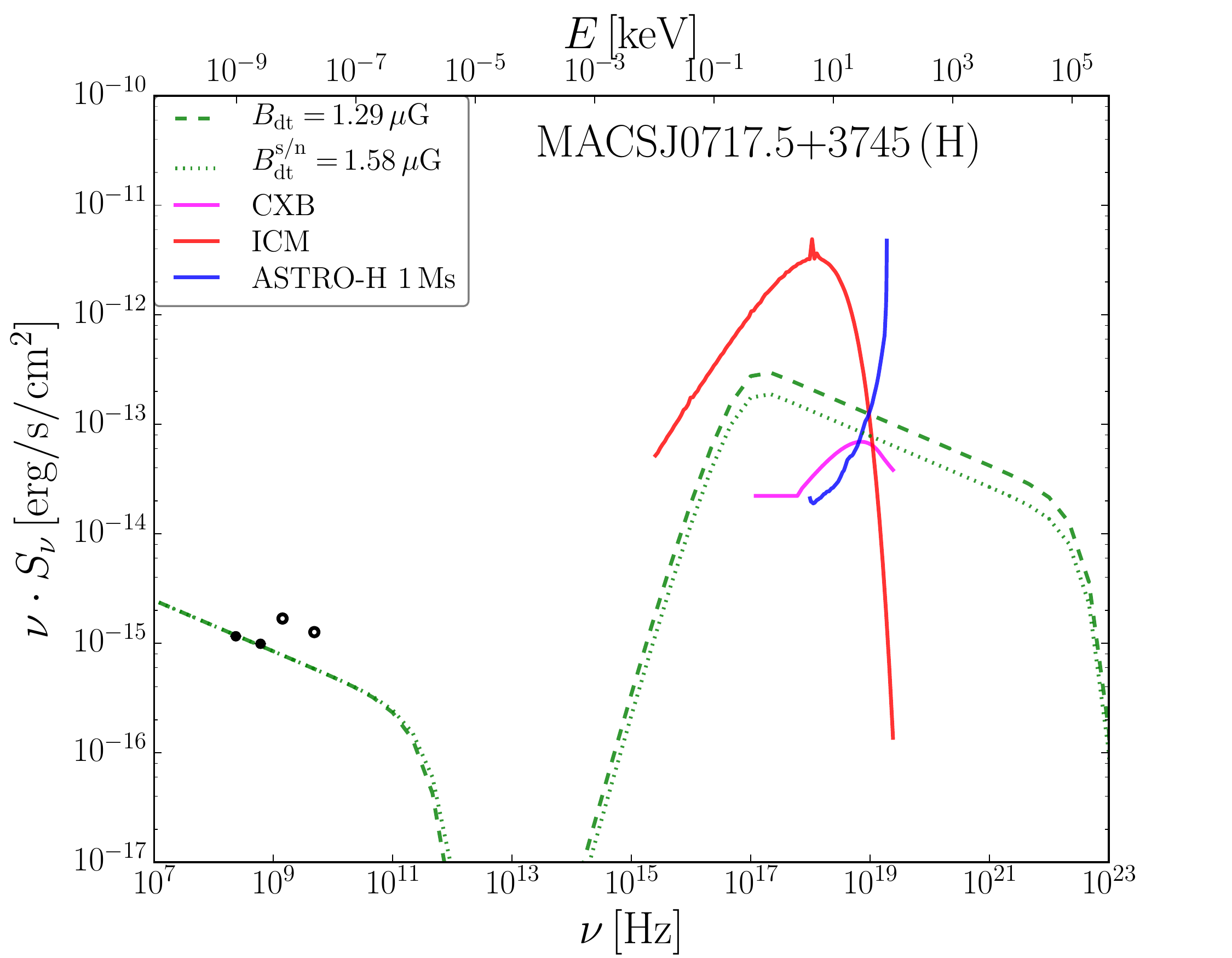}
\includegraphics[width=0.5\textwidth]{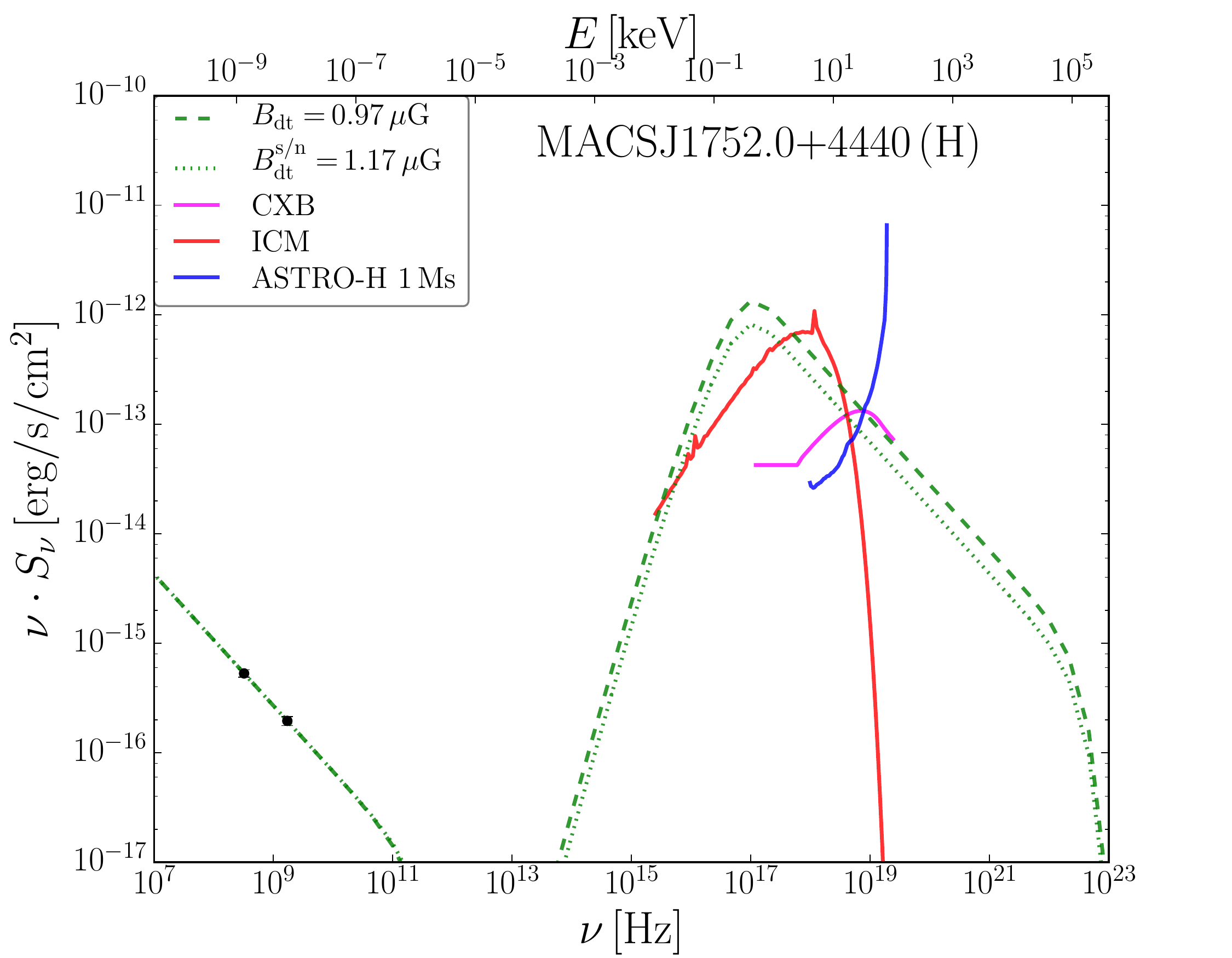}
\includegraphics[width=0.5\textwidth]{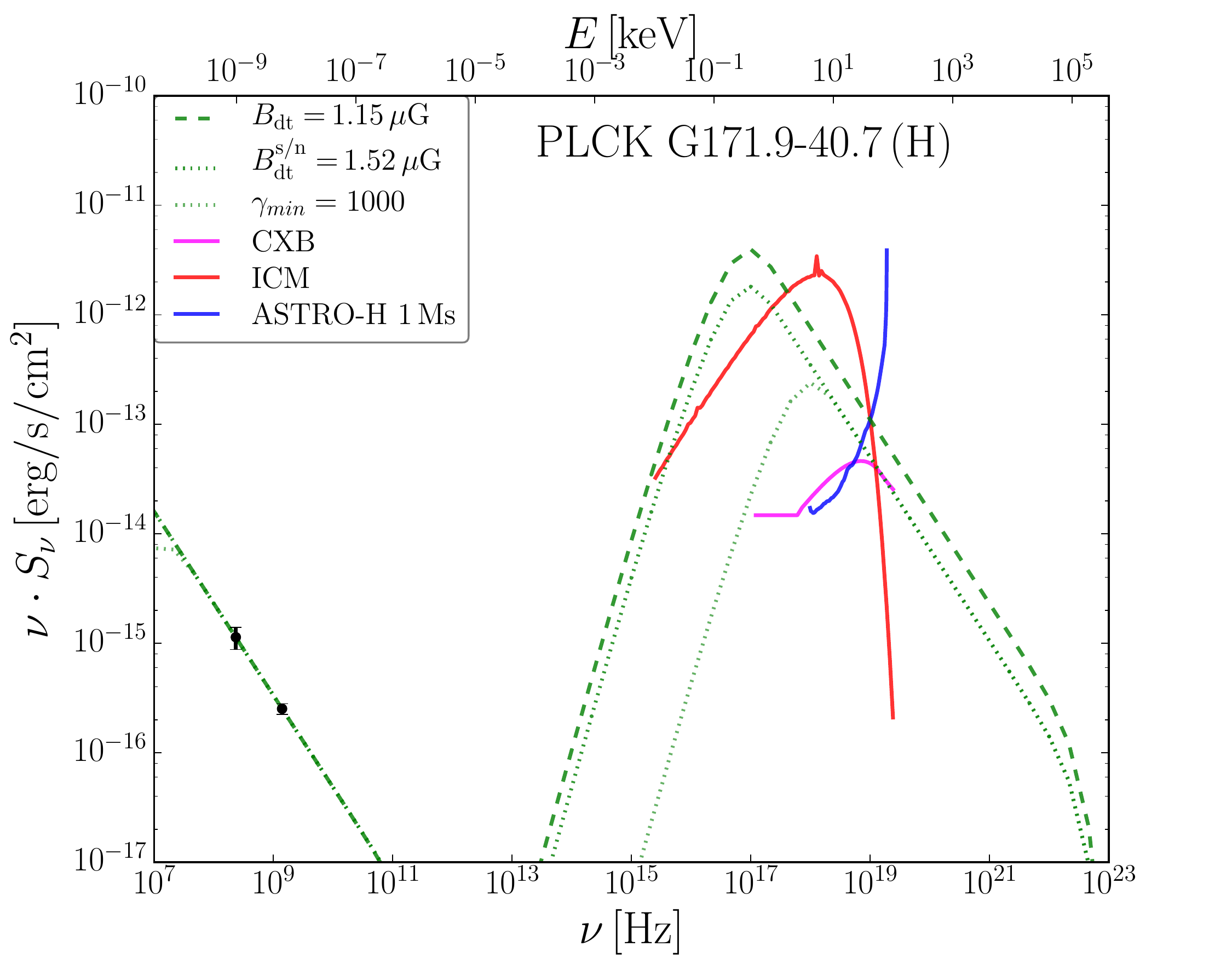}
\includegraphics[width=0.5\textwidth]{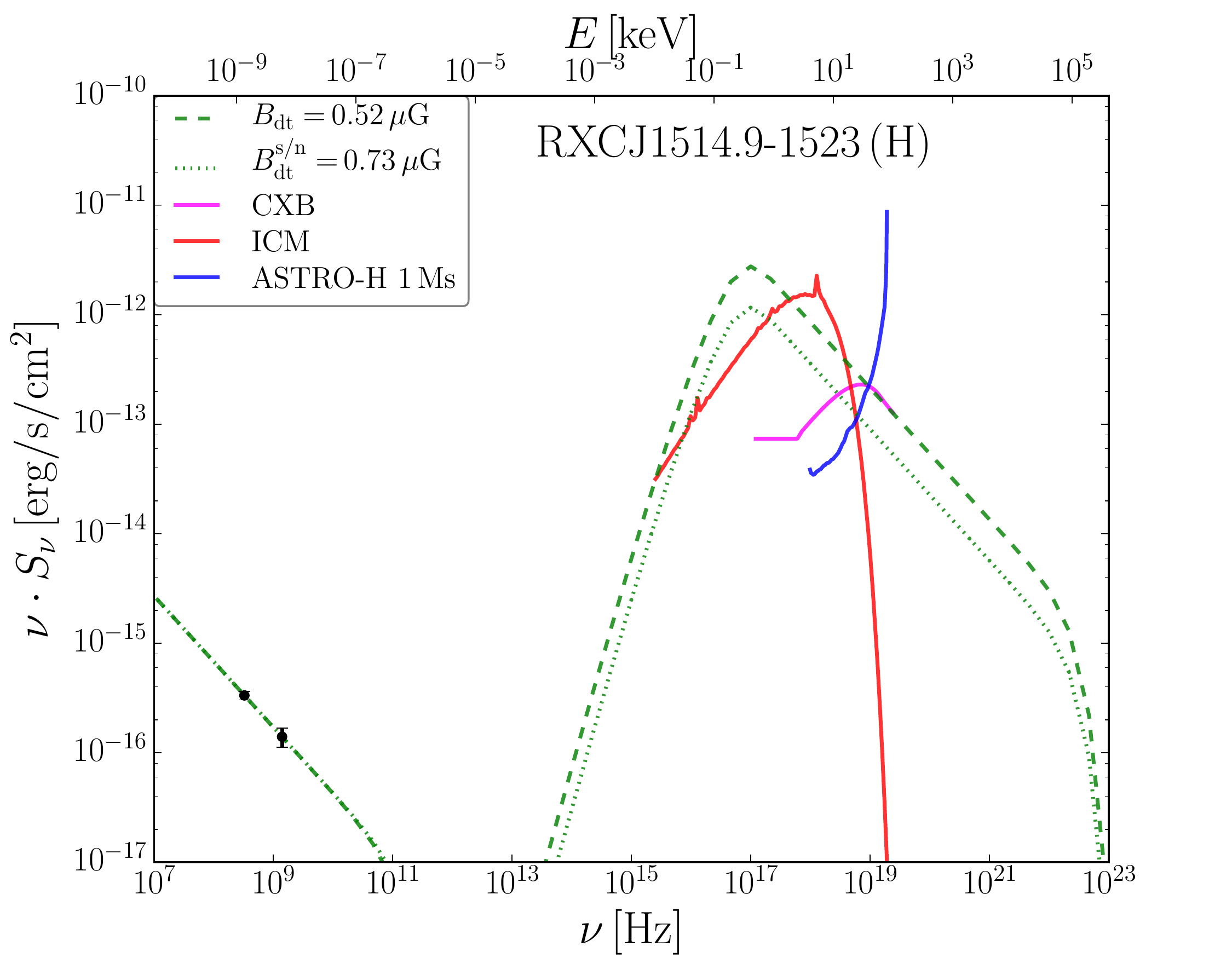}
\includegraphics[width=0.5\textwidth]{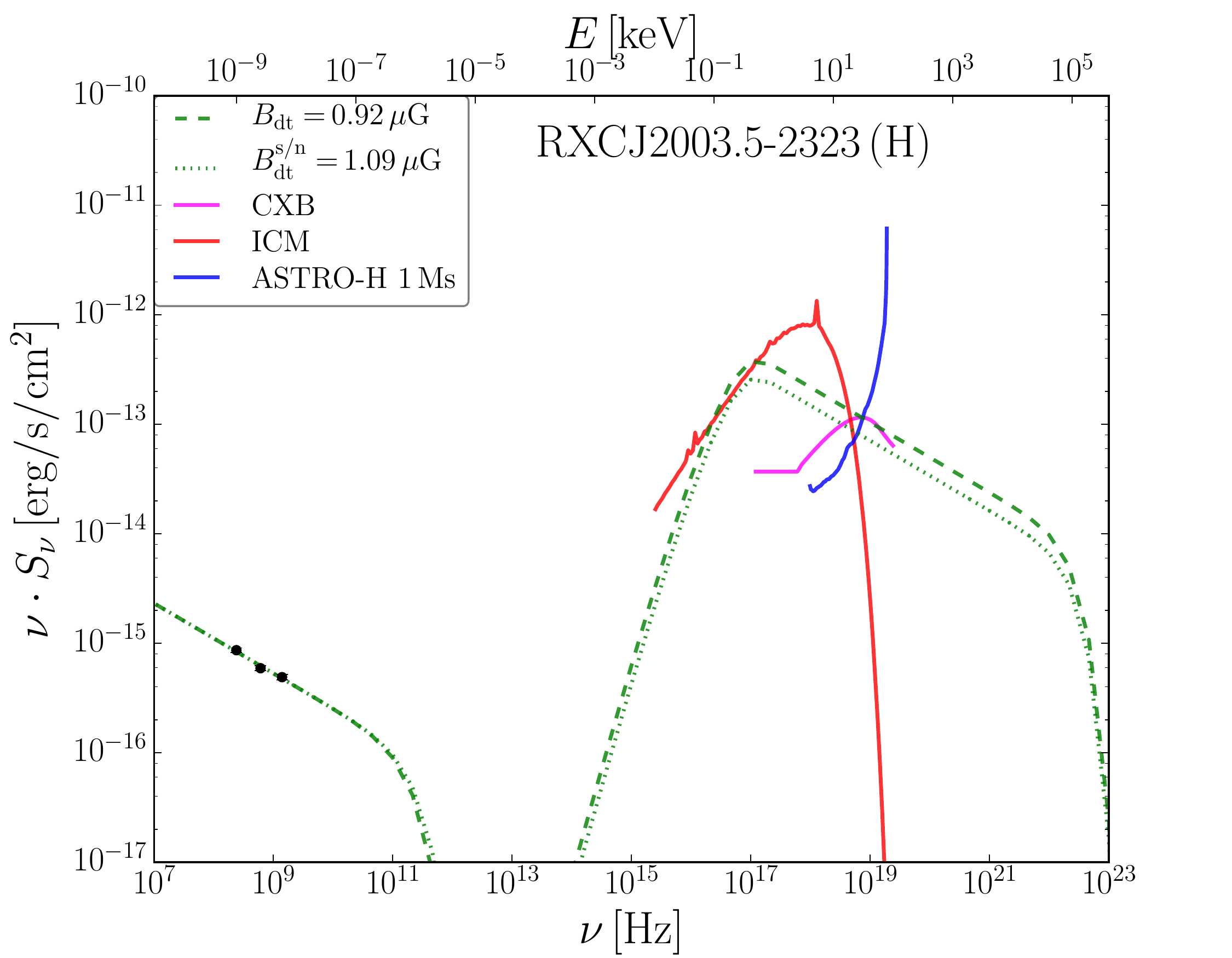}
\end{figure*}
%
\begin{figure*}[hbt!]
\includegraphics[width=0.5\textwidth]{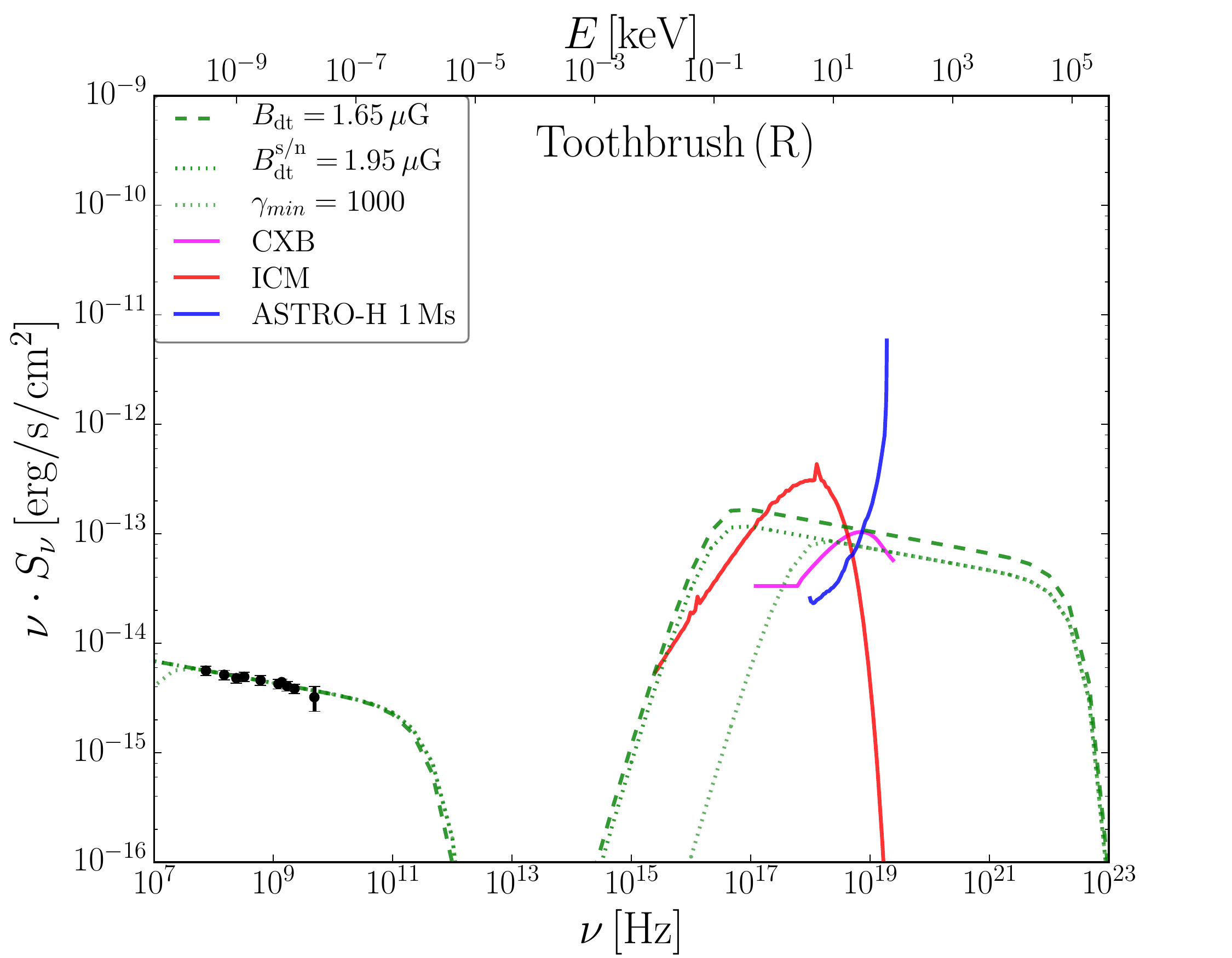}
\includegraphics[width=0.5\textwidth]{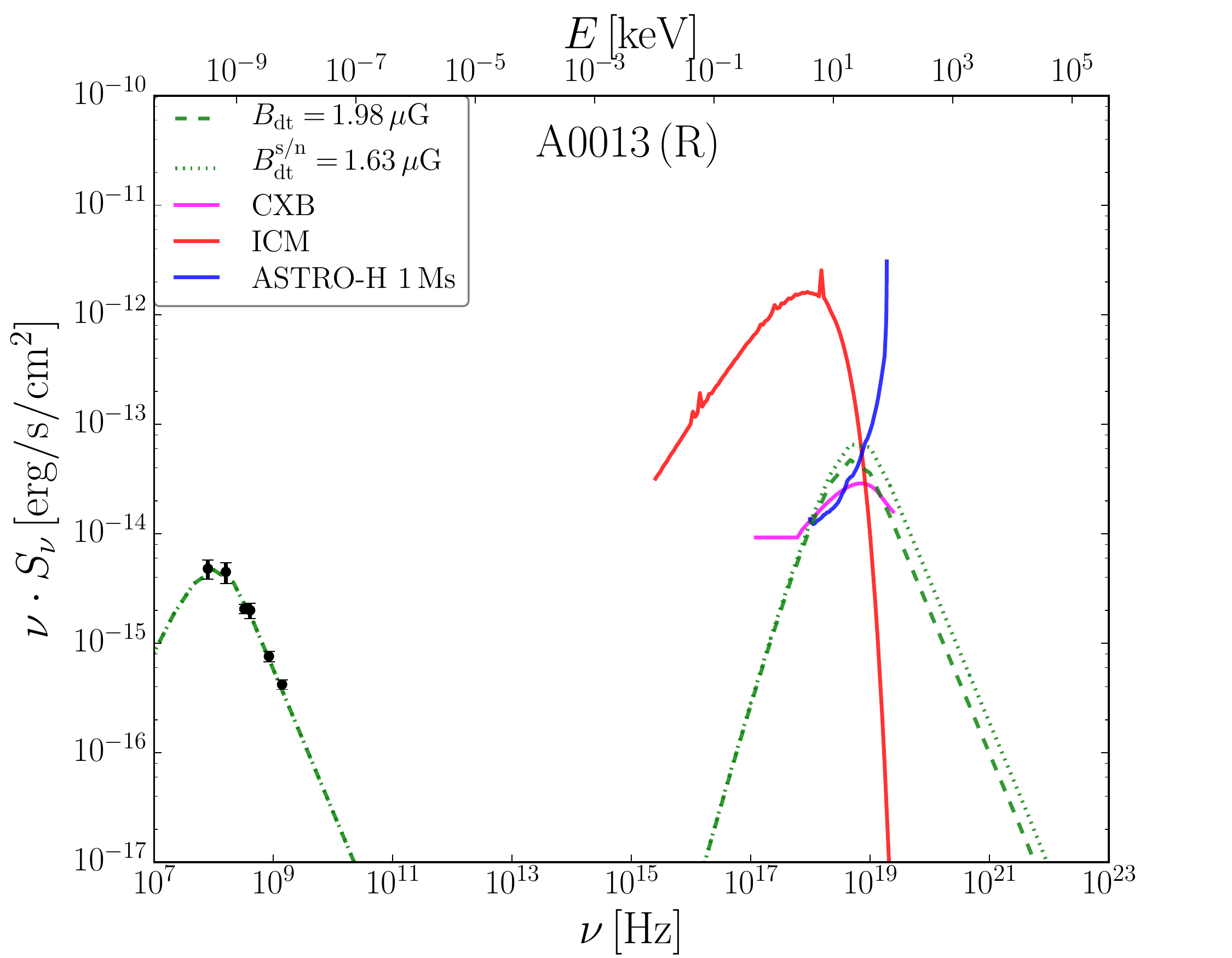}
\includegraphics[width=0.5\textwidth]{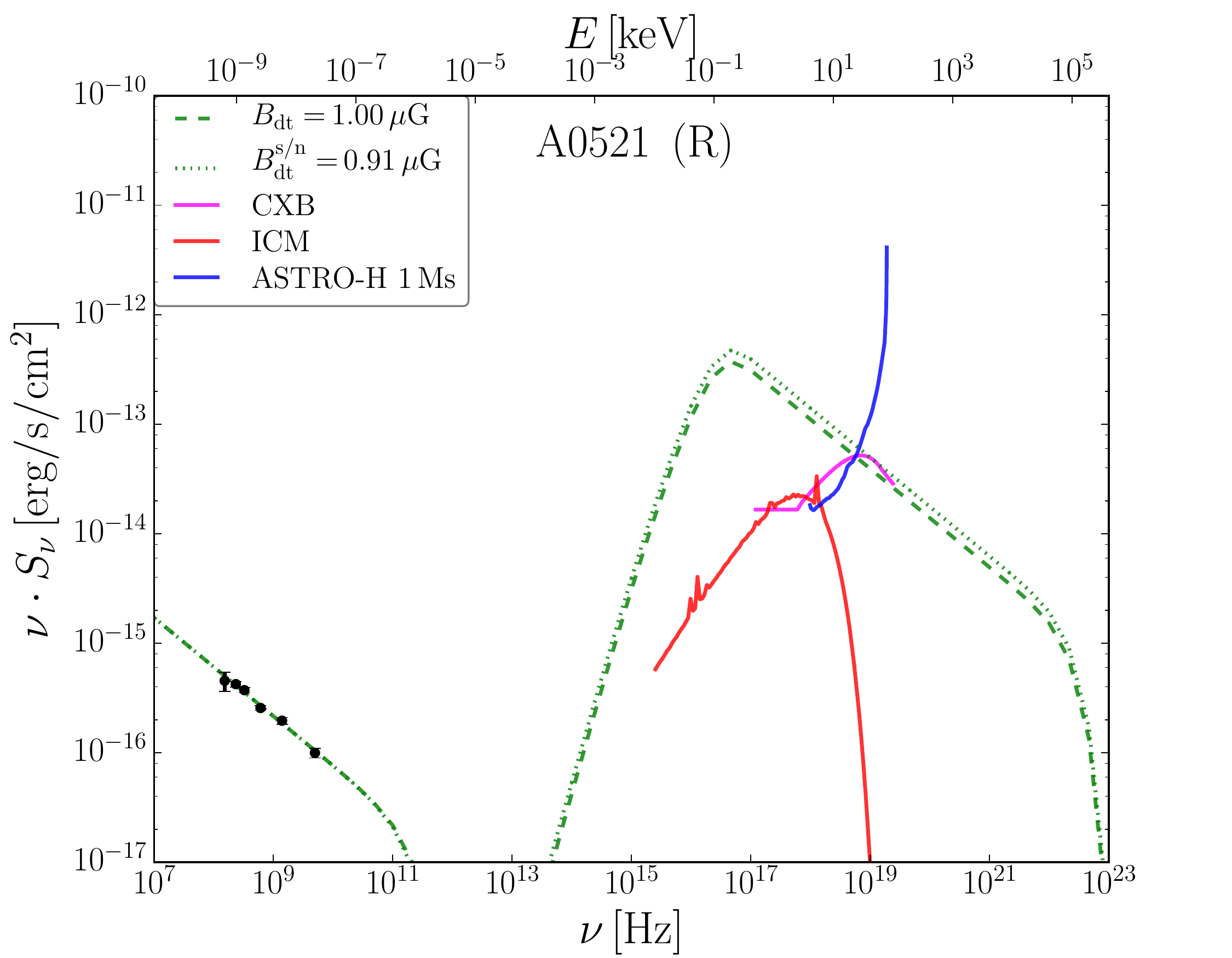}
\includegraphics[width=0.5\textwidth]{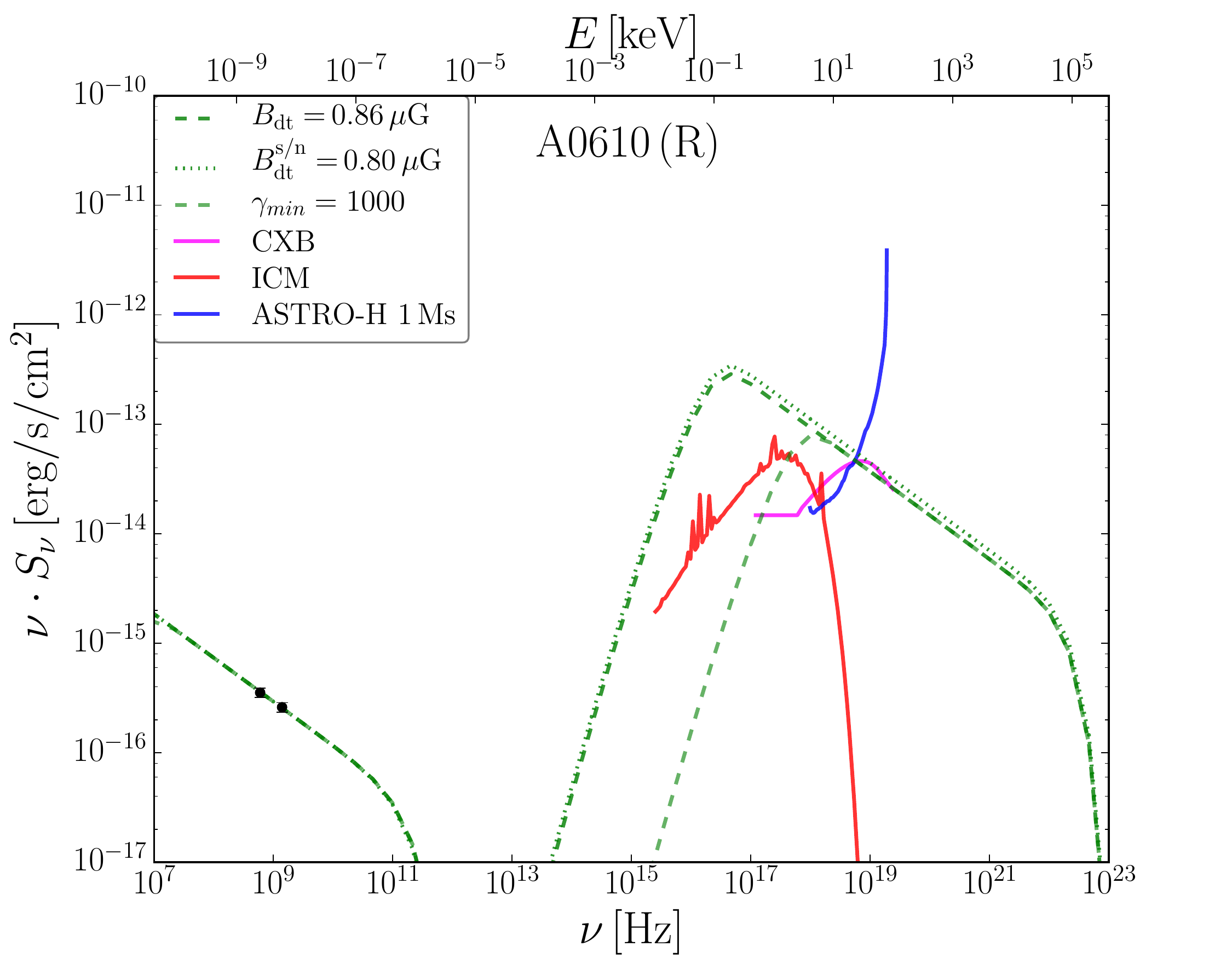}
\includegraphics[width=0.5\textwidth]{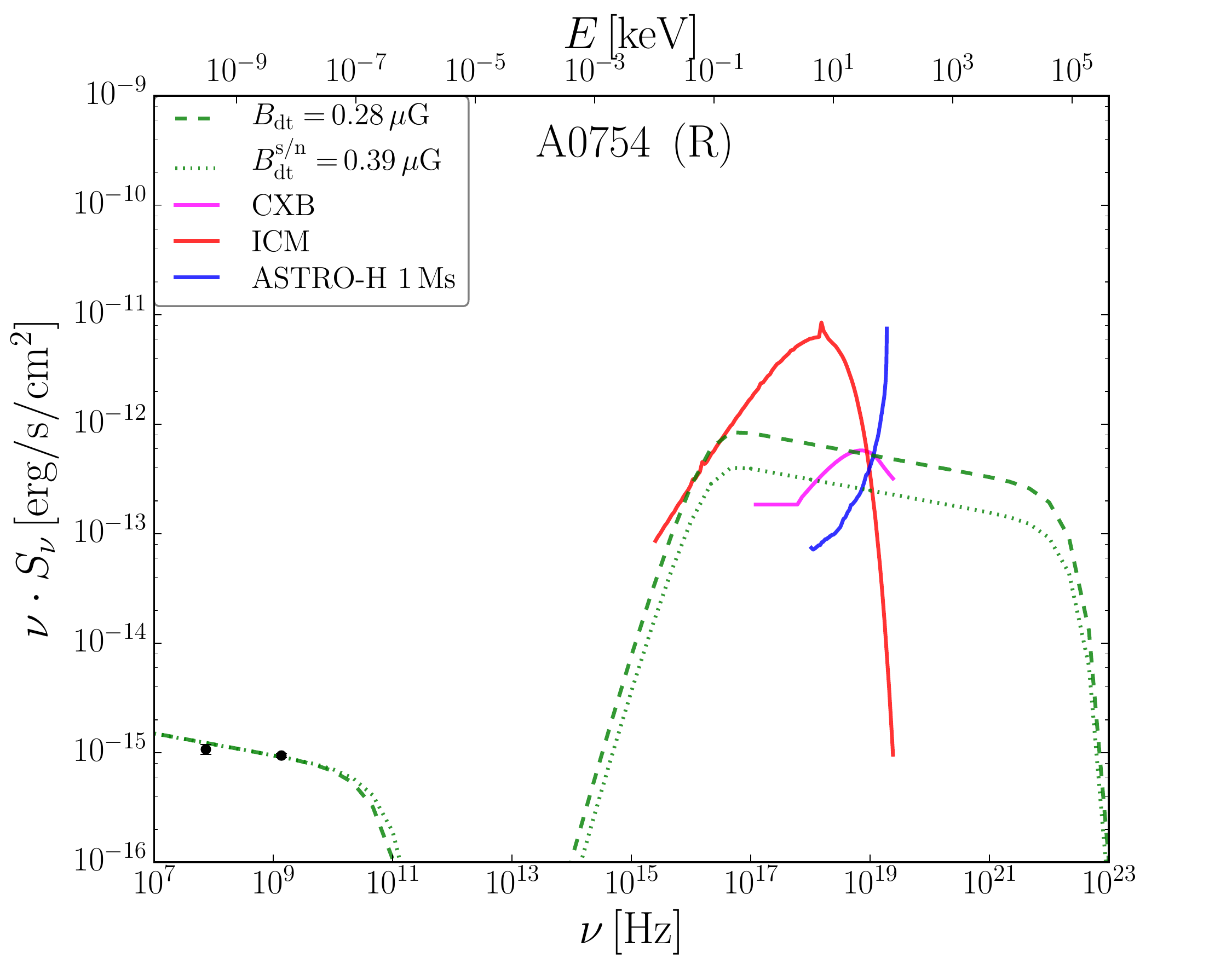}
\includegraphics[width=0.5\textwidth]{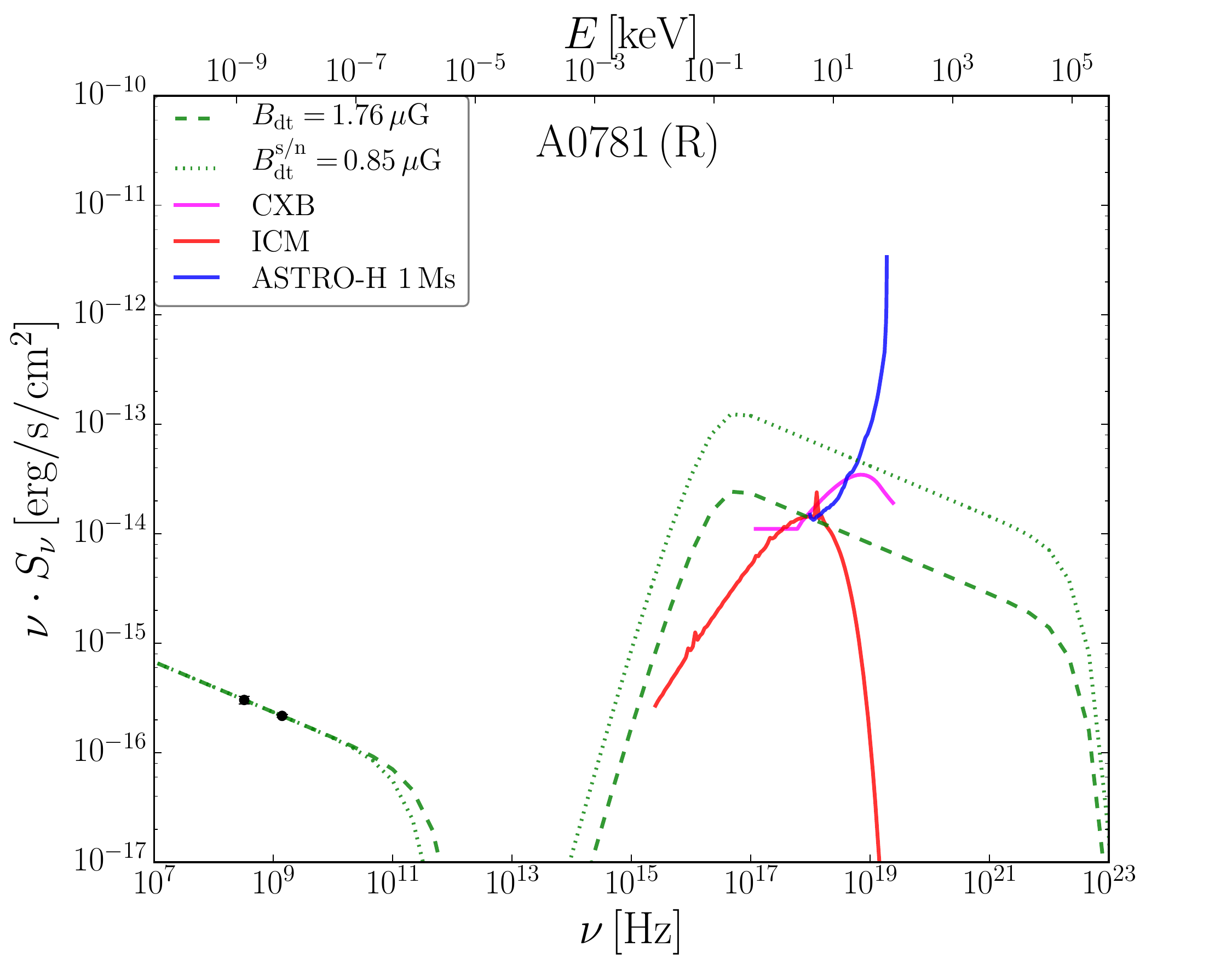}
\end{figure*}
\begin{figure*}[hbt!]
\includegraphics[width=0.5\textwidth]{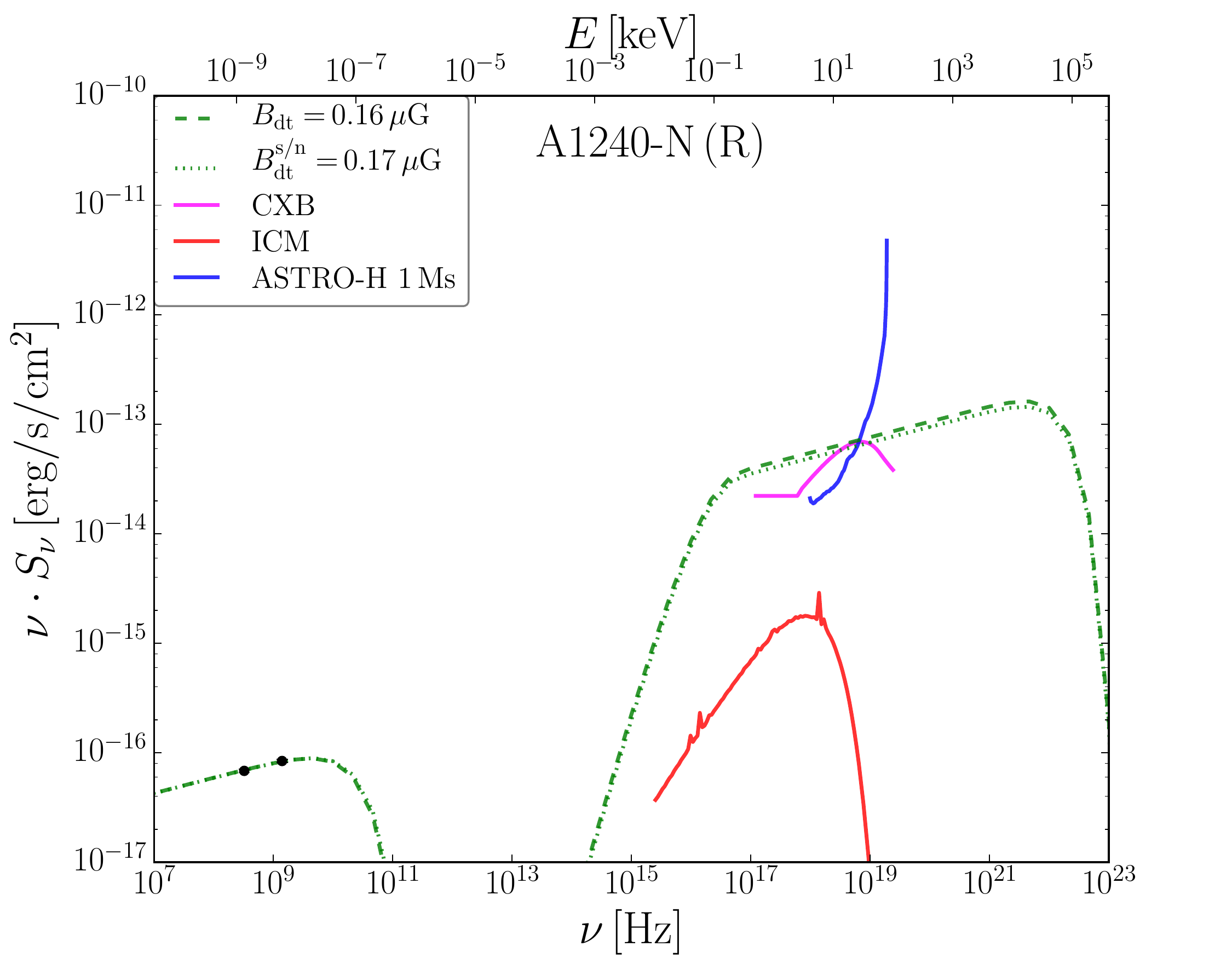}
\includegraphics[width=0.5\textwidth]{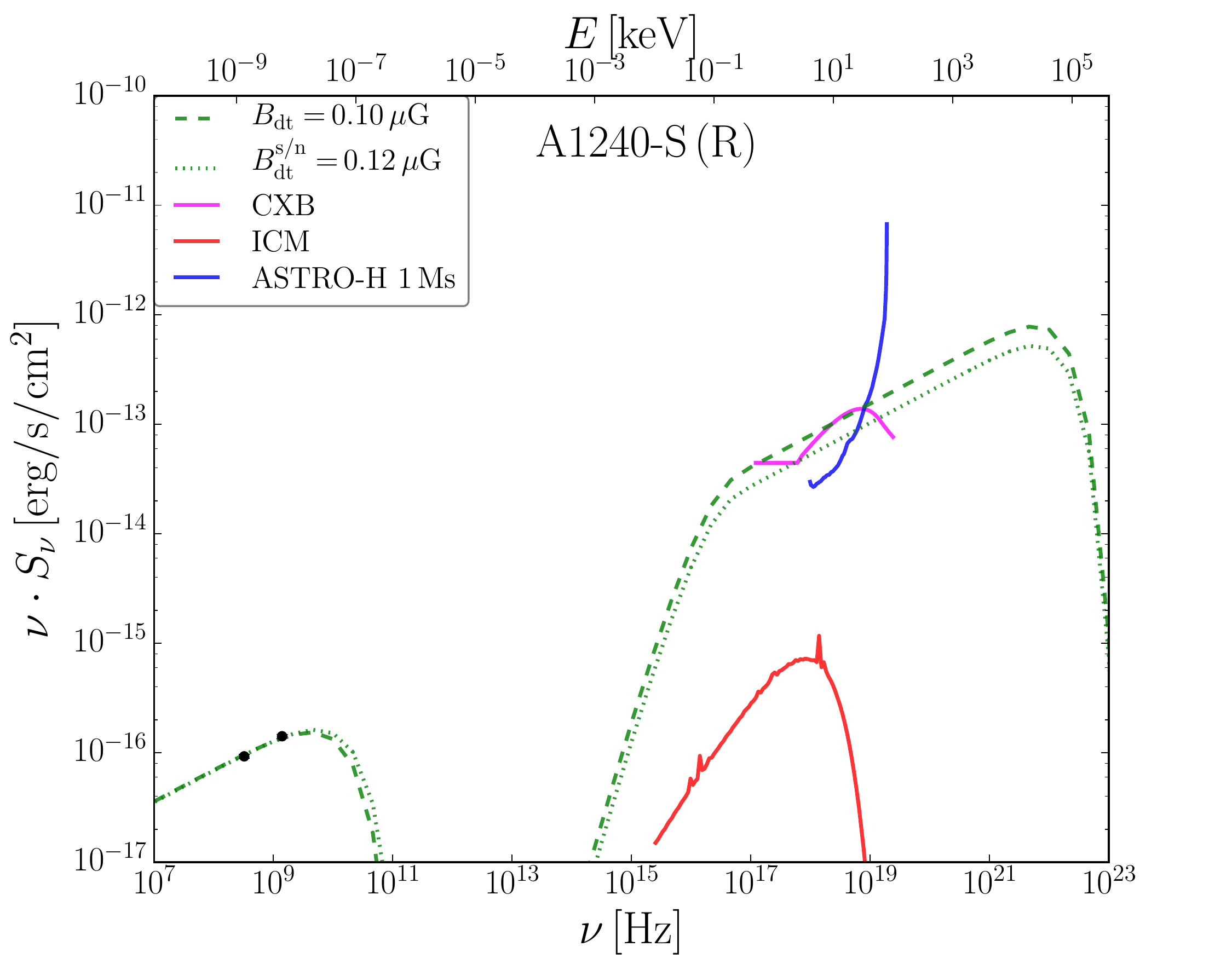}
\includegraphics[width=0.5\textwidth]{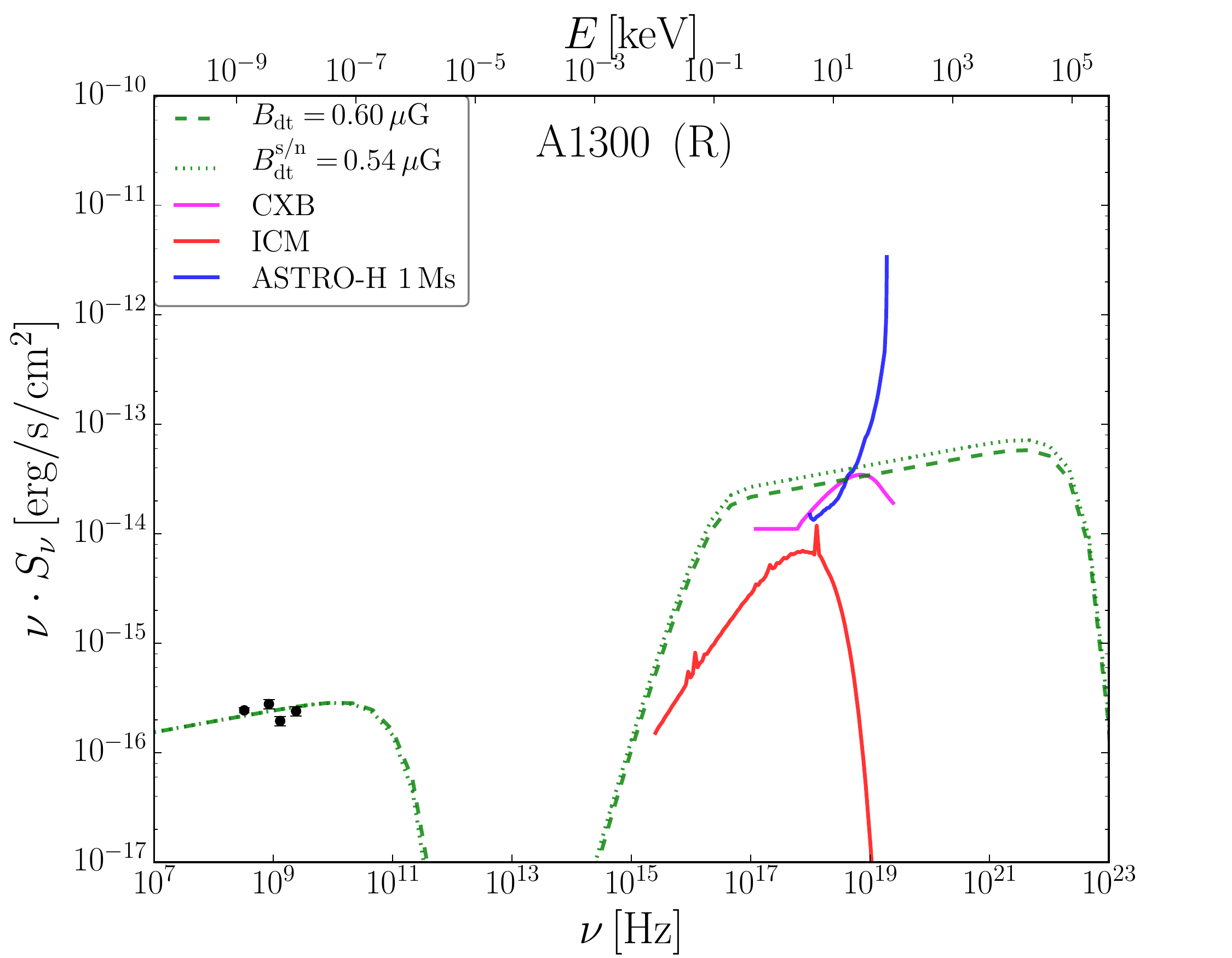}
\includegraphics[width=0.5\textwidth]{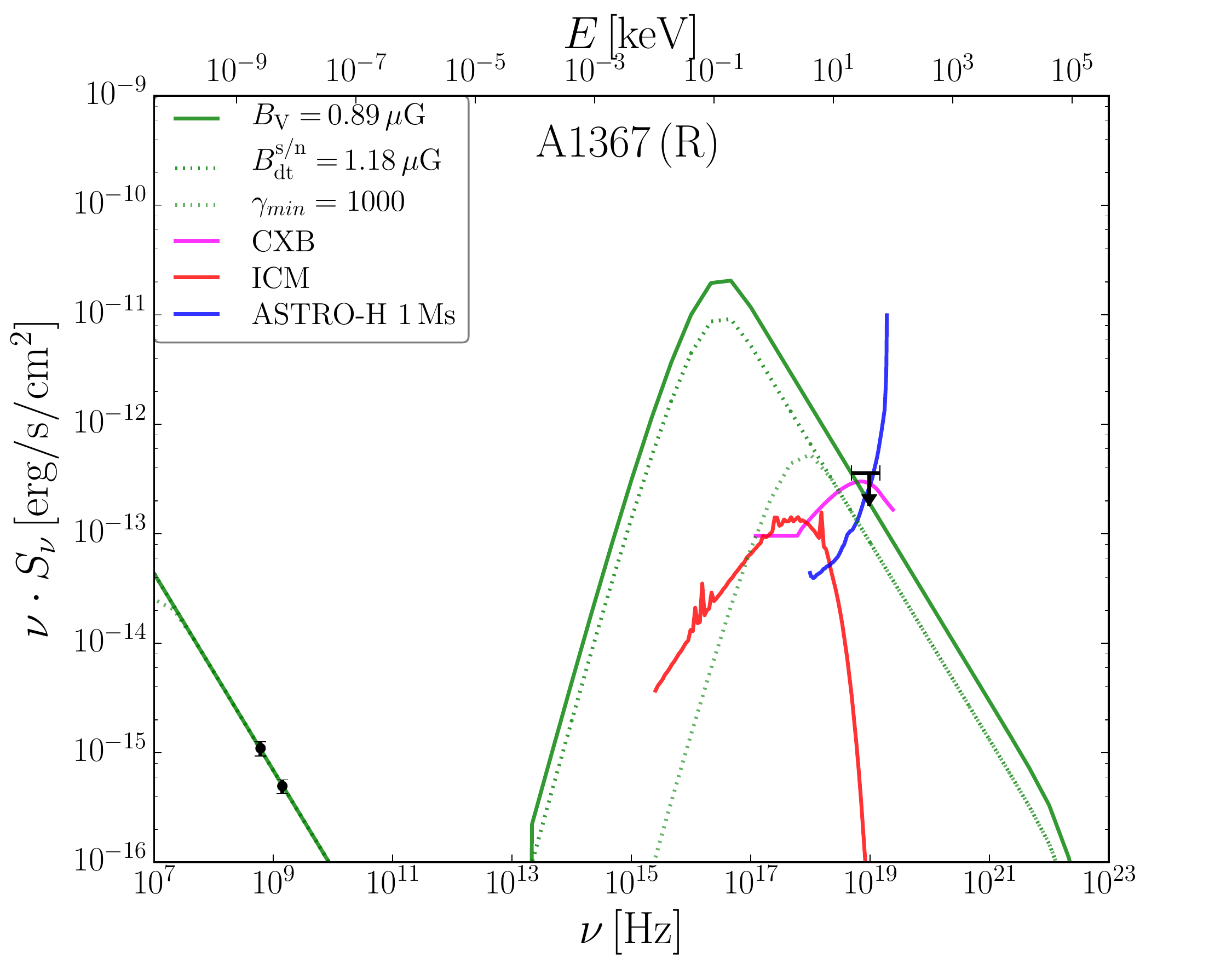}
\includegraphics[width=0.5\textwidth]{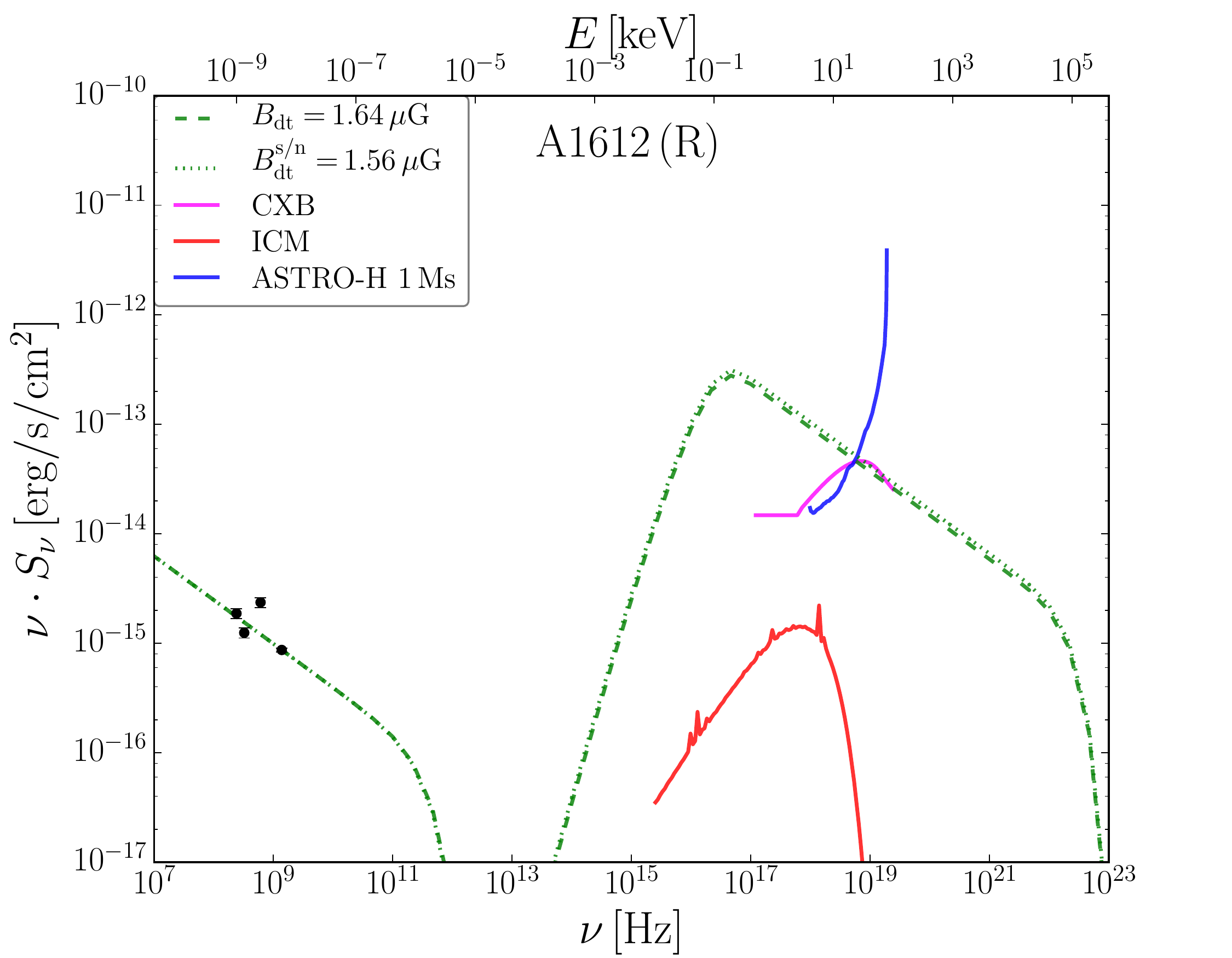}
\includegraphics[width=0.5\textwidth]{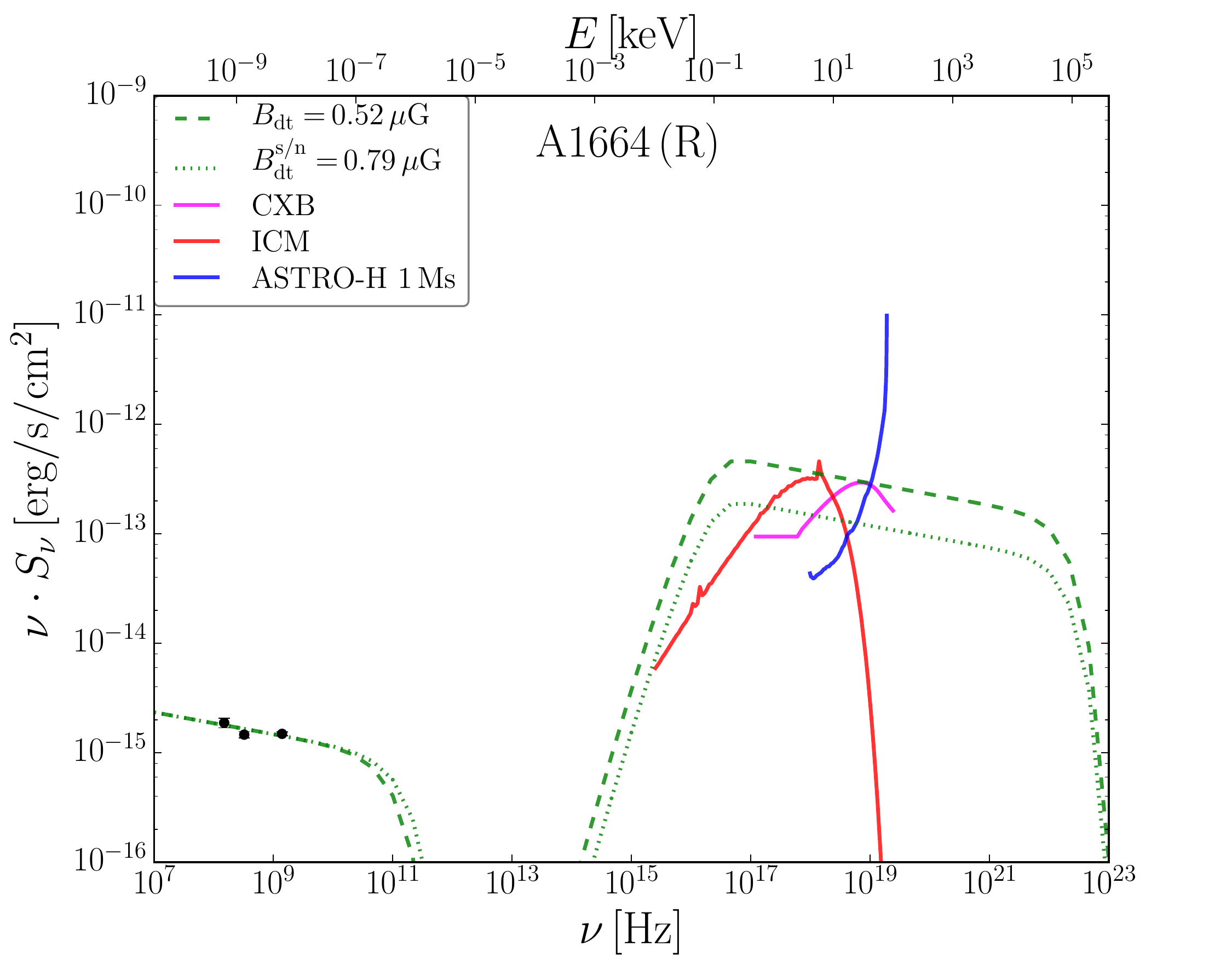}
\end{figure*}
\begin{figure*}[hbt!]
\includegraphics[width=0.5\textwidth]{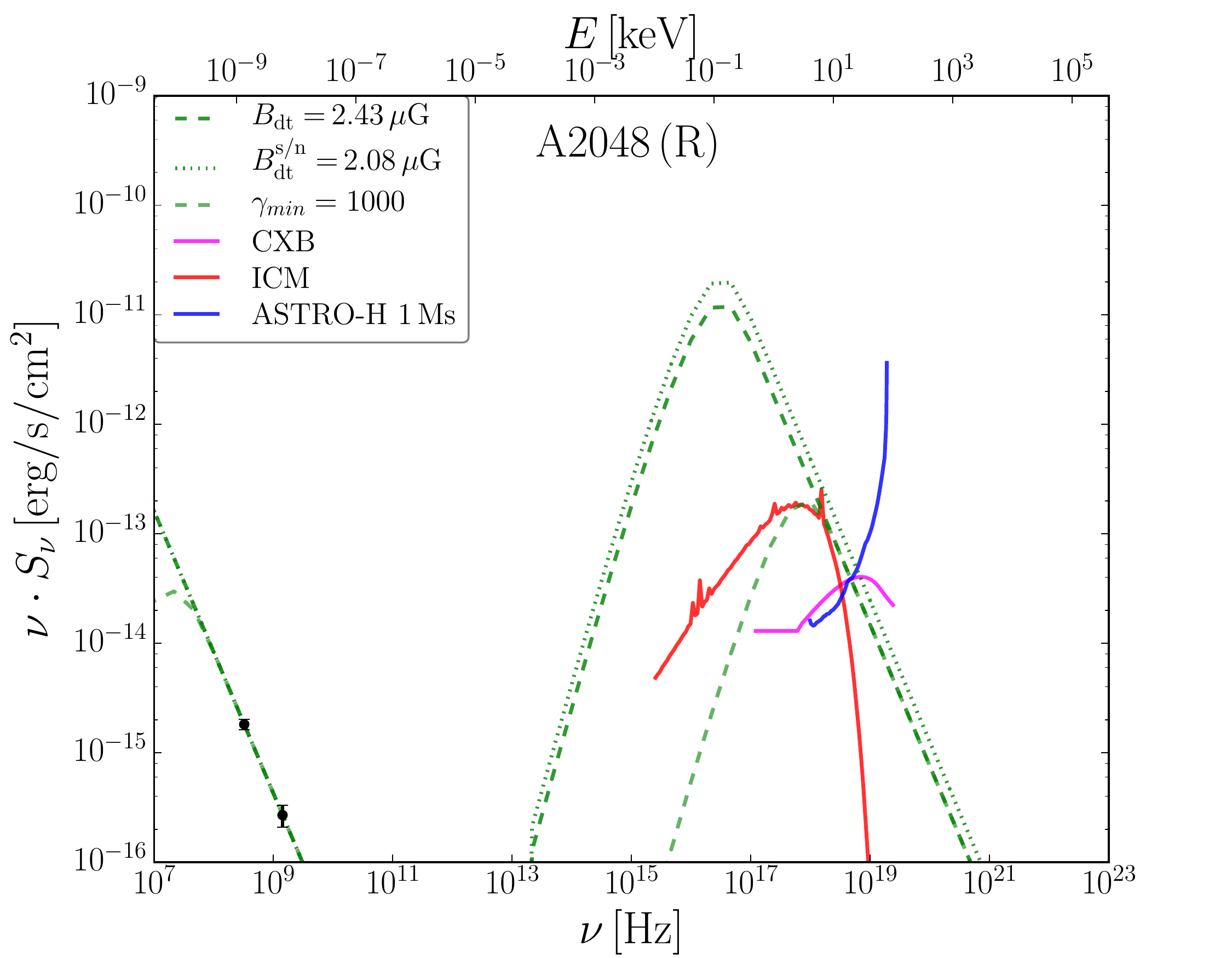}
\includegraphics[width=0.5\textwidth]{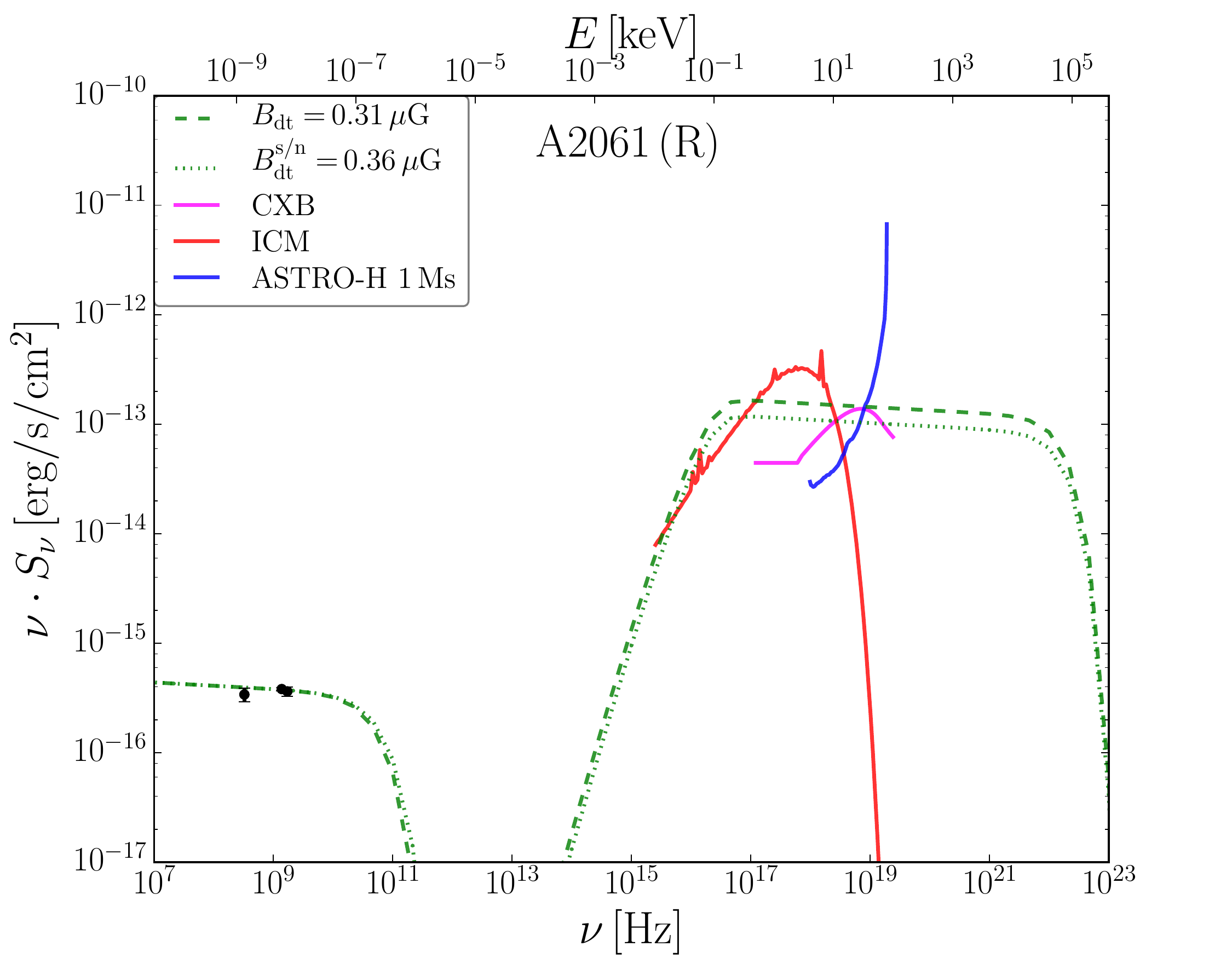}
\includegraphics[width=0.5\textwidth]{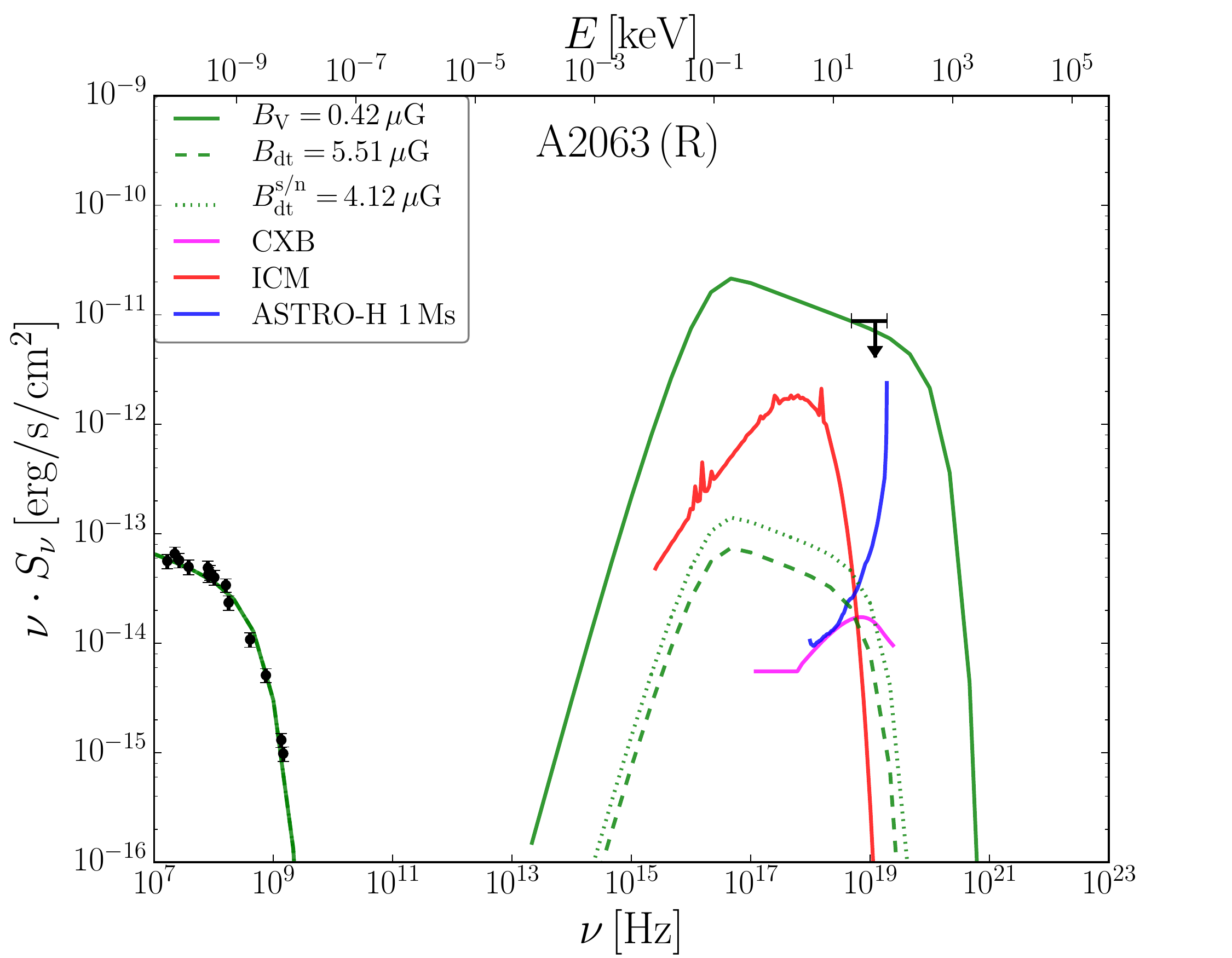}
\includegraphics[width=0.5\textwidth]{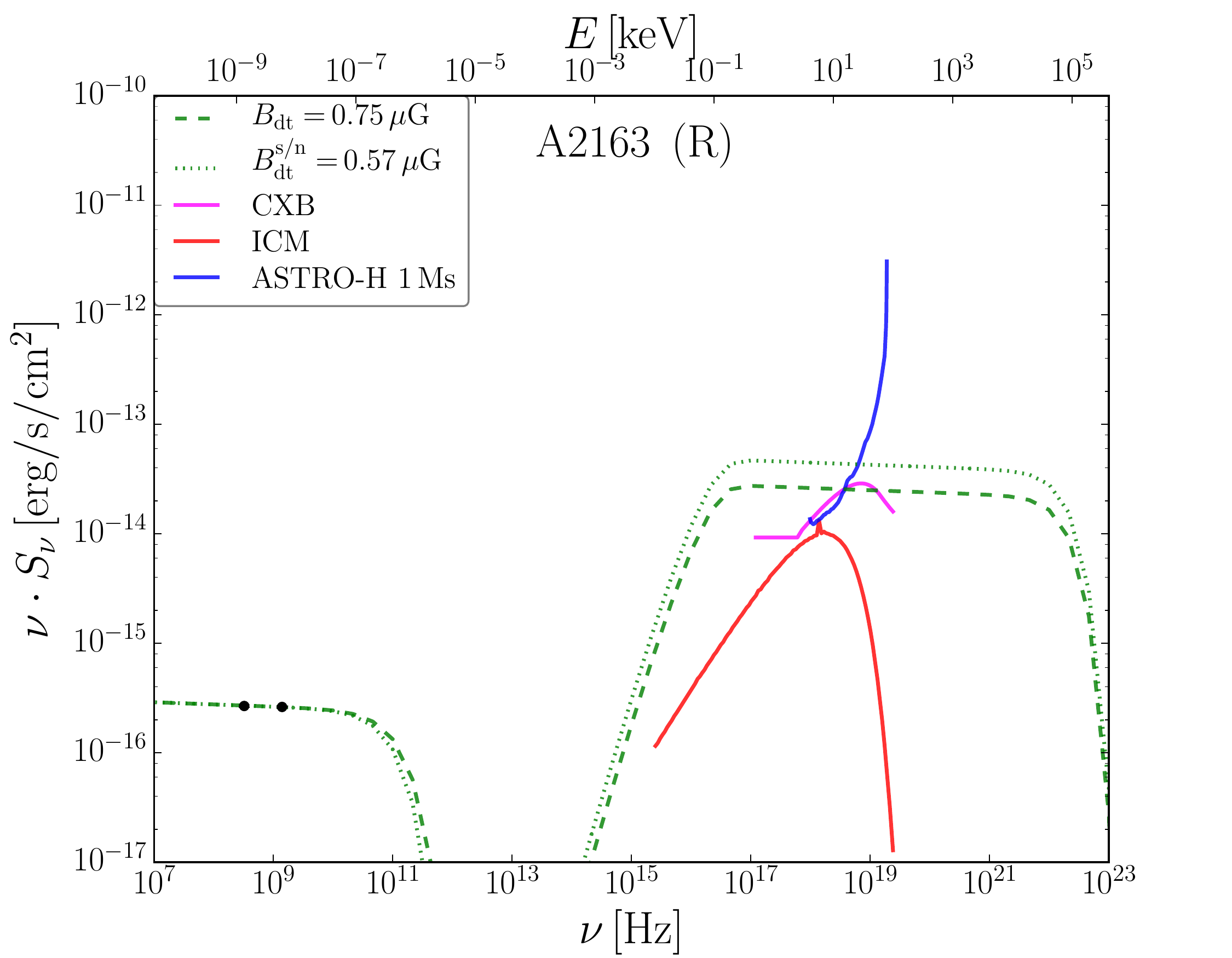}
\includegraphics[width=0.5\textwidth]{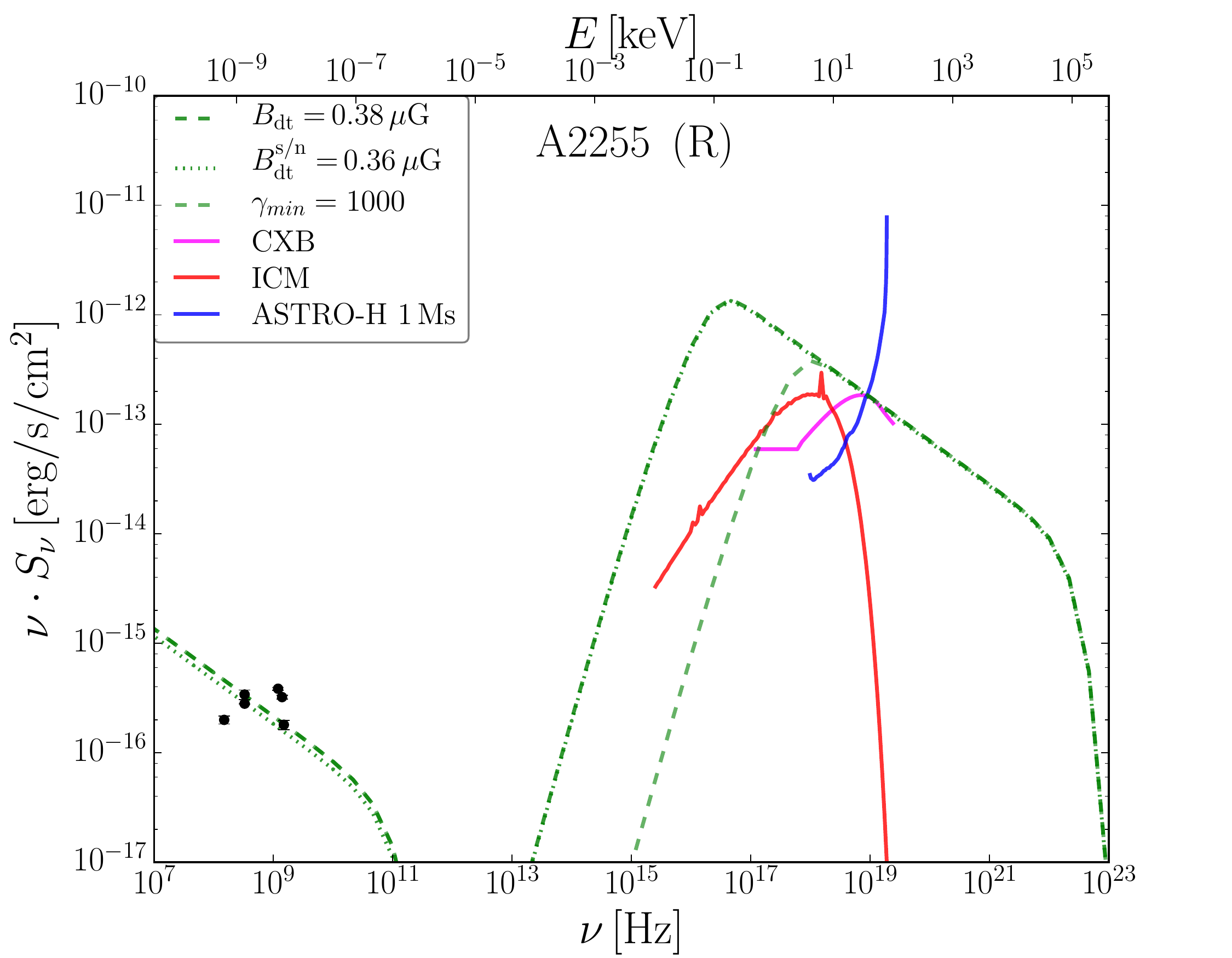}
\includegraphics[width=0.5\textwidth]{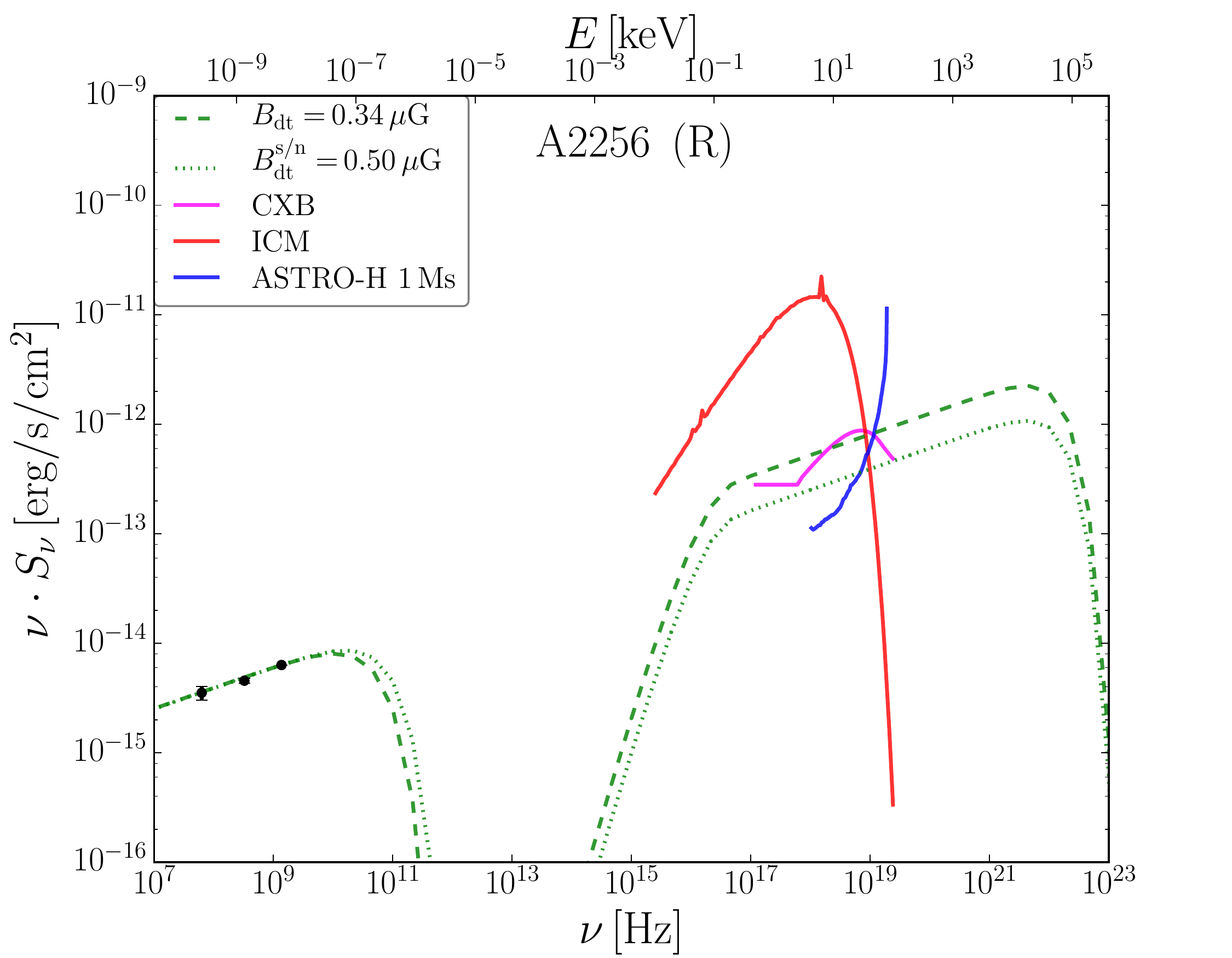}
\end{figure*}
\begin{figure*}[hbt!]
\includegraphics[width=0.5\textwidth]{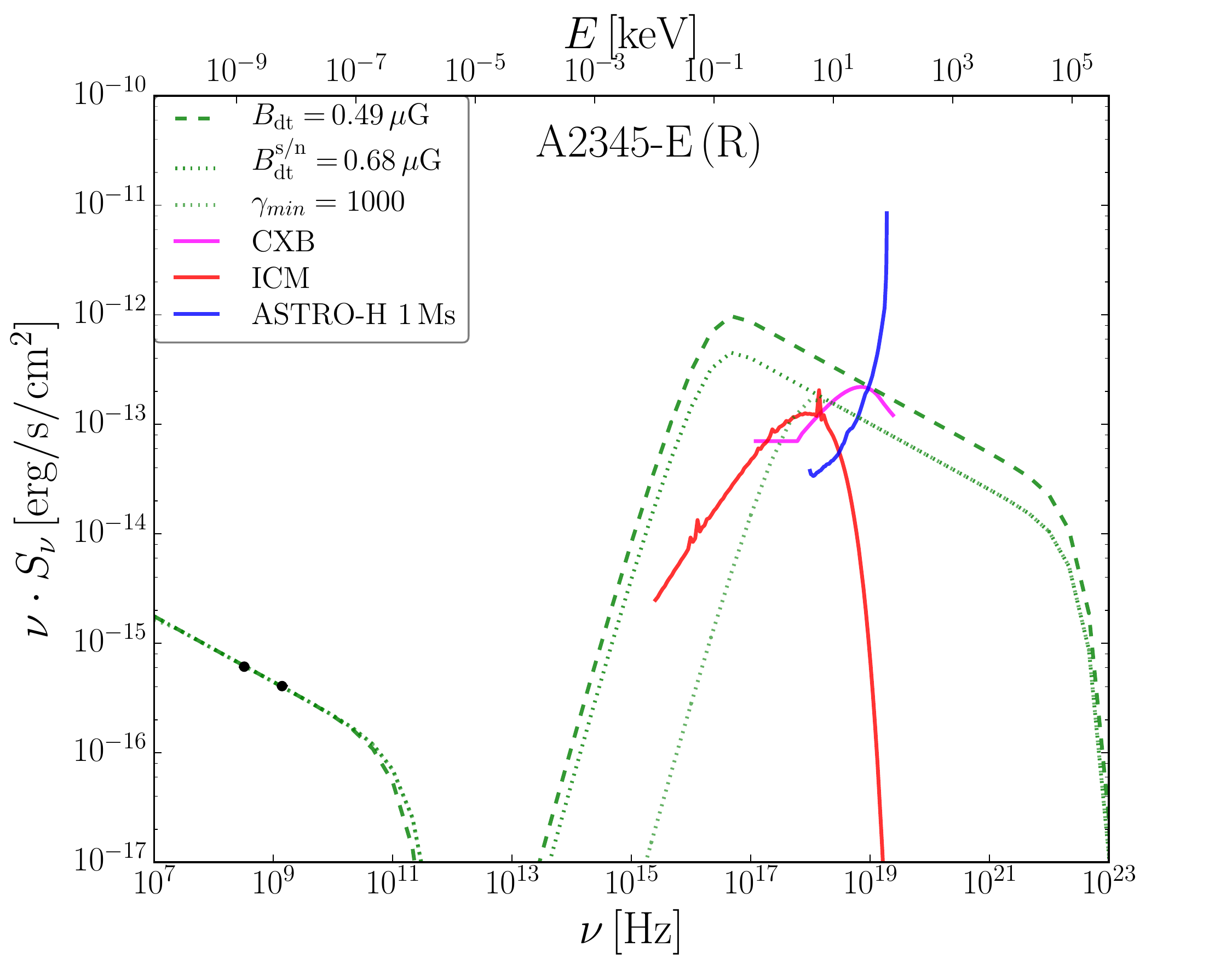}
\includegraphics[width=0.5\textwidth]{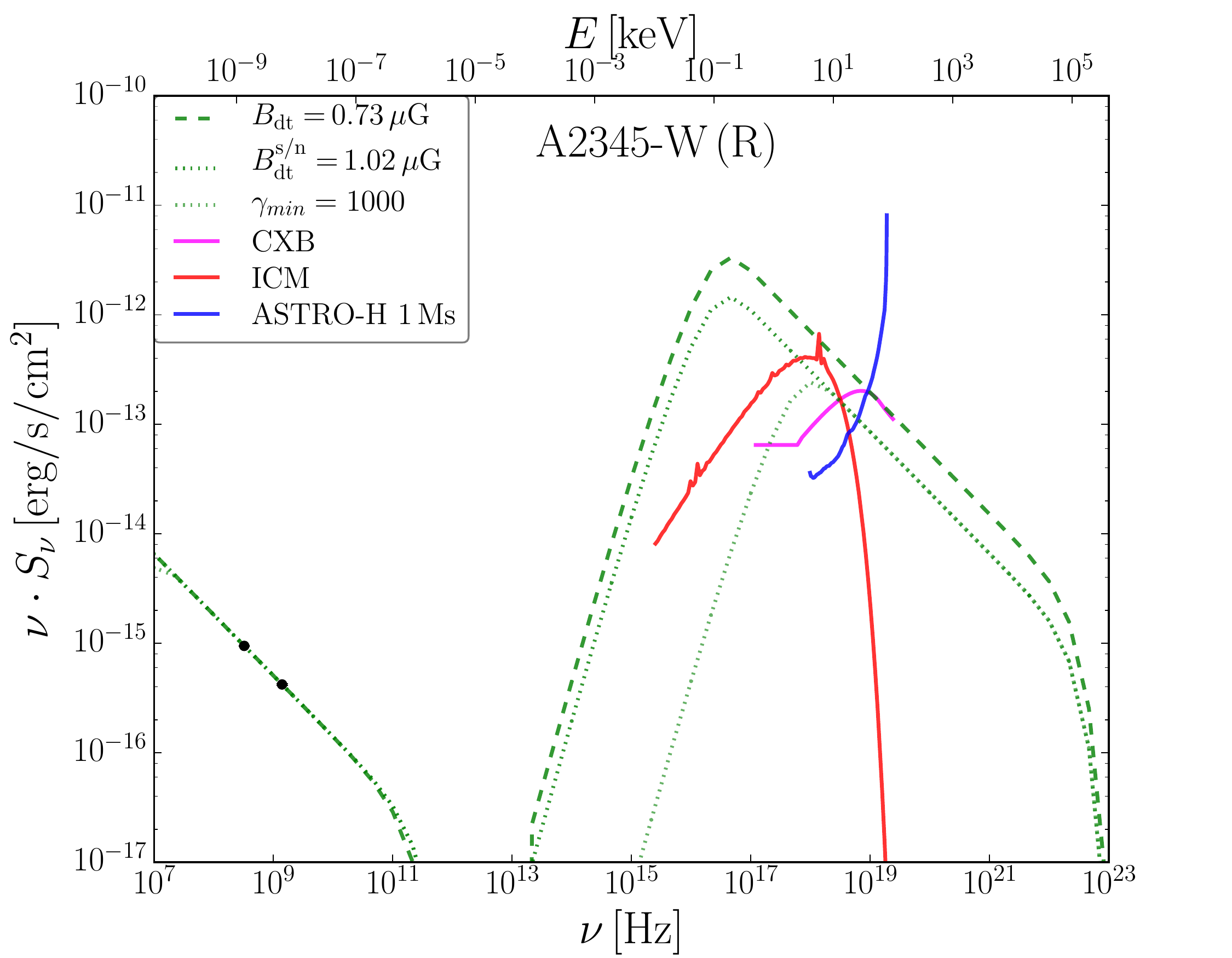}
\includegraphics[width=0.5\textwidth]{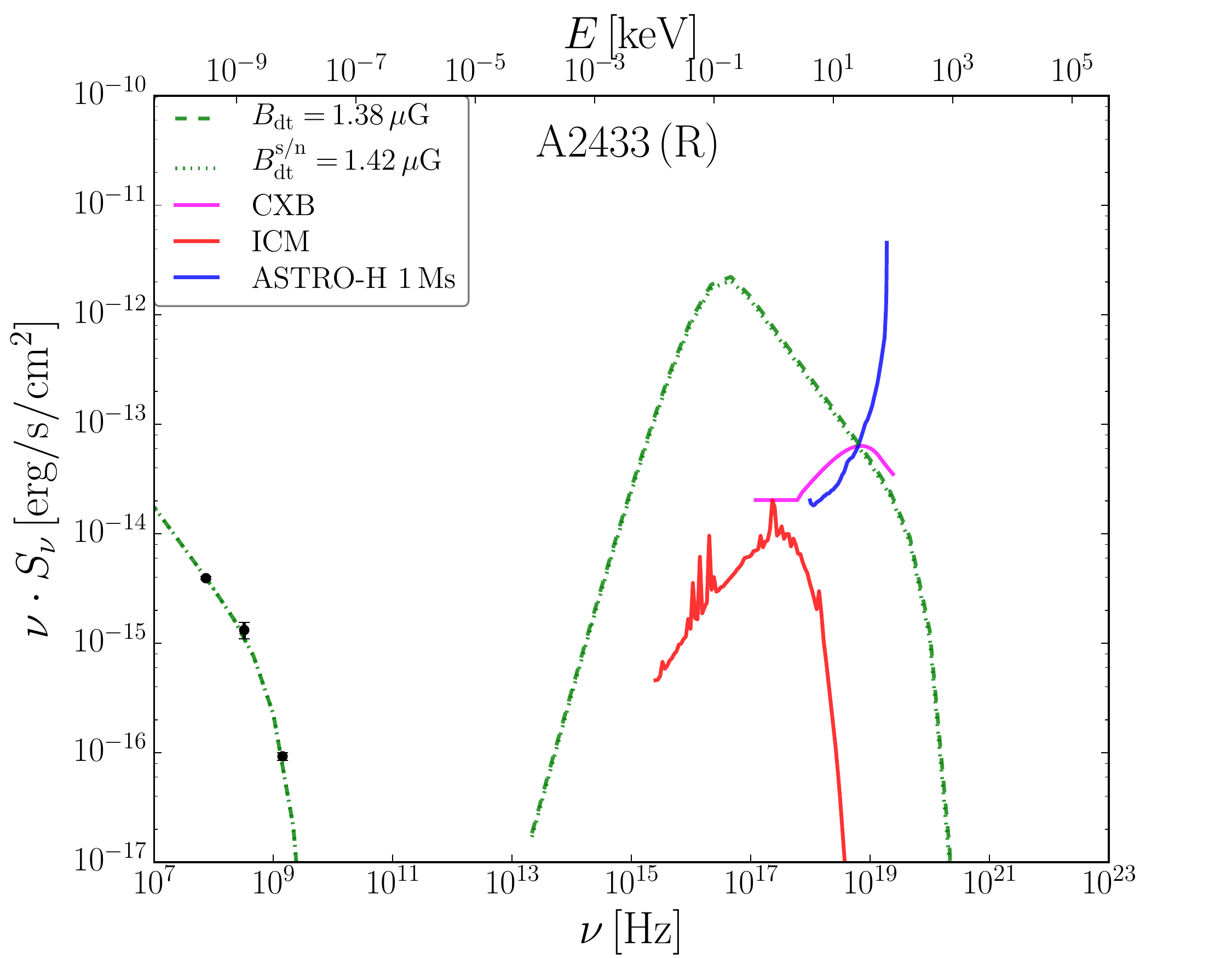}
\includegraphics[width=0.5\textwidth]{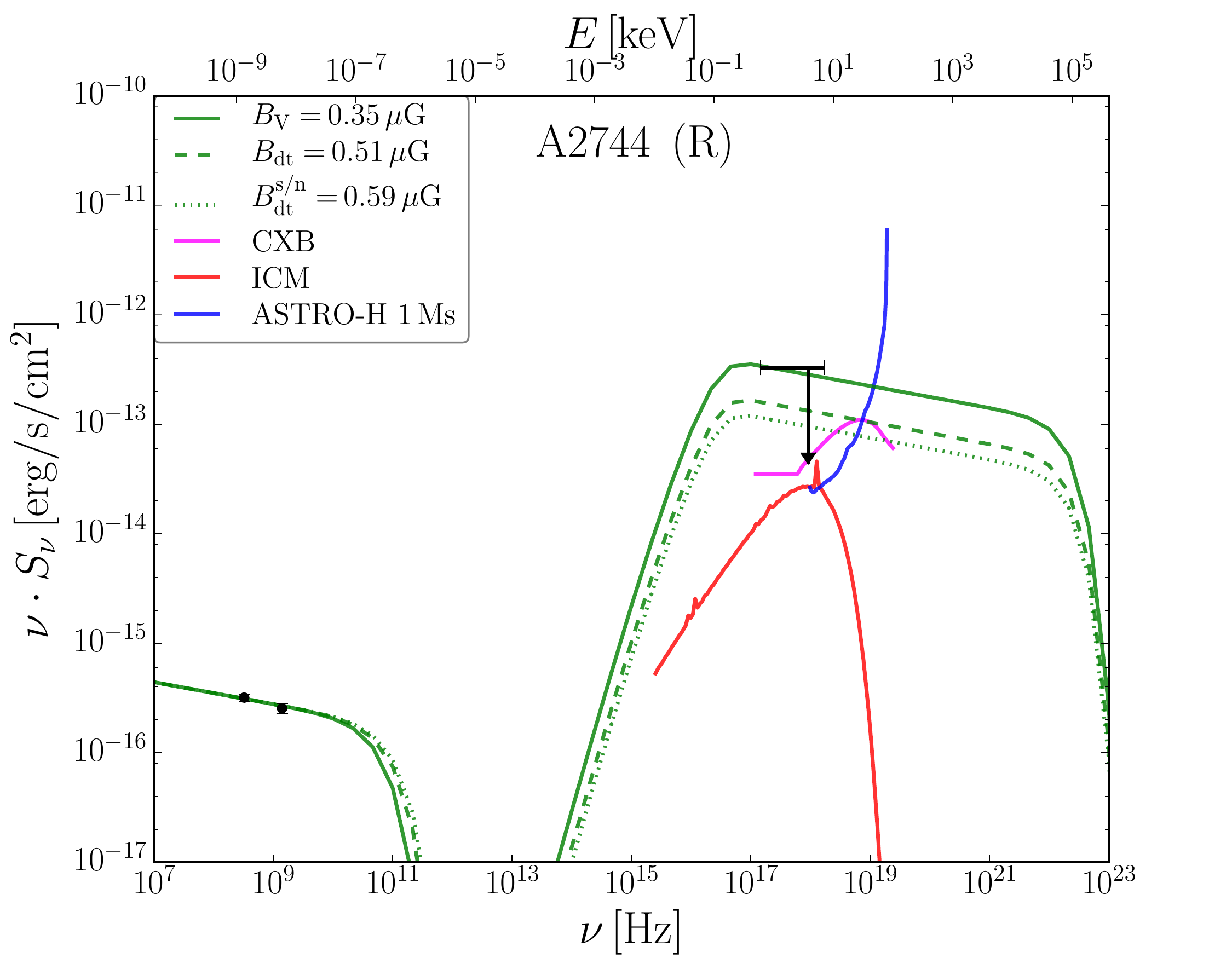}
\includegraphics[width=0.5\textwidth]{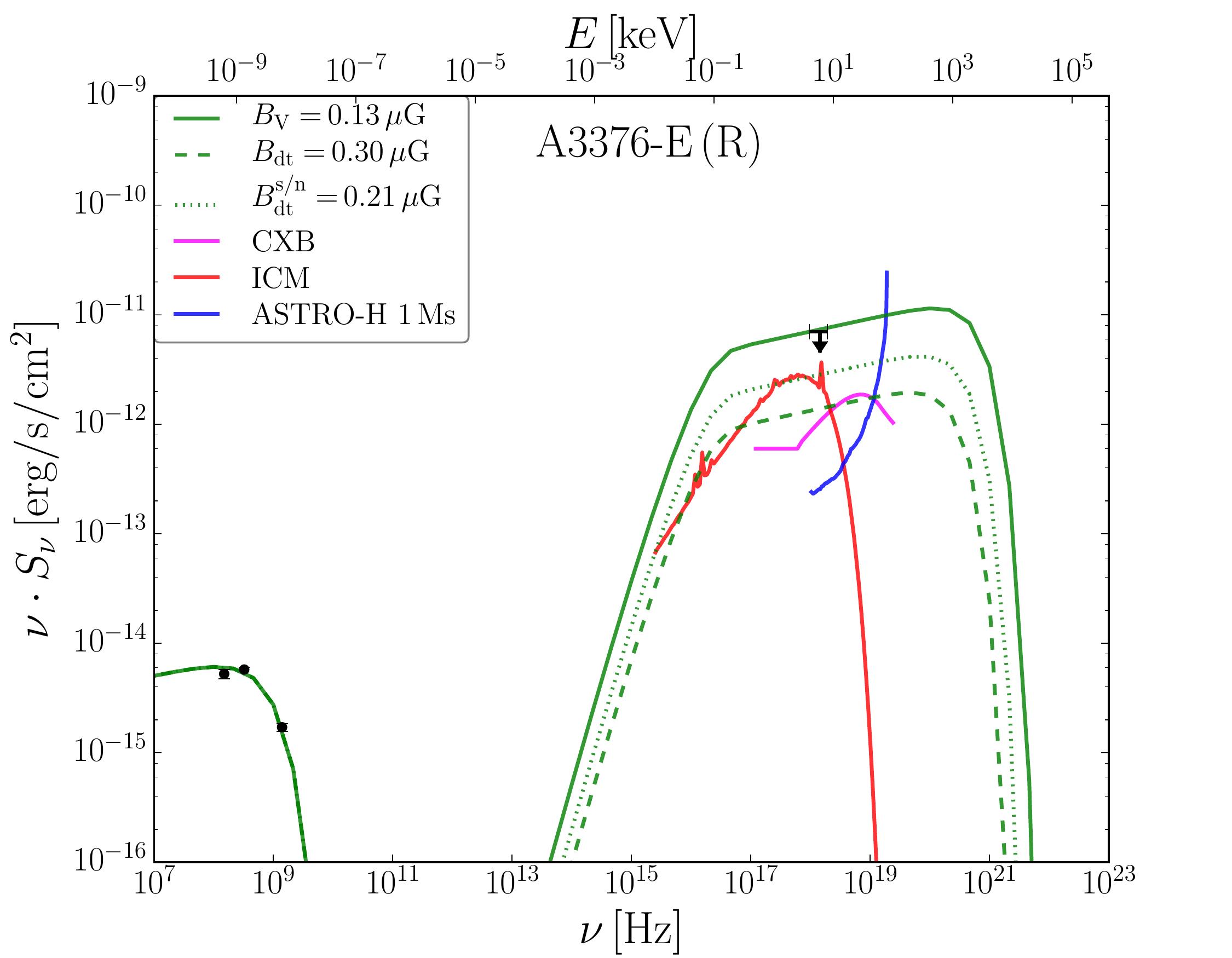}
\includegraphics[width=0.5\textwidth]{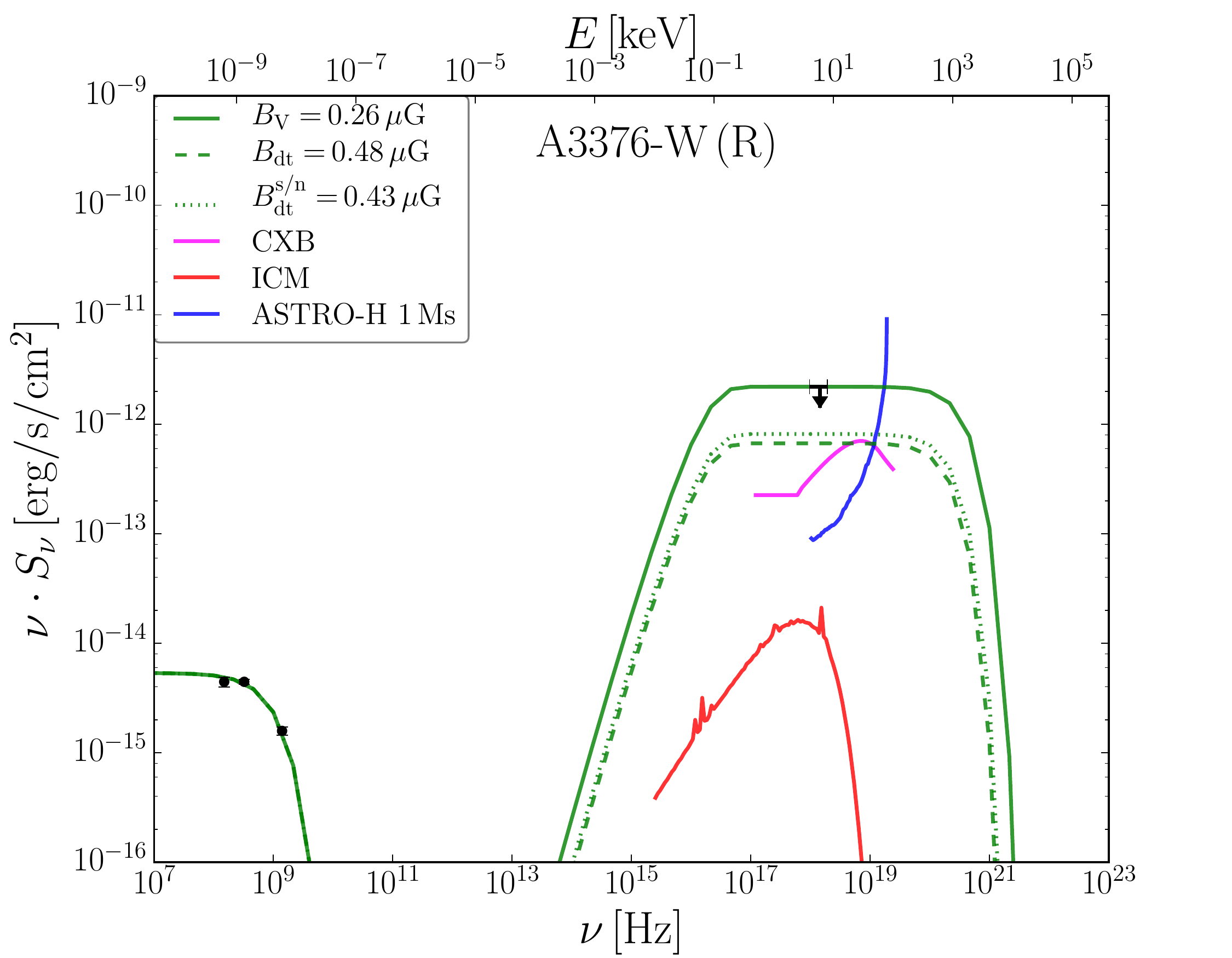}
\end{figure*}
\begin{figure*}[hbt!]
\includegraphics[width=0.5\textwidth]{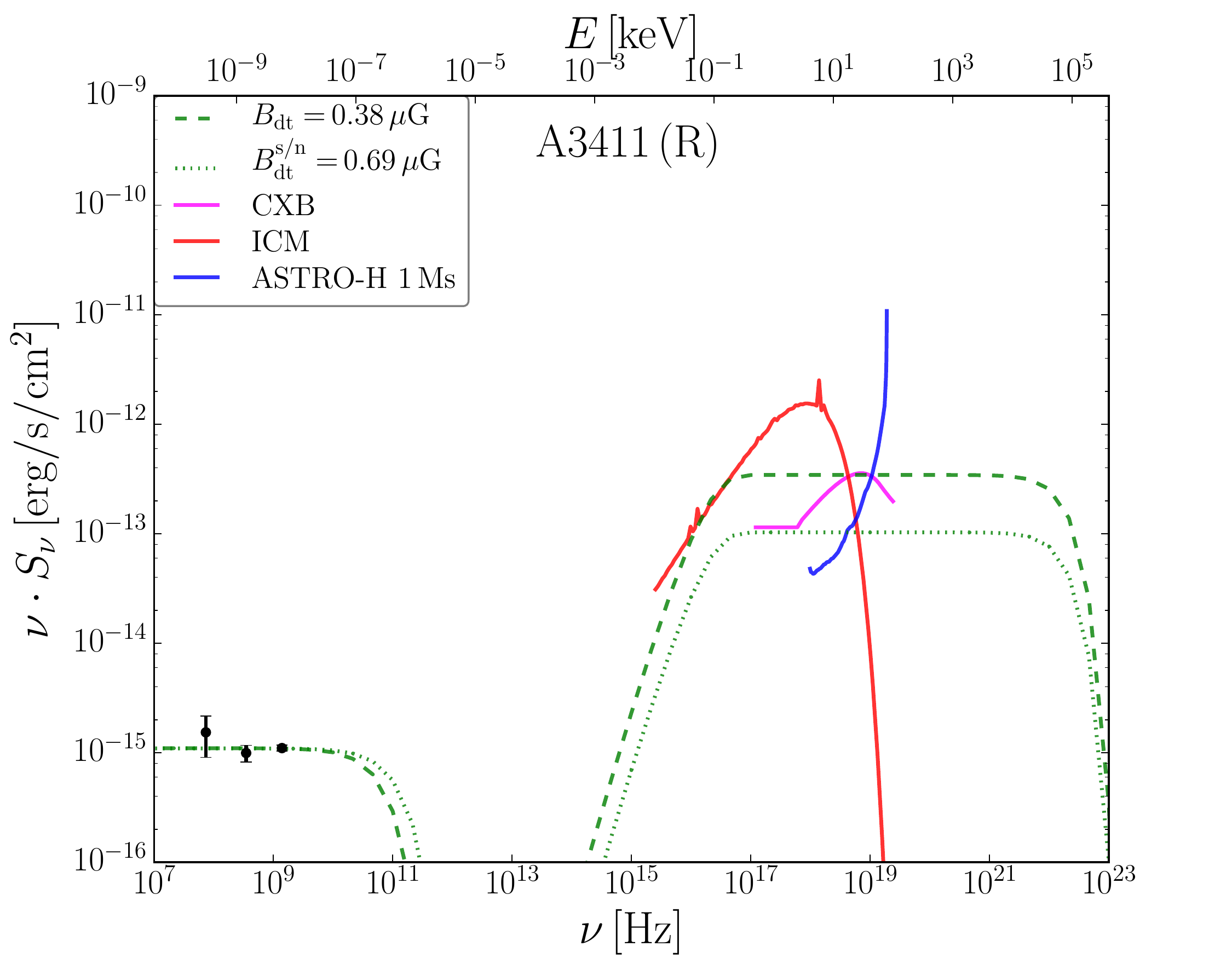}
\includegraphics[width=0.5\textwidth]{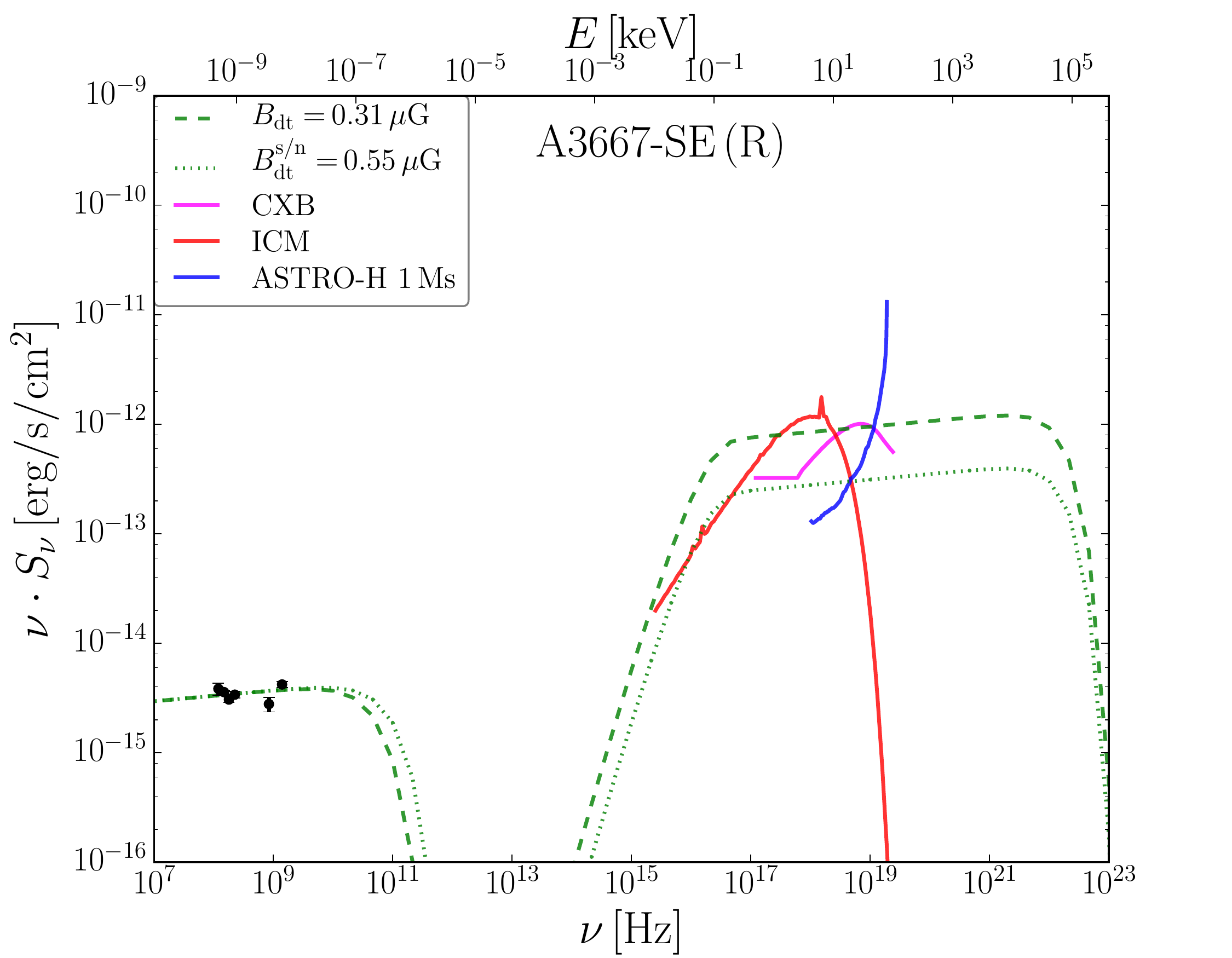}
\includegraphics[width=0.5\textwidth]{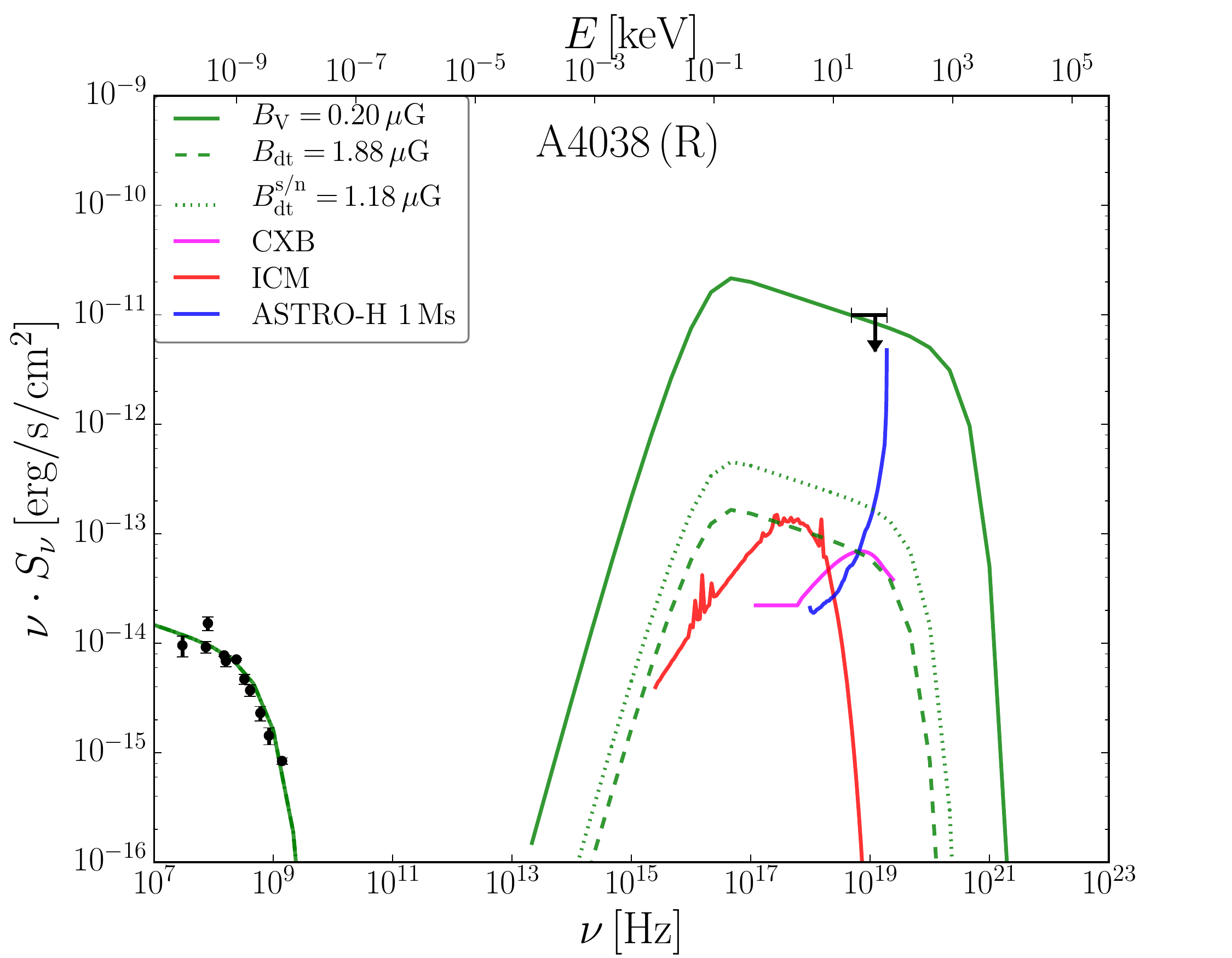}
\includegraphics[width=0.5\textwidth]{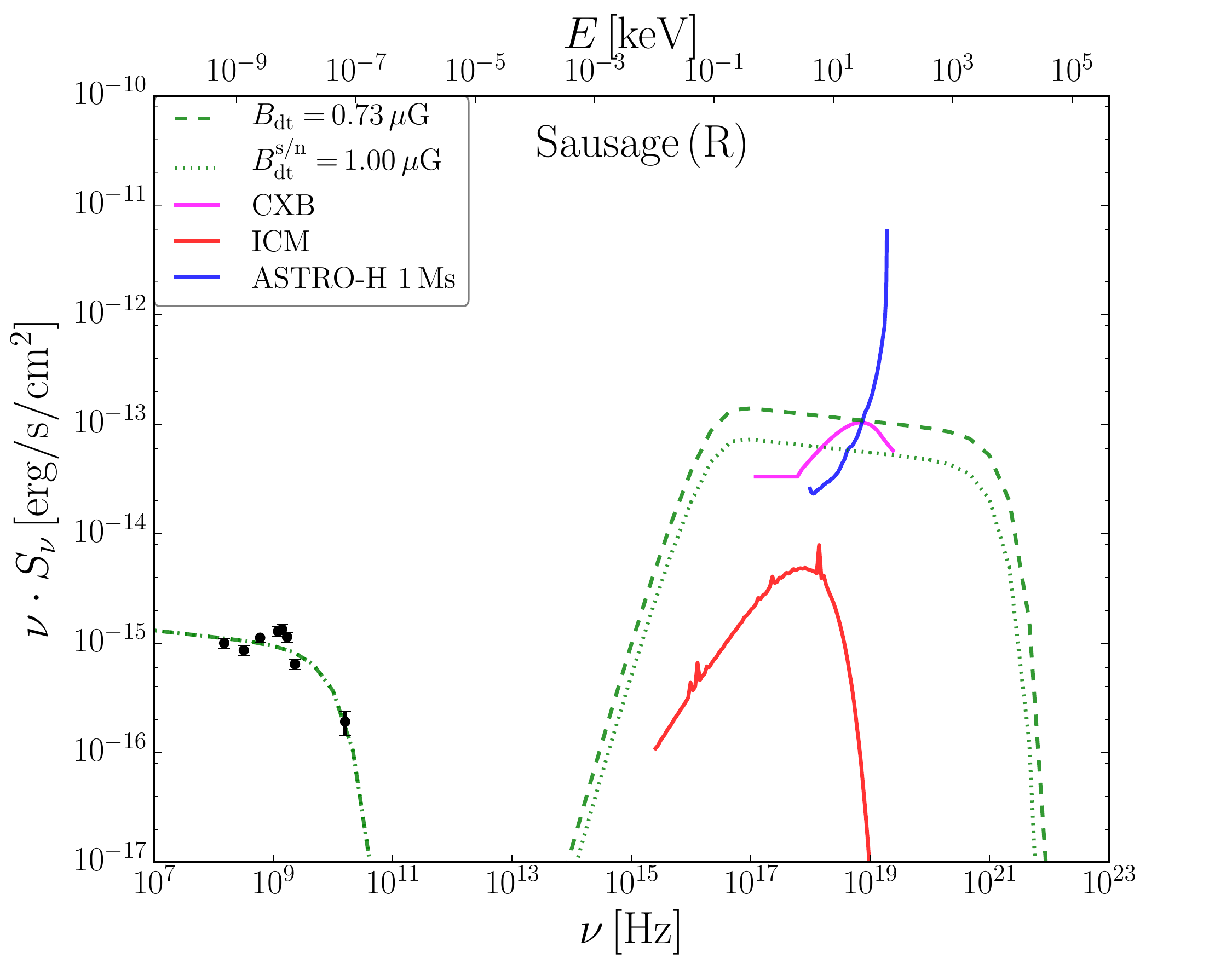}
\includegraphics[width=0.5\textwidth]{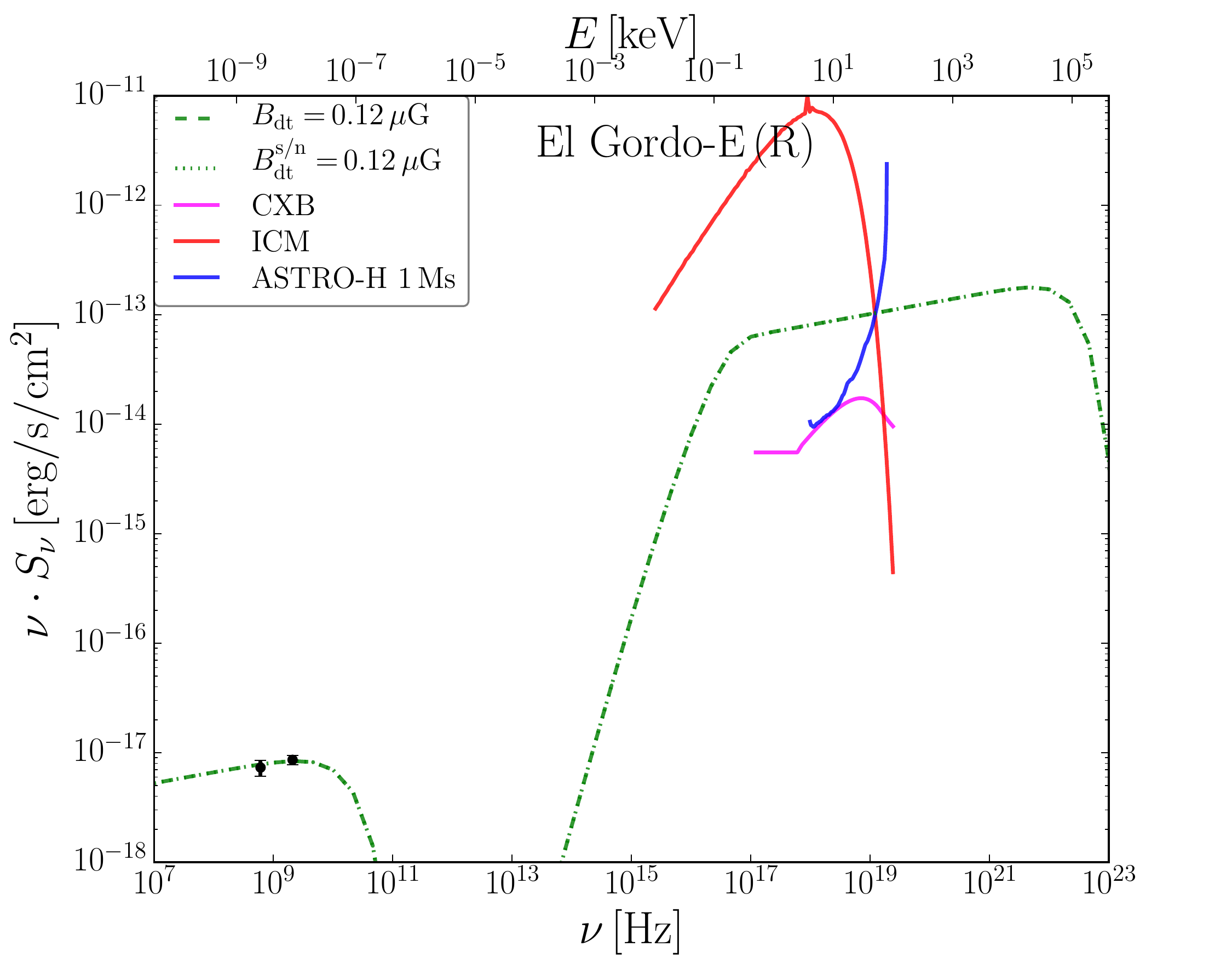}
\includegraphics[width=0.5\textwidth]{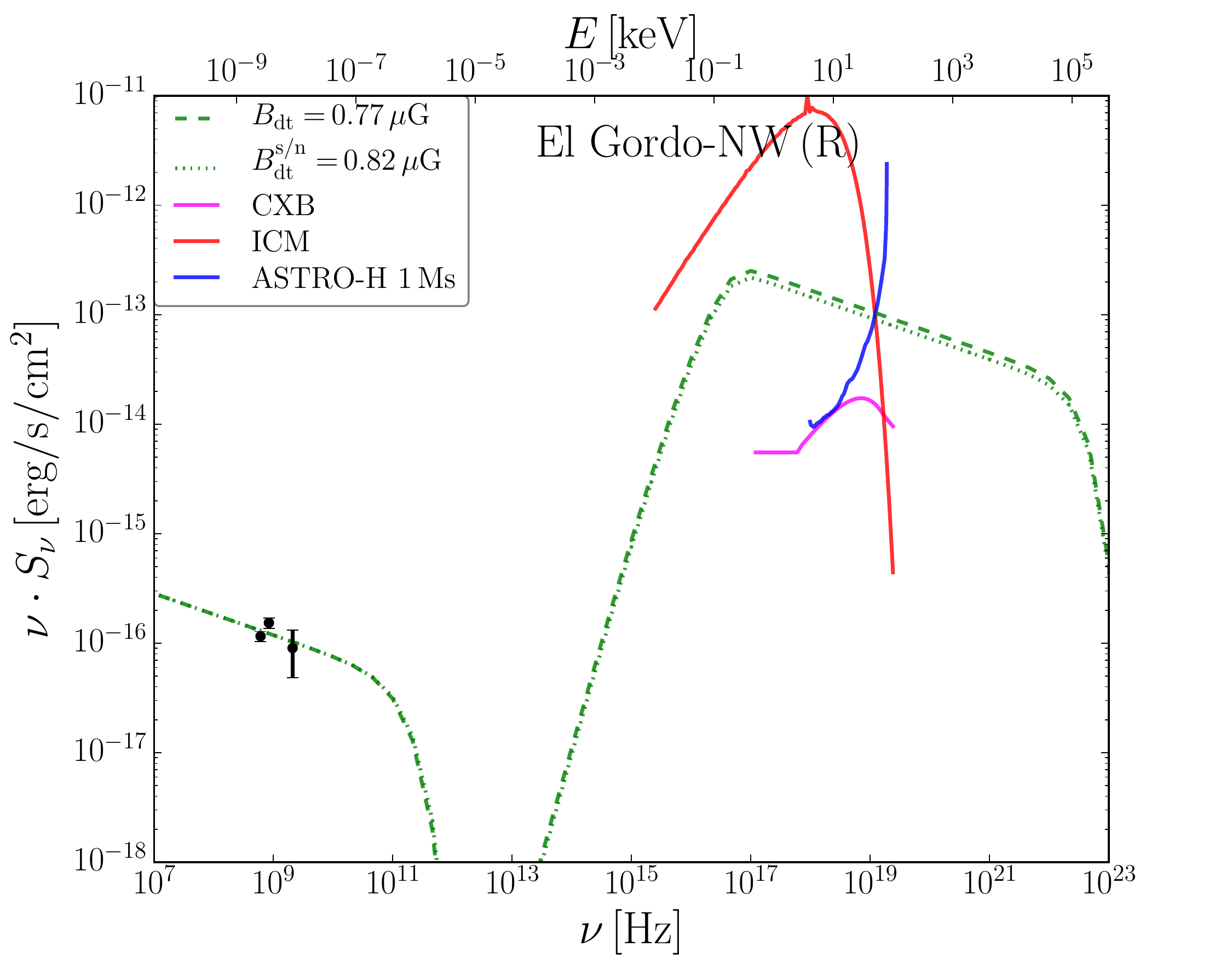}
\end{figure*}
\begin{figure*}[hbt!]
\includegraphics[width=0.5\textwidth]{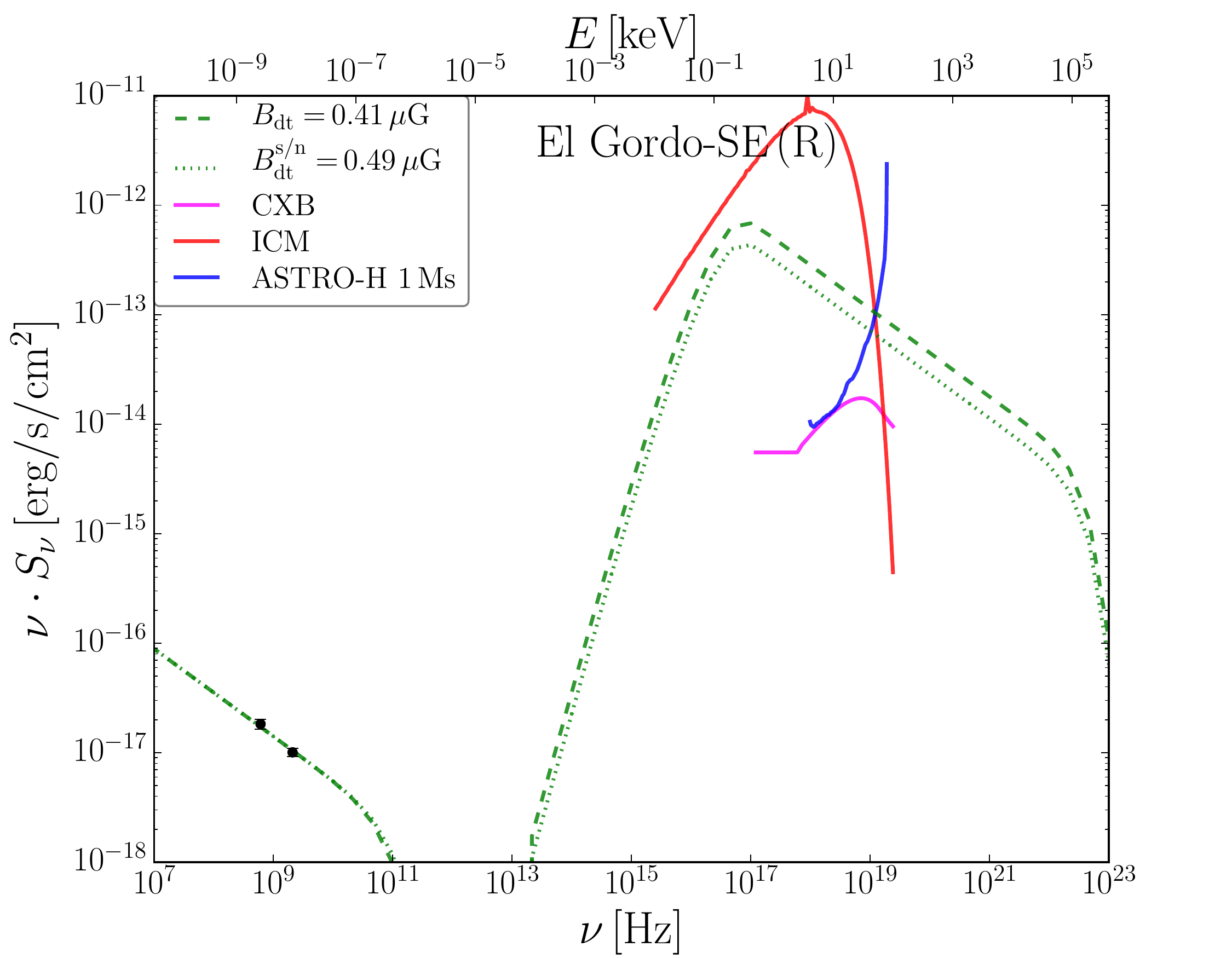}
\includegraphics[width=0.5\textwidth]{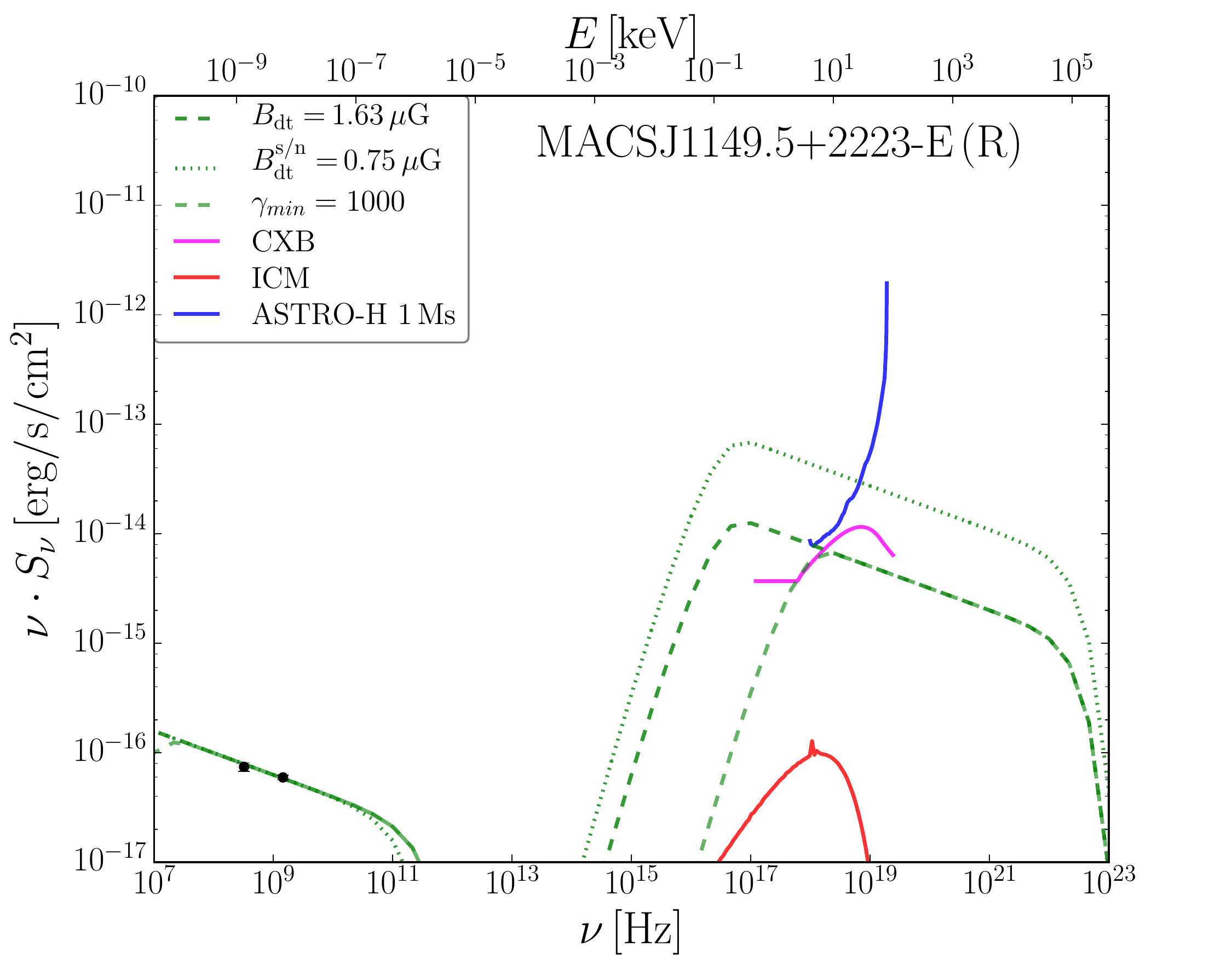}
\includegraphics[width=0.5\textwidth]{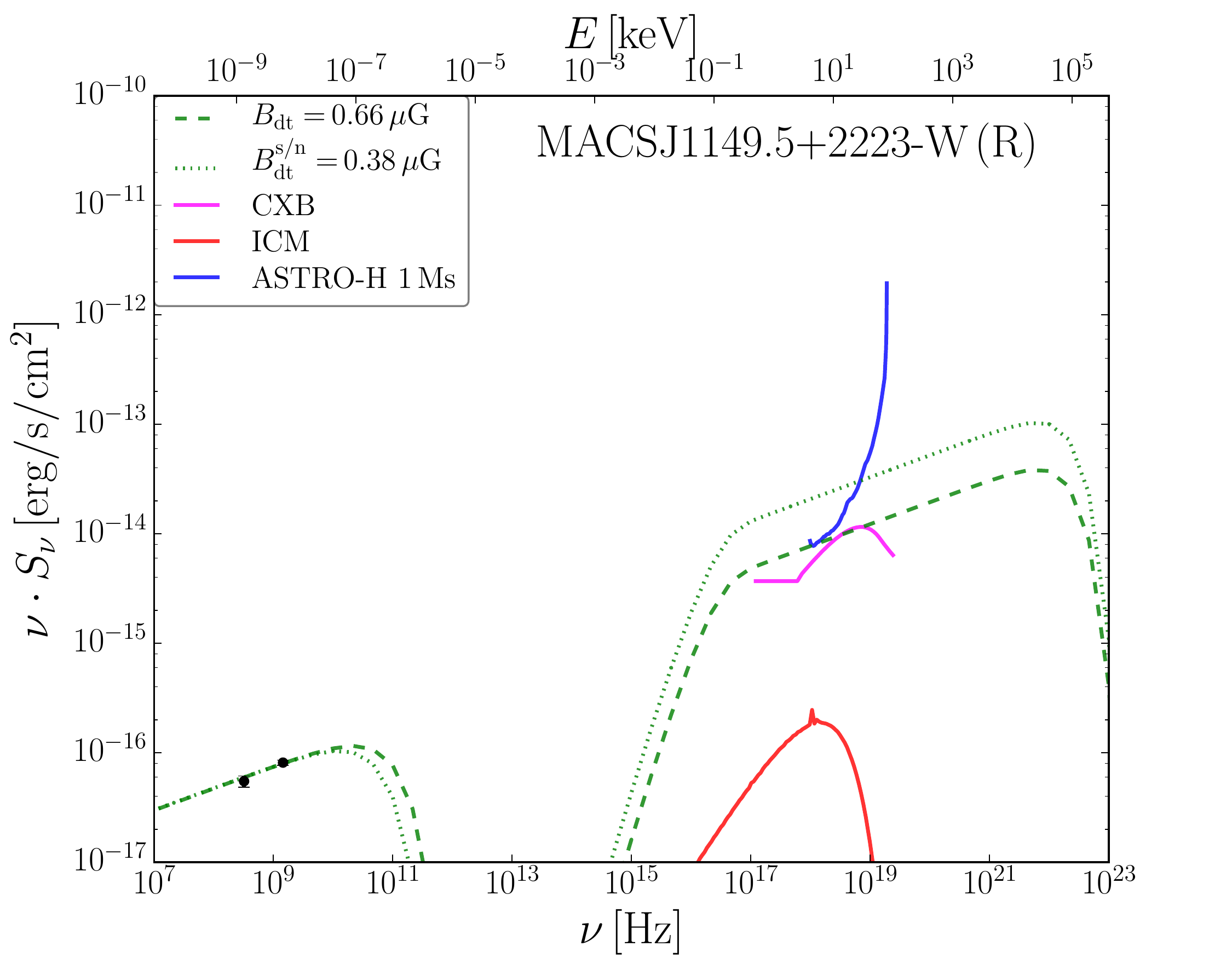}
\includegraphics[width=0.5\textwidth]{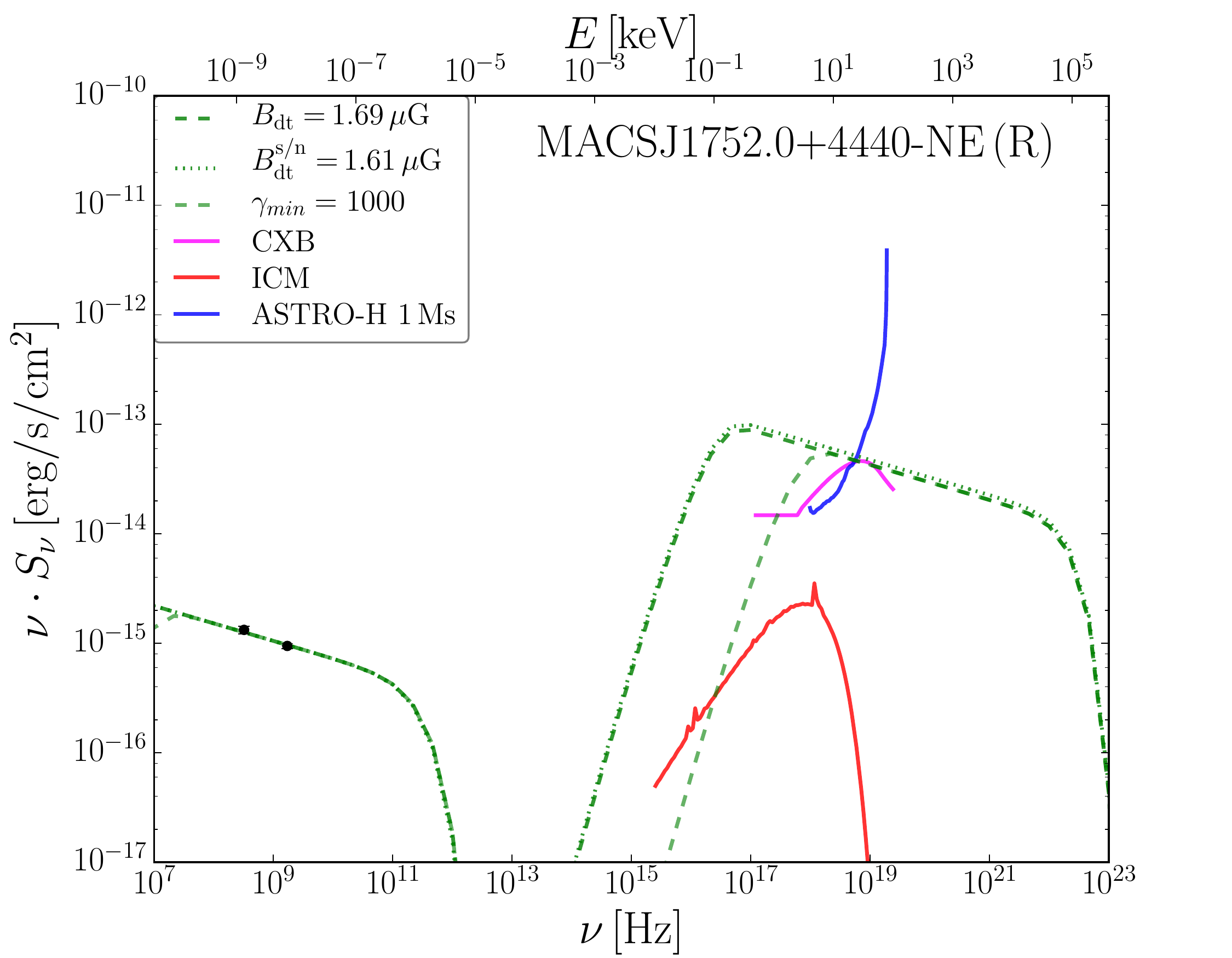}
\includegraphics[width=0.5\textwidth]{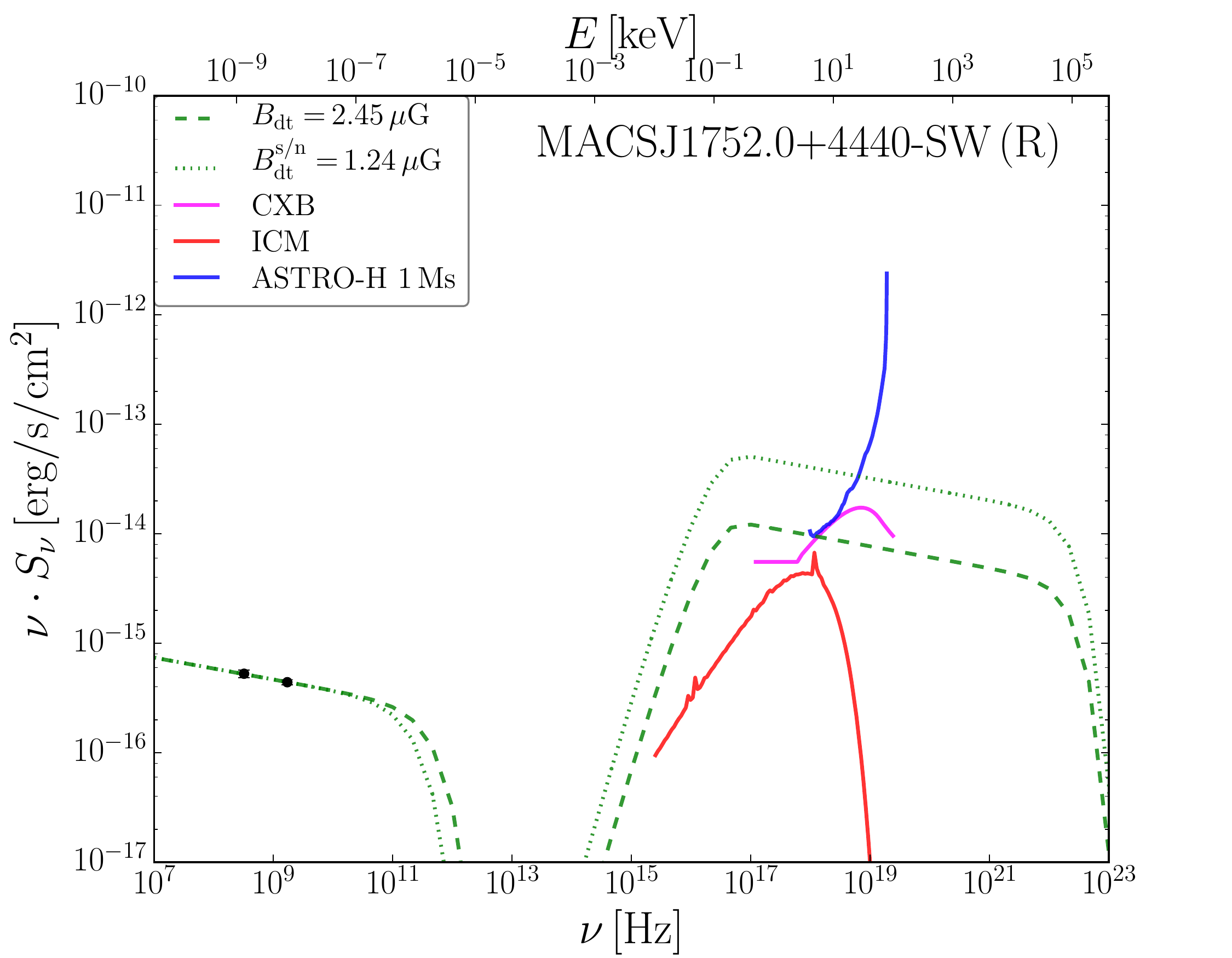}
\includegraphics[width=0.5\textwidth]{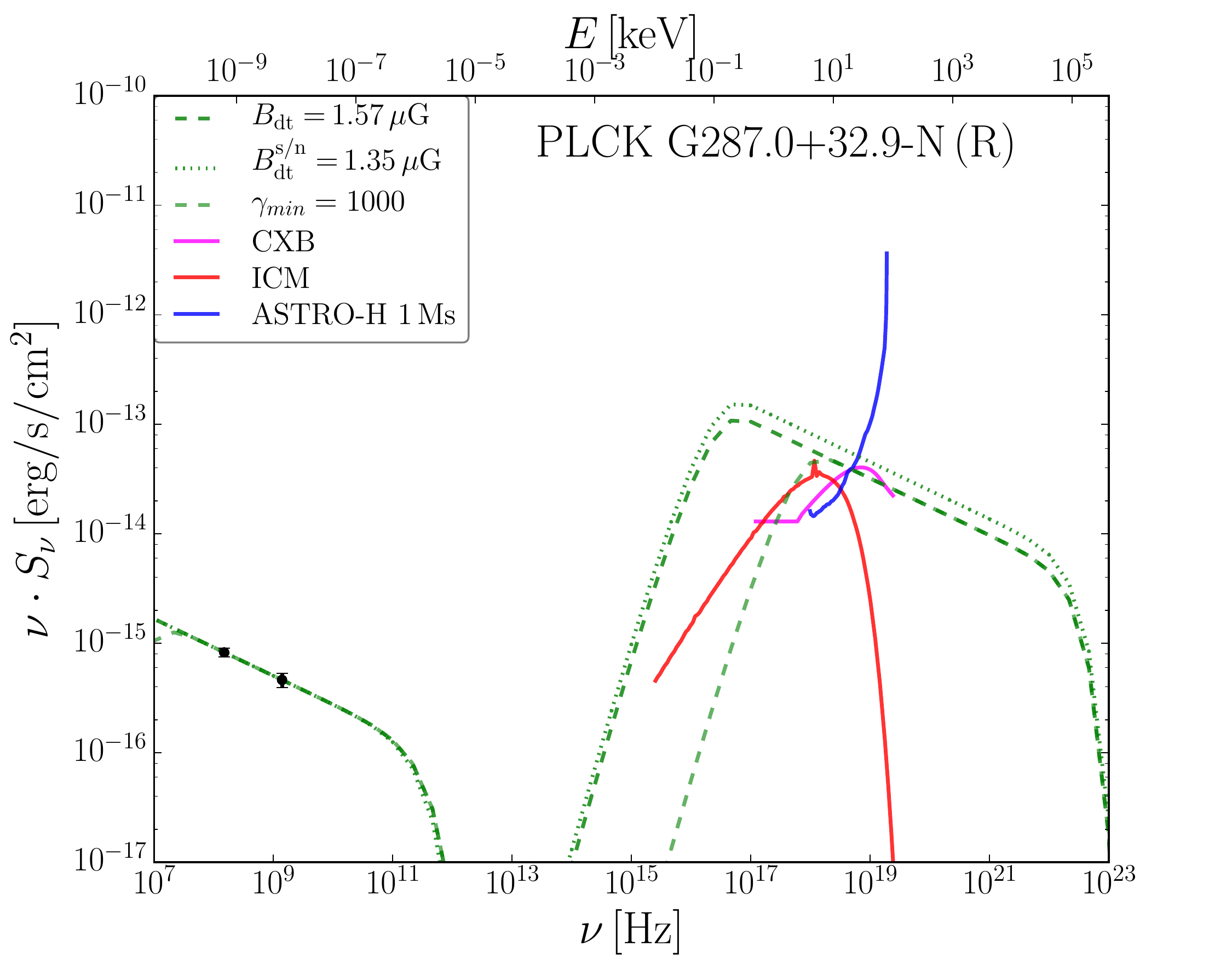}
\end{figure*}
\begin{figure*}[hbt!]
\includegraphics[width=0.5\textwidth]{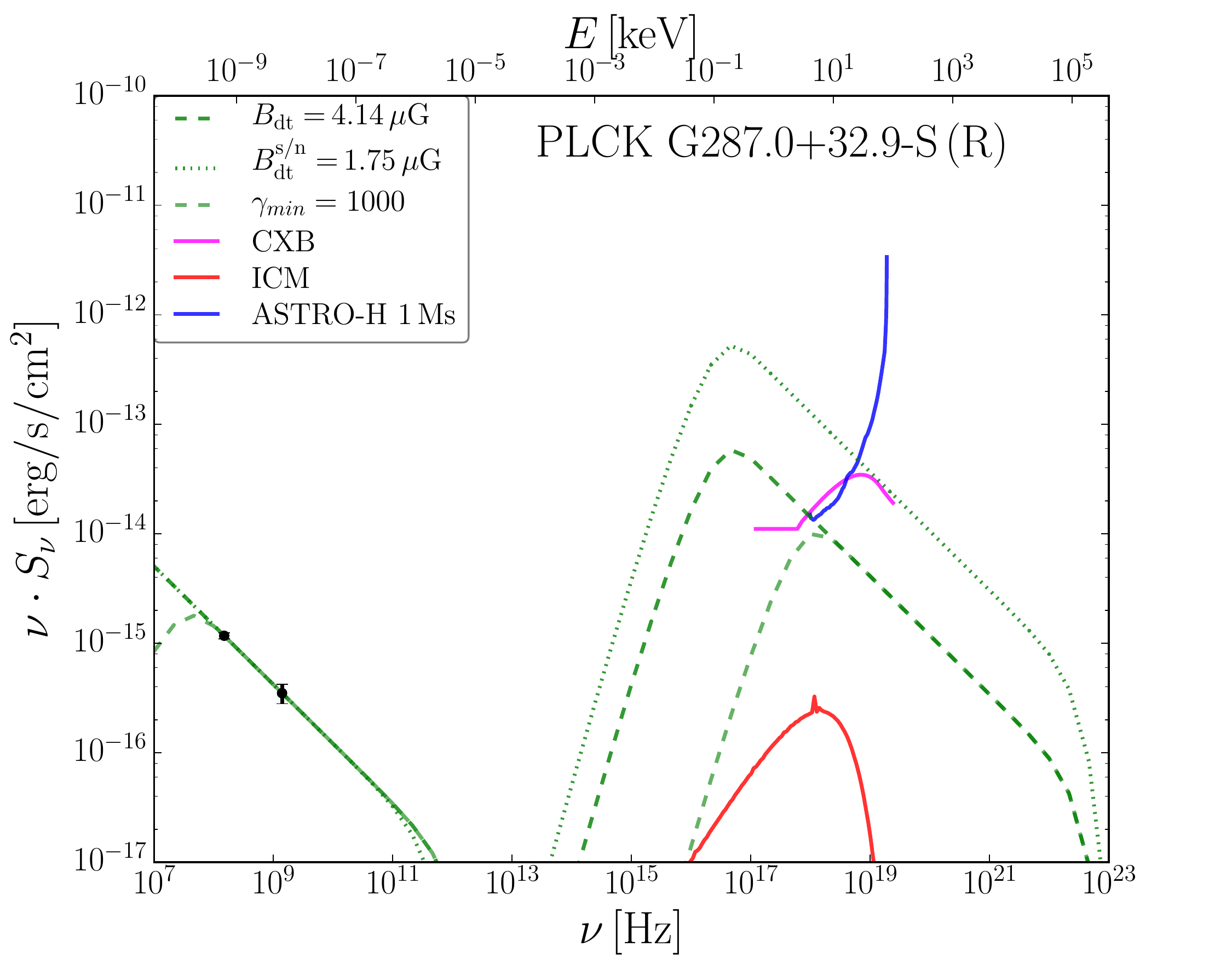}
\includegraphics[width=0.5\textwidth]{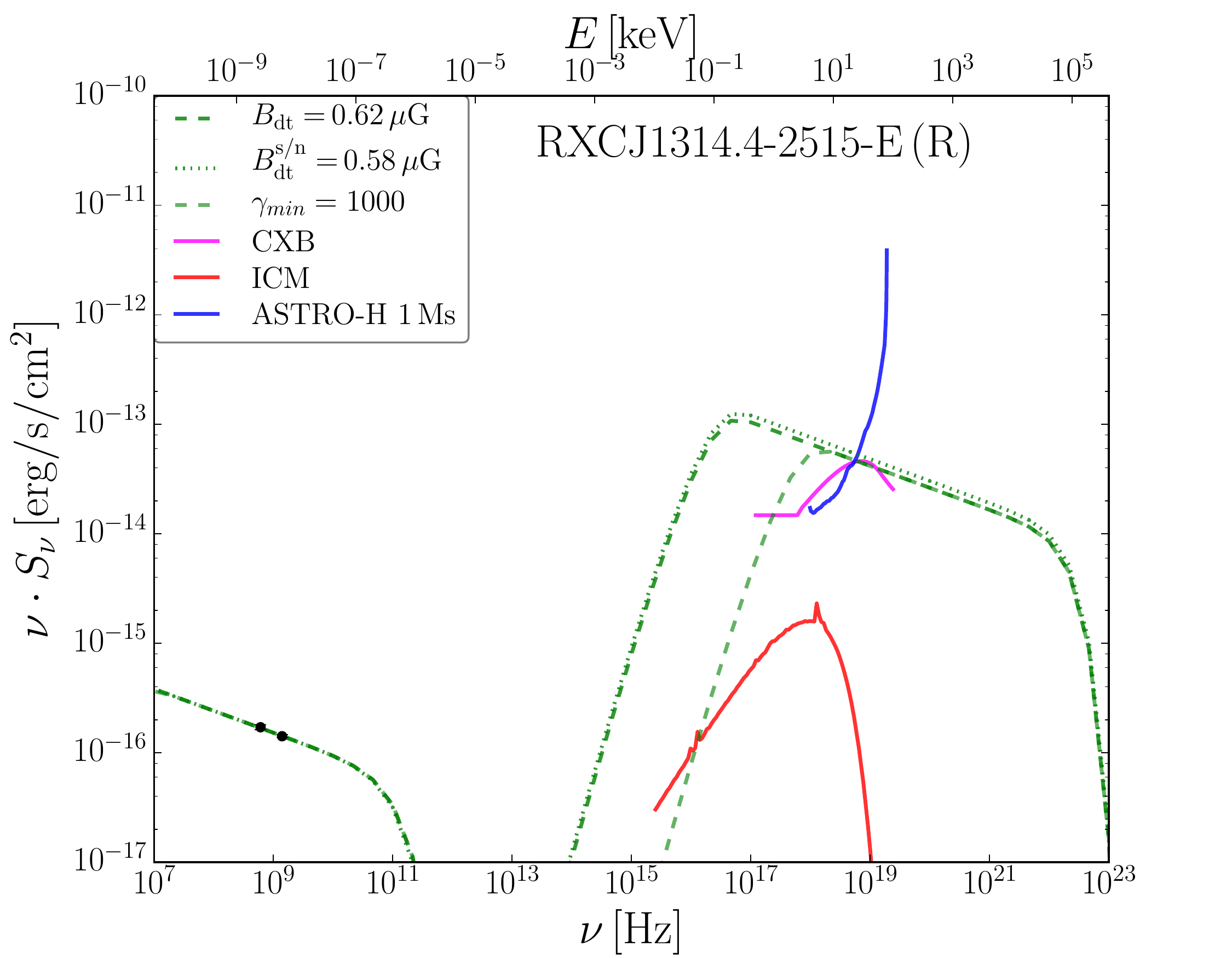}	
\includegraphics[width=0.5\textwidth]{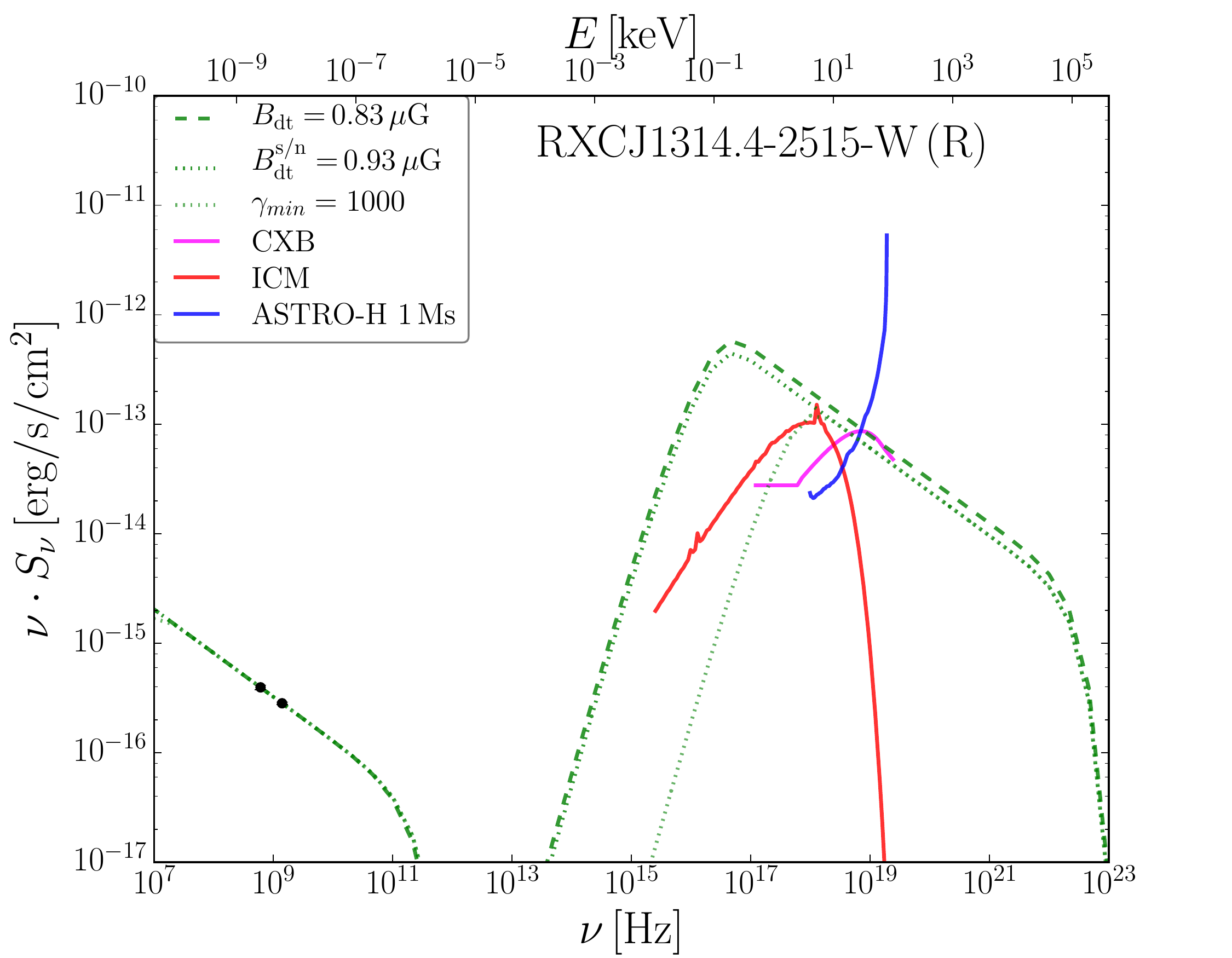}
\includegraphics[width=0.5\textwidth]{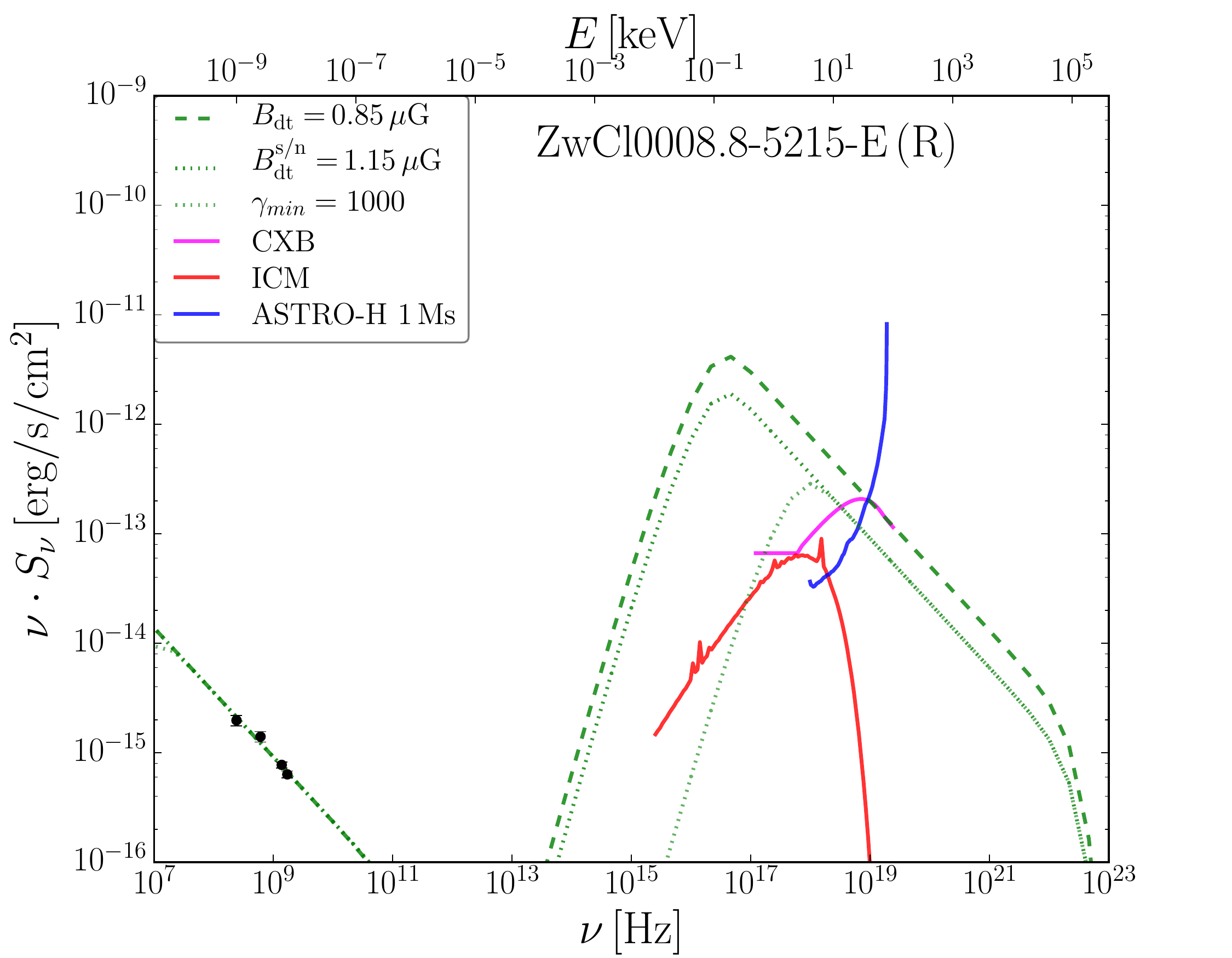}
\includegraphics[width=0.5\textwidth]{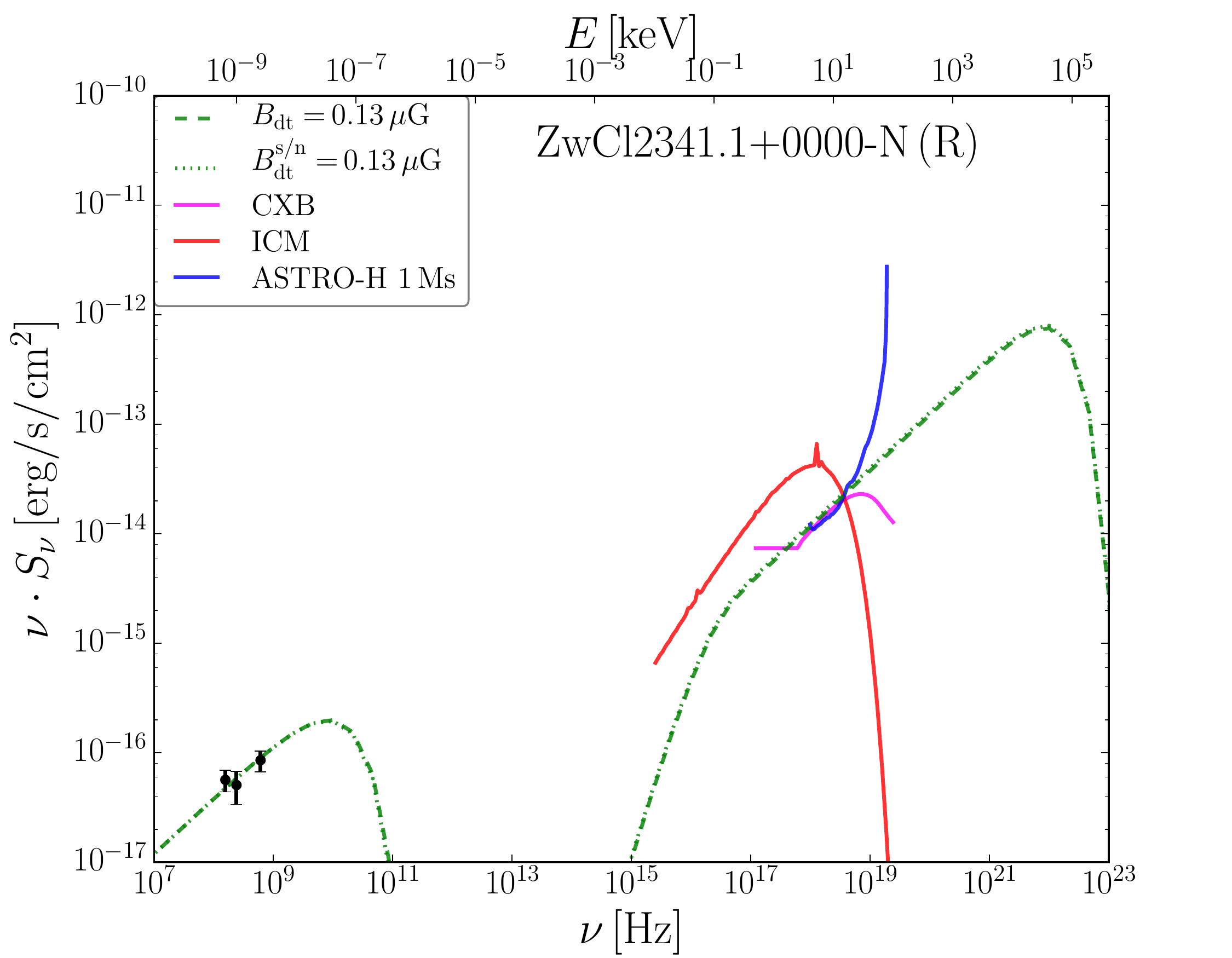}
\includegraphics[width=0.5\textwidth]{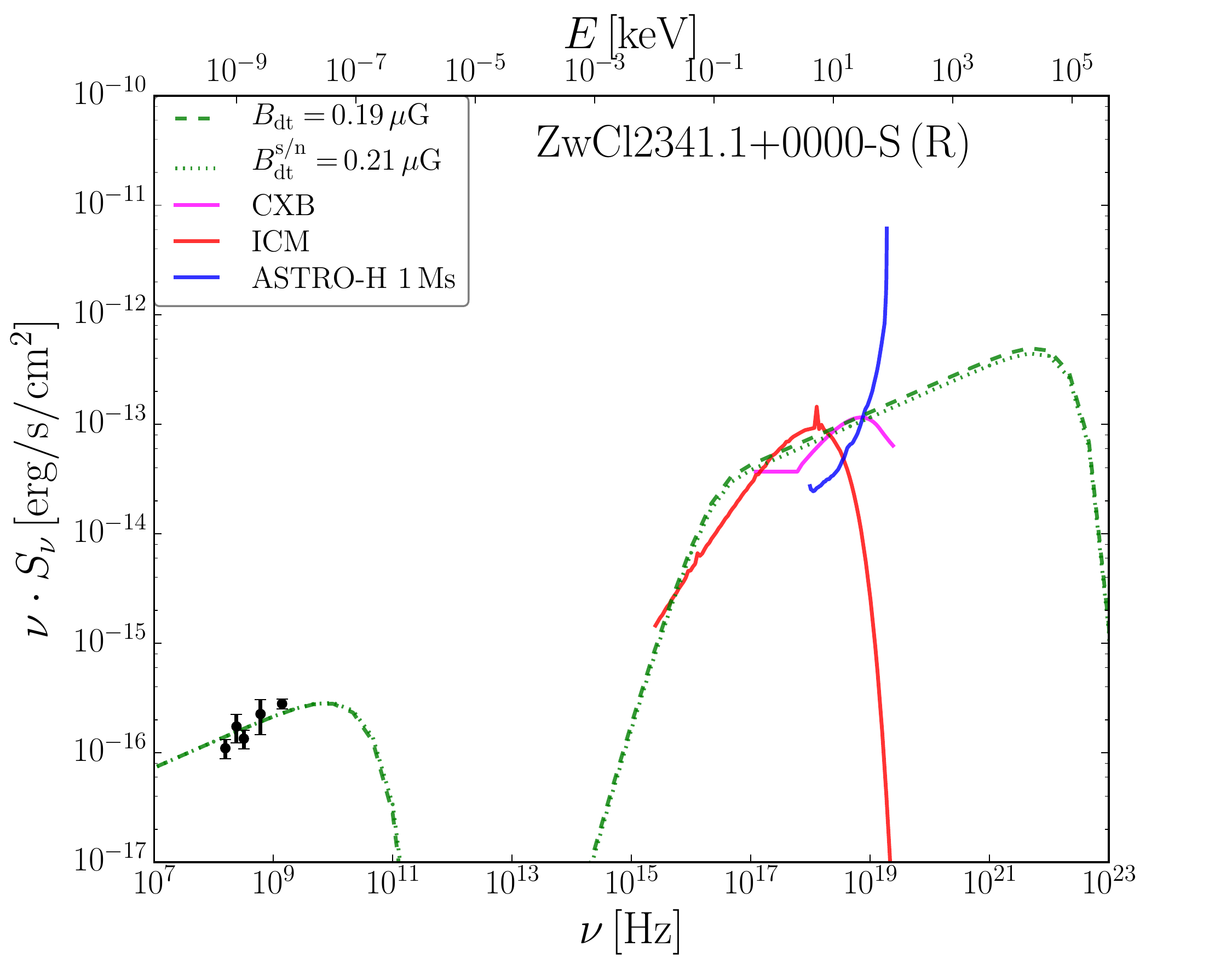}
\end{figure*}

\end{appendix}

\end{document}